\RequirePackage{fix-cm}
\documentclass[smallextended]{svjour3}       % onecolumn (second format)
\smartqed  % flush right qed marks, e.g. at end of proof
\usepackage{graphicx}
\usepackage{afterpage}
\usepackage{array}
\usepackage{booktabs}
\usepackage[sort,compress]{natbib}
\usepackage{enumitem}
\usepackage{float}
\usepackage{listings}
\usepackage{longtable}
\usepackage{multirow}
\usepackage{graphicx}
\usepackage{tcolorbox}
\usepackage{xcolor}
\usepackage{xspace}
\usepackage{lmodern}
\usepackage{xurl}
\usepackage{hyperref}
\usepackage[normalem]{ulem}
\usepackage{amsmath}
\usepackage{subfig}
\usepackage{pgfplots}
\usepackage{tabularx}
\usepackage{booktabs} % For better table formatting
\usepackage{adjustbox} % For adjusting table to fit within page width
\usepackage{pdflscape}
\usepackage{mdframed}
\usepackage{tcolorbox}
\usepackage[table]{xcolor}
\usepackage{longtable}
\usepackage{tikz}
\usepackage[T1]{fontenc}

\tcbuselibrary{skins}

\newcolumntype{C}[1]{>{\centering\arraybackslash}p{#1}}

\newcommand{\supblue}[1]{\textsuperscript{\textcolor{blue}{\textbf{#1}}}}

\newcommand{\rpv}[1]{\textcolor{brown}{remove passive voice}}

\DeclareUnicodeCharacter{202F}{\,}
\extrafloats{500}

\makeatletter
\newcommand{\ie}{\emph{i.e.}\@ifnextchar.{\!\@gobble}{}}
\newcommand{\eg}{\emph{e.g.}\@ifnextchar.{\!\@gobble}{}}
\newcommand{\etc}{etc\@ifnextchar.{}{.\@}}
\makeatother

\newcommand{\secref}[1]{Section~\ref{#1}}
\newcommand{\tabref}[1]{Table~\ref{#1}}

\newcommand{\component}[1]{\mbox{\texttt{\detokenize{#1}}}}
\newcommand{\pattern}[1]{\textit{#1}}

\definecolor{linehl}{gray}{0.90}
\definecolor{hlyellow}{RGB}{255,245,170}

\lstdefinestyle{emsepython}{
    language=Python,
    basicstyle=\ttfamily\small,
    numbers=left,
    numberstyle=\tiny\color{gray},
    stepnumber=1,
    numbersep=6pt,
    breaklines=true,
    breakatwhitespace=false,
    showstringspaces=false,
    columns=fullflexible,
    keepspaces=true,
    frame=none,
    keywordstyle=\color{blue},
    stringstyle=\color{brown},
    commentstyle=\color{teal},
    emphstyle=\color{black}\bfseries\colorbox{linehl}
}

\newtcolorbox{mybox}[2][]{
top=0.15in,left=4pt,right=4pt,bottom=4pt,
fonttitle=\bfseries,
colbacktitle=gray,
colback=gray!5,
colframe=gray!40!black,
enhanced,
attach boxed title to top left={xshift=0em,yshift=-\tcboxedtitleheight/2},
boxed title style={size=small},
drop shadow={black!50!white},
title=#2,#1}

\newtcolorbox{querybox}{
  colback=white,      % background color
  colframe=black,     % frame color
  boxrule=0.5pt,      % line width
  arc=0pt,            % square corners
  left=2pt,           % padding
  right=2pt,
  top=2pt,
  bottom=2pt
}

\lstdefinestyle{mypython}{
    language=Python,
    basicstyle=\ttfamily\small,
    keywordstyle=\color{blue},
    stringstyle=\color{brown},
    commentstyle=\color{teal},
    showstringspaces=false,
    breaklines=true,
    numbers=left,
    numberstyle=\tiny\color{gray},
    escapeinside={(*@}{@*)} % allows inline LaTeX markup
}
\smartqed
\journalname{Empirical Software Engineering}
\pgfplotsset{compat=1.18} 
\begin{document}

\title{An Empirical Study of Testing Practices in Open Source AI Agent Frameworks and Agentic Applications}
%\subtitle{}
\titlerunning{Testing Practices in AI Agent Frameworks and Agentic Applications}

\author{Mohammed~Mehedi~Hasan \and Hao~Li \and Emad~Fallahzadeh \and Gopi~Krishnan~Rajbahadur \and Bram~Adams \and Ahmed~E.~Hassan}
%\authorrunning{Short form of author list}

\institute{
    Mohammed Mehedi Hasan and Hao Li and Emad Fallahzadeh and Gopi Krishnan Rajbahadur and Bram Adams and Ahmed E. Hassan \at
    School of Computing, Queen's University, Kingston, ON, Canada \\
    \email{\{mohammedmehedi.hasan, hao.li,cj79\}@queensu.ca, grajbahadur@acm.org, bram.adams@queensu.ca, ahmed@cs.queensu.ca}
}

\date{Received: date / Accepted: date}

\maketitle

\begin{abstract}
Foundation model (FM)-based AI agents are rapidly gaining adoption across diverse domains, but their inherent non-determinism and non-reproducibility pose testing and quality assurance challenges. While recent benchmarks provide task-level evaluations, there is limited understanding of how developers verify the internal correctness of these agents during development.

To address this gap, we conduct the first large-scale empirical study of testing practices in the AI agent ecosystem, analyzing 39 open-source agent frameworks and 439 agentic applications. We identify ten distinct testing patterns and find that novel, non-determinism-specific methods like \pattern{DeepEval} are seldom used (around 1\%), while traditional patterns like negative and membership testing are widely adapted to manage FM uncertainty. By mapping these patterns to canonical architectural components of agent frameworks and agentic applications, we uncover a fundamental inversion of testing effort: deterministic components like \component{Resource Artifacts} (tools) and \component{Coordination Artifacts} (workflows) consume over 70\% of testing effort, while the FM-based \component{Plan Body} receives less than 5\%. Crucially, this reveals a critical blind spot, as the \component{Trigger} component (prompts) remains neglected, appearing in around 1\% of all tests.

Our findings offer the first empirical testing baseline in FM-based agent frameworks and agentic applications, revealing a rational but incomplete adaptation to non-determinism. To address it, framework developers should improve support for novel testing methods, application developers must adopt prompt regression testing, and researchers should explore barriers to adoption. Strengthening these practices is vital for building more robust and dependable AI agents.
\end{abstract}

\keywords{AI Agent \and LLM Agent \and Mining Software Repositories \and Agentic Application \and Testing}

\section{Introduction}
\label{sec:introduction}
Autonomous agents have long been a foundational concept in AI, typically defined as systems capable of decision-making and action execution in an environment to fulfill given goals and objectives~\citep{liu2023agentbench}. This classical idea has been supercharged by the advent of foundation models (FMs), giving rise to a new paradigm: FM-based agentic applications. These agentic applications augment an FM ``brain'' with capabilities such as tool use, planning, and memory, enabling them to tackle complex, high-level goals with minimal human intervention~\citep{hettiarachchi2025exploring, park2023generative}. Modern agent frameworks further accelerate this trend, enabling developers to build sophisticated single and multi-agent applications that can reason, collaborate, and dynamically adapt their behavior~\citep{wu2024autogen}.

With the rapid proliferation of agent frameworks and agentic applications, and the increasingly consequential tasks being delegated to them, evaluation has become a central organizing practice in the community. Most prior work adopts a benchmark-driven evaluation paradigm that measures agent performance on standardized task suites, enabling reproducible comparison under fixed conditions, e.g., AgentBench, GAIA, WebArena, and MINT~\citep{liu2023agentbench,mialon2023gaia,zhou2023webarena,wang2023mint}. While such benchmarks are effective for objective comparison and assessing task-level capability, they are an incomplete proxy for real-world reliability. In practice, benchmark success does not guarantee robustness to distribution shifts, resilience to tool or environment failures, or stable behavior under underspecified or adversarial user inputs. Consequently, agents that rank highly on leaderboards may still perform poorly when faced with edge cases, get stuck in buggy loops, produce hallucinations, or succumb to unhandled faults when presented with minor variations unseen in the benchmarks~\citep{gevers2025benchmarks}.

This gap highlights the need to distinguish evaluation from testing in the context of FM-based agents. While often used interchangeably, testing and evaluation diverge significantly in the context of FM-based agents. Evaluation of FM-based agents assesses the interaction settings, agent behavior, and performance, including contextual consistency, factual accuracy, bias, and safety~\citep{ma2025rethinking}. In contrast, software testing is an engineering process aimed at verifying that an implemented system behaves as intended, continues to satisfy specified requirements as the code evolves, and exposes newly introduced faults in code chunks that previously worked properly~\citep{niedermayr2016will,mohammadi2025evaluation}. Despite the central role of testing in mature software systems, there remains a lack of systematic study of how practitioners test FM-based agent frameworks and agentic applications in practice.

While research has focused extensively on agent architectures, capabilities, and benchmarks, there is a clear lack of empirical evidence on how practitioners actually test agent frameworks and agentic applications. We found no comprehensive studies examining the state-of-the-art testing practices for open-source agents: how developers structure tests, verify non-deterministic outputs, and which parts of the agent they prioritize while testing. These gaps are concerning, given that the potential impact of untested or unverified agents can be poor in real-world use cases. For example, previous studies have already reported that the performance of FM-based systems can deteriorate during model upgrades. Moreover, we have also observed that different components (e.g., prompts, tools) behave differently in different FMs or even on different versions of the same FM~\citep{ma2024my}. Even with the same version of FM, a slight change in prompt phrasing or order can introduce significant behavioral drift~\citep{razavi2025benchmarking, zhuo2024prosa, rafi2024impact}. Without testing, it is difficult to identify such silent performance degradation. Similarly, prior studies have reported at least seven error patterns while invoking tools in FM-based systems~\citep{kokane2025toolscan}, which can also lead to production issues if not appropriately tested during the development phase.

The rapid evolution of these agent frameworks and agentic applications further amplifies the challenge. Practitioners are confronted with a constant flux of new, implementation-specific components, e.g., from novel tool invocation protocols and multi-agent communication standards~(MCP, A2A, ACP) to evolving memory and planning modules~\citep{hasan2025modelcontextprotocolmcp,  huang2025agentic}. This architectural evolution makes it exceedingly difficult to establish durable testing strategies; what is tested today may become obsolete or change fundamentally tomorrow. As established in software engineering research, the key to managing such complexity is to map these volatile, concrete components to a stable, canonical, or conceptual architectural model, which provides a durable foundation for reasoning about system quality and test coverage~\citep{lukyanenko2024universal}. While prior work has proposed valuable taxonomies that categorize emerging agent components~\citep{handler2023taxonomy}, we have not seen any work grounding the architecture component to any conceptual architecture till date.

% In particular, we lack studies that examine what is tested, how nondeterminism is handled in test oracles, and which agent components receive the most testing attention.

To fill these critical gaps, we conduct the first large-scale empirical study of unit testing practices in the agent ecosystem. We analyze 39 agent frameworks and 439 agentic applications, examining how practitioners arrange tests and assert correctness. Then we map the state-of-the-art agent architecture component to an existing conceptual architecture framework~\citep{boissier2020multi} and identify component-specific testing patterns. Our study answers the following research questions (RQs):

\smallskip
\noindent\textbf{RQ1: What testing practices are commonly adopted by practitioners to evaluate open-source AI agent frameworks and agentic applications?}

\textbf{Motivation}: Adapting traditional testing practices to FM-based agent frameworks and agentic applications is challenging due to the inherent non-determinism and non-reproducibility of the underlying models~\citep{hassan2024rethinking}. Recent studies show that such non-determinism can persist even with fixed low temperature and Top-P settings in practical deployments~\citep{atil2024non}. Moreover, in many agentic contexts (e.g., creative assistants), this variability is expected and supports natural behavior, rather than indicating a defect~\citep{krakowski2025human, hexmoor2026behaviour}. While testing methodologies are well studied in traditional software engineering~\citep{zhu2025understanding, zamprogno2022dynamic} and machine learning systems~\citep{openja2024empirical}, the FM-based agent ecosystem remains underexplored. This study addresses this gap by presenting the first large-scale empirical investigation of unit testing practices in open-source agent frameworks and applications.

\textbf{Findings}: We find a crucial split in practitioner strategy. On the one hand, novel patterns designed explicitly for this paradigm, such as \pattern{DeepEval} and \pattern{Hyperparameter Control}, see very low adoption, suggesting a significant knowledge or awareness gap. On the other hand, to compensate for this, practitioners strategically adapt familiar, battle-tested, less strict verification patterns, e.g., Membership Testing, Mock Assertion, and Negative Testing, to gracefully handle non-determinism and avoid the test fragility caused by strict assertions.

\smallskip
\noindent\textbf{RQ2: How do these testing practices map to the architectural components of the agent frameworks and agentic applications?}

\textbf{Motivation}: Agent ecosystem is a complex ecosystem of interacting components, from memory and tools to planning and coordination artifacts~\citep{hettiarachchi2025exploring}. Understanding how testing effort is distributed across this heterogeneous architecture is vital for assessing whether critical components receive sufficient attention and for supporting the development of more systematic and effective testing practices. To create a durable analysis that transcends rapid evolution of agent frameworks and agentic applications, we opt to map observed components to a stable, canonical architectural model, e.g., JaCaMo framework~\citep{boissier2020multi}, which will allow us to assess whether testing effort is distributed effectively across fundamental roles. Then we investigate the relationship between the testing patterns identified in RQ1 and the underlying components of agent frameworks and applications.

\textbf{Findings}: We identify 13 canonical components that receive unit-testing attention in agent frameworks and agentic applications. First, we observe a strategic inversion of testing efforts in agent frameworks and agentic applications compared to traditional ML applications: instead of focusing on the model itself, developers concentrate on deterministic infrastructure such as \component{Resource Artifacts} (tools, parsers), which account for 29.7\% of tests in frameworks and 40.1\% in applications. Second, we identify a critical testing blind spot: the \component{Trigger} component (prompts) is dangerously under-tested, appearing in around 1\% of test functions, which can introduce significant risks of silent failures and performance degradation as the underlying foundation models evolve.

The main contributions of our paper are as follows:
% \todo{update contribution later}
\begin{itemize}

    \item\textbf{A Curated Dataset}: The first large-scale, curated dataset of 39 agent frameworks, 439 agentic applications, and their corresponding test functions, providing a foundational resource for future research.
    \item\textbf{A Taxonomy of Agent Testing Patterns}: A comprehensive taxonomy of 10 testing patterns, including novel and adapted traditional techniques, used to manage the challenges of testing agentic systems.  
    \item\textbf{First Comparative Baseline}: The first empirical baseline of testing practices in the agent ecosystem in comparison with traditional software engineering and ML applications' testing patterns.
    \item \textbf{Canonical Architectural Component List}: An extensible catalog of 13 canonical components built on the classical JaCaMo framework \citep{boissier2020multi} and recent FM-based agent taxonomies. Twelve components correspond directly to JaCaMo, while one additional component \linebreak(i.e., \component{Registry}) is introduced based on contemporary practice. This catalog provides a reusable reference model for analyzing and testing future agent frameworks and agentic applications.
    \item \textbf{Component-wise Testing Patterns}: A comprehensive mapping of testing patterns to canonical components of agent frameworks and agentic applications, which can be leveraged by practitioners to select appropriate testing strategies for existing components and to design tests for new components.
\end{itemize}

%\begin{acknowledgements}
%\end{acknowledgements}


\section{A Motivational Example}
\label{sec:motivational-example}

Alex, a senior developer with a background in traditional software engineering, recently joined an ambitious AI startup. His first assignment is to lead the development of ``HomeBot’’, an autonomous Smart Home Management Agent. The agent's core mandate is to manage IoT devices (e.g., thermostats, locks, and lights) and assist users with technical troubleshooting via natural language. To achieve this, Alex builds the system on a popular open-source agent framework that orchestrates a state-of-the-art Foundation Model (FM).

\textbf{Phase 1: It works.} In the initial phase, using a popular open-source agent framework, Alex quickly builds a prototype. The framework provides seamless integration with the FM. For example, when any user says \textit{``It’s too dark to read.''}, the agent consistently interprets this intent and executes the \textsc{turn\_on\_lights(brightness=100)} tool. The initial results are impressive, and the team moves forward with deployment, confident in the power of their new AI-driven architecture.

\textbf{Phase 2: The Unseen Decay.} Weeks after launch, user feedback reveals a degradation in quality. Agents are becoming chatty, requiring increasing number of  instructions to accomplish a task. Alex investigates and discovers a baffling pattern: for the exact same inputs, e.g., \textit{``It’s too dark to read''}, the agent now exhibits non-deterministic behavior. 
\begin{itemize}

 \item \textit{Behavior A (Desired):} The agent silently executes the tool to turn on the lights. 

 \item \textit{Behavior B (Drift):} The agent responds conversationally, \textit{``I can help with that. Would you like me to turn on the reading lamp?''} 

\end{itemize} 

Both behaviors are plausible and expected from a conversational standpoint, yet they represent materially different system actions. Alex investigates and finds the root cause: a silent, unannounced update to the underlying FM has altered its interpretation of their prompts. The application is performing, yet the user-facing quality has declined. Alex realizes he has no automated regression suite capable of detecting this semantic drift; his standard unit tests expect deterministic string outputs, while the agent’s real-world behavior is intentionally non-deterministic.

\textbf{Phase 3: The Component Explosion.} Things become more complicated with the product team introducing two new features:

\begin{itemize} 

    \item \textbf{Contextual Troubleshooting (RAG):} Users can ask, \textit{``What does the red blinking light on my heater mean?''} This requires integrating a VectorDB to perform Retrieval-Augmented Generation (RAG) over technical manuals~\citep{lewis2020retrieval}. 

    \item \textbf{Real-World Integration (MCP):} Users can ask, \textit{``How much is my current electricity bill?''} This requires the agent to interface with a third-party utility API via a Model Context Protocol (MCP) server~\citep{hasan2025modelcontextprotocolmcp}. 

\end{itemize}

Alex’s testing challenges have now multiplied. Not only does he need to validate the non-deterministic output of the FM, but he must now also evaluate distinct components:

\begin{itemize} 
  \item \textbf{Testing RAG:} How can he write a test to verify that the retrieved manual chunk is relevant to the blinking light error? A simple keyword match fails if the manual uses synonyms (e.g., flashing indicator). 

   \item \textbf{Testing MCP:} The utility API is a black box. If the agent fails to answer the billing question, is it because the FM failed to call the tool, or because the third-party MCP server changed its schema? \end{itemize}

\textbf{Phase 4: The Testing Gap}. Drawing on his traditional software engineering background, Alex employs familiar techniques such as mocking and component-level testing. However, he quickly finds these practices difficult to apply in agentic systems where non-deterministic FM behavior is an expected design choice and components evolve independently. To seek guidance, he examines the testing suites of popular open-source agent frameworks and agentic applications. Instead of established practices, Alex finds a fragmented landscape of ad hoc scripts and coarse-grained end-to-end tests that offer little insight into component-level failures. There is no clear consensus on how practitioners test agentic systems, how they handle non-deterministic outputs, or which components receive focused testing. This gap motivates two fundamental questions:

\begin{enumerate}
\item What testing practices are commonly adopted by practitioners to evaluate open-source AI agent frameworks and agentic applications? \textbf{(explored in RQ1)}
\item How are testing efforts distributed across the components of AI agent frameworks and agentic applications? \textbf{(explored in RQ2)}
\end{enumerate}

\section{Background}
\label{sec:background}

This section establishes the foundational concepts underpinning our study. We first define the core principles of software testing, then describe the emerging agent ecosystem, and finally, present a canonical agent architecture that provides a stable model for analyzing testing practices in this rapidly evolving domain.

\subsection{Foundations of Software Testing}
Testing has been an integral part of software quality assurance. Software testing is typically defined as a process that systematically executes a program or system with the explicit objective of identifying and isolating defects~\citep{myers2006art}. The focal point of this process is the Subject Under Test (SUT), which represents the specific software artifact being validated, ranging from atomic units like functions or classes to complex modules and entire services~\citep{meszaros2007xunit}. These SUTs are exercised by test functions, also known as test cases, which serve as the atomic units of verification. A test function acts as an executable procedure that initializes a specific state or fixture, executes a command on the SUT, and verifies the outcome against specified requirements~\citep{van2008exploring, tao2009introduction}.

To structure these test functions effectively, developers often employ the Arrange-Act-Assert (AAA) pattern. This convention divides a test into three distinct phases: arranging the preconditions and inputs, acting on the SUT, and asserting that the resulting state matches expectations~\citep{wei2022automatically}. Within this structure, testing patterns offer recurring solutions to common design challenges. For example, Test Doubles (such as mocks and stubs) are frequently used during the arrangement phase to isolate the SUT from external dependencies, ensuring that tests remain deterministic and focused solely on the logic under verification~\citep{zhu2025understanding, meszaros2007xunit}.

\subsection{The Agent Ecosystem}
An agent is an autonomous software entity that perceives its environment and performs actions to achieve specified objectives~\citep{liu2023agentbench}. Classical agent models, such as the Belief-Desire-Intention (BDI) framework, conceptualize agent behavior as a reasoning cycle over internal states, goals, and plans~\citep{rao1995bdi}. Contemporary agent implementations extend this paradigm by incorporating foundation models as the primary reasoning mechanism, supported by memory systems, planning logic, and interfaces for interacting with external tools and services~\citep{park2023generative, hettiarachchi2025exploring}. This integration enables agents to address open-ended tasks while introducing new forms of statefulness, non-determinism, and environmental interaction.

From a software engineering standpoint, the agent ecosystem can be conceptualized as having two distinct layers: (1) agent frameworks and (2) agentic applications. Agent frameworks provide reusable infrastructure for defining agents, managing state and memory, orchestrating tool usage, and coordinating interactions among agents~\citep{wu2024autogen, li2023camel}. Agentic applications are concrete software systems built on top of these frameworks, where domain-specific workflows, prompts, and tool compositions are implemented. This distinction is important for testing because frameworks expose generic execution primitives and abstractions, whereas applications encode task-specific logic and integration behavior. Consequently, the nature and granularity of unit tests differ across these layers.

\subsection{Canonical Agent Architecture}\label{sec:background:canonical-agent-architecture}

To enable consistent analysis across heterogeneous agent frameworks and applications, we adopt a canonical agent architecture as an abstract reference model. This architecture is grounded in established multi-agent system models that decompose systems into agents, environments, and organizational structures~\citep{boissier2020multi}. The Agent dimension captures internal state and control constructs, such as beliefs, goals, and plans. The Environment dimension encompasses the artifacts, resources, and observable signals that agents interact with to effect change. The Organization dimension defines the coordination structures, roles, and communication mechanisms that govern interactions among autonomous entities.

This classical model is further extended to account for architectural elements that have become central in modern FM-based ecosystems, particularly \textit{Registries}. These components facilitate the dynamic discovery and invocation of tools or agents across system boundaries, bridging the gap between static definitions and runtime capabilities~\citep{liu2025agent, hasan2025modelcontextprotocolmcp}. By serving as a unifying vocabulary, this canonical architecture allows for the identification of Subjects Under Test and the categorization of test targets independent of framework-specific abstractions. We provide the formal definitions of these architectural components and their detailed mappings to contemporary agent systems in Appendix~\ref{sec:appendix-canonical-architecture}.

\section{Related Work}
\label{sec:related-work}
\subsection{Testing in OSS Ecosystem}
Software testing is a foundational software engineering practice aimed at verifying quality and finding defects~\citep{myers2006art}. Researchers have extensively studied how developers write and maintain tests in traditional software projects (e.g., Java, Python, JavaScript ecosystems), yielding common empirical findings on test design patterns, structure, and effectiveness~\citep{meszaros2007xunit}.

One widely observed technique is the AAA (Arrange-Act-Assert) pattern, which appears in 77\% of Java unit tests to enhance readability~\citep{wei2025developers}. Within the arrange phase of this pattern, developers adopt recurring testing patterns in the OSS domain in the literature. For example, the use of test doubles (e.g., Mocks, Stubs, Fakes) enables isolation of the Subject Under Test from external dependencies, e.g., databases and remote services. For instance, prior work demonstrates how a data-access dependency can be stubbed to raise an exception on the first invocation and return a valid domain object on subsequent calls, allowing the login and recovery logic of a service to be verified without creating real database records~\citep{zhu2023stubcoder}. Similarly, parameterized testing allows practitioners to execute the same test logic with multiple inputs and has been proven to run test functions 30 times faster than the traditional approach~\citep{kampmann2019carving}.

In the assertion or verification phase,  studies have cataloged 12 assertion patterns, with equality checks being common (39.3\% in JavaScript projects)\linebreak~\citep{zamprogno2022dynamic}. Mock assertion techniques have also gained popularity alongside the use of test doubles. In these tests, practitioners assert whether specific calls were made on a mock rather than asserting a returned value. Such assertions appear in 41\% of test functions, and prior work shows that 50\% of the defects detected by mock assertions are missed by other assertion techniques~\citep{zhu2025understanding}. Recent works have further explored the use of ML and FMs to automatically generate assertions~\citep{fontes2023integration, wang2024software}. Despite the extensive work of testing patterns, it remains unclear how these well-established testing structures and patterns translate to agent frameworks and agentic applications that integrate non-deterministic model calls with procedural code.

\subsection{Testing in ML Ecosystem}
Moving beyond traditional code-centric testing, the Machine Learning (ML) ecosystem introduces the challenge of validating systems whose behavior is learned from data rather than explicitly programmed. This marks a fundamental shift in focus from verifying static code logic to testing dynamic model behavior, data pipelines, and performance over time~\citep{zhang2020machine}. In response, researchers have proposed various testing frameworks tailored to ML contexts, including tools like DeepXplore~\citep{pei2017deepxplore}, TEST4Deep~\citep{nishi2018test}, and FairTest~\citep{tramer2017fairtest}. These frameworks support the validation of models under diverse inputs and conditions, addressing concerns such as fairness, robustness, and interpretability.

Apart from this, a new suite of validation patterns and frameworks emerged. Techniques like \pattern{Oracle Approximation} compare a model’s output against a simpler, trusted heuristic~\citep{nejadgholi2019study}, while \pattern{Value Range Analysis} ensures predictions fall within plausible bounds~\citep{sahoo2021reliable}.  Recent empirical work confirms this evolution, identifying nine high-level testing patterns used by practitioners to verify 16 distinct ML-specific properties, from model performance to data quality~\citep{openja2024empirical}. However, current studies have not explored how the tests are being organized in the ML ecosystem, e.g., from \texttt{Arrange} point of view. Also, while prior work in ML testing has identified verification patterns like Value Range Analysis, it remains unknown how these patterns are being adapted to test the unique, composite architectures of FM-based agent frameworks and agentic applications, which blend traditional code with non-deterministic components.

\subsection{Evaluating Agent Frameworks and Agentic Systems}
Current approaches of evaluating AI agents predominantly focus on task-oriented performance, relying on standardized benchmarks like AgentBench, GAIA, and WebArena to measure an agent's problem-solving capabilities~\citep{liu2023agentbench, mialon2023gaia, zhou2023webarena}. These benchmarks offer structured tasks for assessing agent capabilities such as planning, tool use, and decision-making. However, recent literature has increasingly emphasized the limitations of benchmark-driven evaluation, particularly in addressing the practical challenges posed by foundation model (FM)-based agents~\citep{hassan2024rethinking}. Specifically, issues such as non-determinism, non-reproducibility, and prompt sensitivity have drawn attention to the need for more robust quality assurance practices~\citep{hassan2024rethinking, ma2024my}. Moreover, recent efforts have highlighted vulnerabilities such as tool failures~\citep{kokane2025toolscan}, further underscoring the limitations of relying solely on high-level benchmarks for ensuring system reliability.

Despite these growing concerns, to the best of our knowledge, there is no comprehensive research on the use of quality assurance activities, e.g., unit testing, in agent frameworks and agentic applications till date. This study aims to bridge that gap by conducting the first large-scale empirical investigation of unit testing practices in open-source agent frameworks and the agentic applications built upon them.

% !TeX root = 0-main.tex

\section{Methodology}
\label{sec:methodology}
\begin{figure*}[!t]
    \centering
    \includegraphics[width=1.0\linewidth]{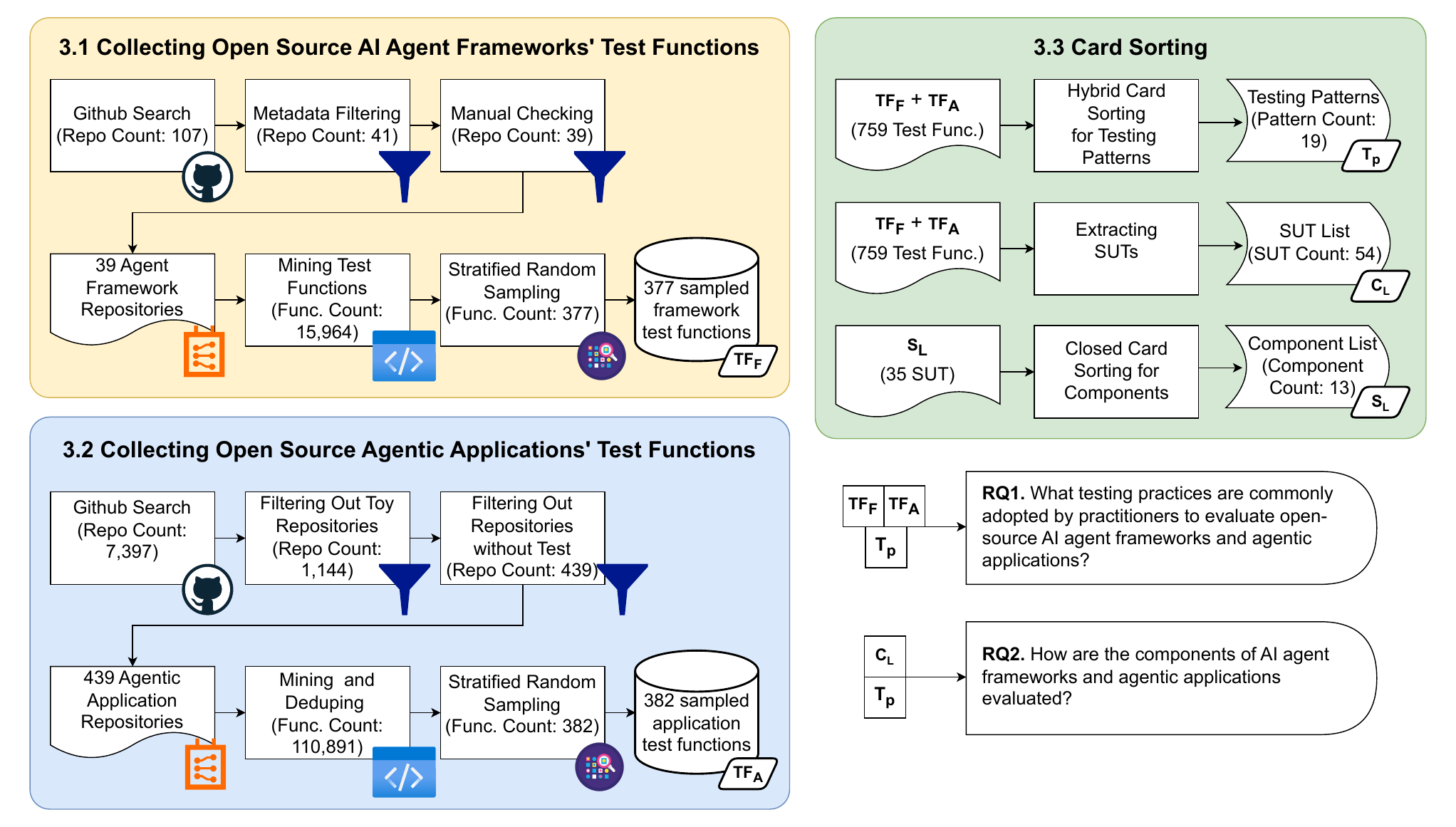}
    \caption{Overview of the Research Method}
    \label{fig:methodology:full}
\end{figure*}

In this section, we describe the methodology used to investigate testing practices in AI agent frameworks and agentic applications. In this study, we adopt a multi-stage empirical approach consisting of three structured phases: (i) collecting test functions from open-source AI agent frameworks, (ii) collecting test functions from open-source agentic applications, and (iii) conducting a systematic qualitative analysis of the test functions through manual card sorting to identify different testing practices and components under test. An overview of the methodological workflow is presented in Figure~\ref{fig:methodology:full}. In the following, we provide a detailed explanation of each step.

\subsection{Collecting Open-Source AI Agent Frameworks' Test Functions}\label{method:framework:function-collection}

To analyze the testing practices of AI agent frameworks, we first identify representative agent framework repositories and systematically extract their test functions. To accomplish these objectives, we employ multiple sub-steps, which are discussed below in chronological order.

\subsubsection{GitHub Search}
Following prior empirical studies on open-source software curation in emerging ML/AI domains~\citep{gonzalez2020state, li2025bridging}, we leverage GitHub’s keyword-based repository search API using domain-specific keywords: 

\begin{querybox}
\begin{center}
``\texttt{AI AND agent AND framework}'', ``\texttt{LLM-based AND agent AND framework}'', ``\texttt{LLM AND agent AND library}'', ``\texttt{multi-agent orchestration framework}'', ``\texttt{LLM powered agents AND framework}''
\end{center}
\end{querybox}

\noindent We conducted the search on Jun 14, 2025, and retrieved a total of 107 agent framework repositories from this search step.

\subsubsection{Metadata Filtering} \label{sec:method:framework:metadata-filtering}

GitHub's repository search API retrieves repositories based on matches with the repository name, description, or README file which can often surface non-representative or experimental repositories~\citep{kalliamvakou2014promises, munaiah2017curating}. 
To address this, we filter repositories using the GitHub Repository and Contributor APIs.\footnote{\url{https://docs.github.com/en/rest}} The filtering steps, shown in Table~\ref{table:frameworks:filter-wise-repository}, are as follows:

\begin{table}[t]
\centering
% \captionsetup{justification=justified, singlelinecheck=false}
\caption{AI Agent Framework: Repository Metadata Filtering}
\begin{tabular}{lr}
\toprule
\textbf{Filter Condition}  & \textbf{Repository Count} \\
\midrule
Mined from GitHub                               & 107        \\             
Contributor count $\geq$ 2            & 104\\
Star count $\geq$ 1000               & 67                       \\ 
Language: Python      & 46                       \\ 
Number of test files $\geq$ 1            & 41                       \\ 
Manual Checking             & 39     \\  
\bottomrule
\end{tabular}
\label{table:frameworks:filter-wise-repository}
\end{table}

\paragraph{Contributor and Popularity Filtering.}
To focus on well-maintained frameworks, we first filter repositories based on contributor count and popularity metrics. Prior research has highlighted that the number of contributors is a strong indicator of repository sustainability and adoption~\citep{han2019characterization}. Therefore, we only retain repositories with at least two contributors. While prior work often adopts higher thresholds (e.g., three contributors~\citep{zou2019branch}) to remove small-scale projects, we intentionally relax this criterion to reflect the early stage of the agent ecosystem, which has only emerged in recent years and where many widely used frameworks are still being developed by small teams.

Additionally, GitHub stars are commonly used as a proxy for repository popularity in empirical studies~\citep{gonzalez2020state, munaiah2017curating, openja2024empirical}. We have observed the median star count of our dataset is 2,015 and choose the repositories with at least 1,000 stars following similar prior studies~\citep{jebnoun2020scent} to preserve coverage of widely used and practically relevant agent frameworks. This step reduces the dataset to 67 repositories.

% \paragraph{Activity-Based Filtering.}
% Repositories that are no longer maintained are unlikely to be relevant for analyzing current AI agent frameworks. Following prior research~\citep{openja2024empirical}, we exclude repositories that had been archived. This step results in the removal of one repository, leaving 25 repositories.

\paragraph{Language-Specific Filtering.}
Since the majority of AI agent frameworks are implemented in Python, we restrict our selection to Python-based repositories. This decision aligns with previous empirical studies in the ML domain, which focused on Python-based projects due to their prevalence in AI research \citep{openja2024empirical}. Consequently, 21 non-Python repositories are removed, leaving 46 repositories.

\paragraph{Ensuring Presence of Test Files.}
As the primary objective of this study is to analyze testing practices, it is essential to ensure that all selected repositories contain at least one test file. In our study, we consider a file to be a test file if its filename starts with the \textbf{test\_} prefix, following conventions from previous studies~\citep{jebnoun2020scent, openja2024empirical}. Repositories without files following standard test naming conventions (e.g., without starting with test\_ prefix) may require function-level analysis to identify test code, potentially complicating test extraction and analysis. Hence, we apply this prefix-based test file detection technique and remove 5 repositories that do not contain any test files. This step results in 41 repositories.
\subsubsection{Manual Checking}
Although metadata-based filtering helps to remove many irrelevant repositories, we cannot completely rely on this step to identify agent frameworks. For instance, some repositories may include agent-related keywords without implementing an AI agent framework. Therefore, we conduct a manual verification step to ensure that only genuine agent frameworks are retained.

% \hao{specify: the first and the x author?} 
The first and second authors evaluate each repository independently based on the following criteria:

\begin{itemize}
    \item The repository publishes a PyPI package and includes documentation or examples demonstrating how to build agents using the framework.
    \item If no PyPI package is available, the README file must explicitly describe agent-related capabilities and provide usage instructions.
    \item In the absence of detailed documentation in the README, the repository must reference an official website with verifiable information about the framework’s agent-supporting features.
\end{itemize}

Two repositories fail to meet those criteria, resulting in a final curated set of 39 AI agent frameworks.~\tabref{tab:framework:repo-stats} consolidates the names and statistics of these frameworks.% We completed this mining process on Jun 14, 2025.  
% While cross-verifying with existing agent framework lists, we identify that 10 out of 13 (77\%) of Python-based frameworks mentioned in the most comprehensive survey~\citep{yu2025survey} are available in our agent framework list, which indicates the robustness of this dataset.
% \hao{cut-off date and how many major frameworks (golden set) are here}

\subsubsection{Mining Test Functions}\label{methodology:subsection:framework:mining-test-function}
To systematically analyze the testing patterns, we need to extract the test functions from the 39 curated AI agent frameworks. Given the volume and complexity of these repositories, manual extraction was infeasible. We therefore develop an automated parser that follows standard Python testing conventions~\citep{okken2022python} and prior research \citep{bodea2022pytest}. 

We define a test function as a Python function that meets both of the following criteria:
\begin{itemize}
    \item The function is defined in a Python file intended for testing, as indicated by the filename pattern \texttt{test\_*.py}.
    \item The function is explicitly marked as a test through its name, which starts with the prefix \texttt{test\_} and is discoverable by common Python testing frameworks.\footnote{\url{https://docs.pytest.org/en/stable/how-to/unittest.html}}
\end{itemize}
We leverage \textbf{Abstract Syntax Tree}~(AST), a structured representation of source code that preserves its syntactic hierarchy \citep{cui2010code}, to extract test functions by traversing the source code of each agent framework programmatically~\citep{spirin2021psiminer}. We extract a total of 15,964 test functions from 1,981 test files. While these test functions are unstructured data, we need a database that can facilitate storing and analyzing unstructured data efficiently. We use Elasticsearch,\footnote{\url{https://github.com/elastic/elasticsearch}} an open-source search engine database, to store and manage these test functions efficiently for further analysis and searching.

% \todo{We share the dataset in our replication package.}
\subsubsection{Stratified Random Sampling}\label{methodology:subsection:framework:stratified-random-sampling}

Given the scale of our dataset, i.e., 15,964 test functions across 39 agent frameworks, manually analyzing every function is infeasible. Additionally, applying simple random sampling can result in over-representation from larger repositories and under-representation from smaller ones. To ensure balanced coverage, we employ \textit{stratified random sampling}, where each repository is treated as a distinct stratum and contributes a proportionate number of samples~\citep{zhao2024empirical}.

To determine the appropriate sample size, we followed the statistical practices used in prior empirical software engineering studies and calculated the required sample size using Cochran’s formula with finite population correction~\citep{chaokromthong2021sample}, which is defined as:
\[
n = \frac{\dfrac{Z^{2}p(1-p)}{\varepsilon^{2}}}{1 + \left(\dfrac{Z^{2}p(1-p)}{\varepsilon^{2}N}\right)}
\]
where $N$ is the population size, $Z$ is the $z$-score for the desired confidence level, $p$ is the estimated population proportion, and $\varepsilon$ is the margin of error. Following prior empirical studies~\citep{openja2024empirical}, we set $Z = 1.96$ for a 95\% confidence level, $p = 0.5$ as a conservative estimate, and $\varepsilon = 0.05$. Substituting $N = 15{,}964$ yields an adjusted sample size of 377 (rounded up).

We then apply stratified random sampling, whereby these 377 test functions are randomly selected within each agent framework repository in proportion to the total number of test functions that repository contains. This ensures that the final sample is both representative of the overall dataset and preserves the distribution of test functions across different agent frameworks.

\begin{table}[H]
\small
% \captionsetup{justification=justified, singlelinecheck=false}
\caption{Overview of AI agent frameworks and their GitHub statistics, including thousand lines of code (KLoC) as a proxy for repository size and complexity following prior studies~\citep{barb2014statistical}.}
\label{tab:framework:repo-stats}
\begin{tabularx}{\textwidth}{>{\raggedright\arraybackslash}p{3.4cm} r r r r r}
\toprule
\textbf{AI agent framework} & 
\textbf{\begin{tabular}[c]{@{}r@{}}\# Stars\\ (k)\end{tabular}} & 
\textbf{\begin{tabular}[c]{@{}r@{}}\#\\Contributors\end{tabular}} & 
\textbf{\begin{tabular}[c]{@{}r@{}}\# Test \\ files\end{tabular}} & 
\textbf{\begin{tabular}[c]{@{}r@{}}\# Test \\ functions\end{tabular}} & 
\textbf{KLoC} \\
\midrule
FoundationAgents/MetaGPT & 56.4 & 146 & 240 & 964 & 53.9 \\
microsoft/autogen & 45.9 & 551 & 80 & 1041 & 59.7 \\
crewAIInc/crewAI & 32.9 & 248 & 50 & 588 & 43.5 \\
agno-agi/agno & 28.2 & 227 & 158 & 1697 & 142.5 \\
huggingface/smolagents & 20.1 & 157 & 18 & 463 & 25.1 \\
openai/swarm & 19.9 & 13 & 2 & 11 & 1.2 \\
letta-ai/letta & 16.8 & 135 & 30 & 650 & 68.0 \\
eosphoros-ai/DB-GPT & 16.8 & 146 & 73 & 728 & 115.3 \\
TransformerOptimus/SuperAGI & 16.4 & 73 & 114 & 466 & 21.6 \\
raga-ai-hub/RagaAI-Catalyst & 16.2 & 26 & 17 & 87 & 6.4 \\
langchain-ai/langgraph & 14.3 & 229 & 43 & 906 & 68.1 \\
camel-ai/camel & 12.9 & 148 & 221 & 1507 & 92.8 \\
openai/openai-agents-python & 11.4 & 96 & 61 & 563 & 22.2 \\
pydantic/pydantic-ai & 10.2 & 161 & 74 & 1096 & 43.3 \\
QwenLM/Qwen-Agent & 9.7 & 28 & 22 & 48 & 12.7 \\
Upsonic/Upsonic & 7.5 & 24 & 5 & 24 & 3.2 \\
crestalnetwork/intentkit & 6.4 & 15 & 2 & 17 & 31.9 \\
livekit/agents & 6.3 & 150 & 15 & 137 & 40.1 \\
lavague-ai/LaVague & 6.1 & 25 & 2 & 1 & 7.3 \\
awslabs/agent-squad & 6.0 & 23 & 22 & 312 & 19.5 \\
superduper-io/superduper & 5.1 & 48 & 65 & 318 & 18.9 \\
kyegomez/swarms & 4.9 & 44 & 54 & 645 & 40.2 \\
MervinPraison/PraisonAI & 4.8 & 20 & 33 & 241 & 9.0 \\
AgentOps-AI/agentops & 4.5 & 42 & 28 & 337 & 22.8 \\
VRSEN/agency-swarm & 3.7 & 19 & 5 & 78 & 9.5 \\
SylphAI-Inc/AdalFlow & 3.3 & 26 & 37 & 299 & 22.5 \\
cheshire-cat-ai/core & 2.8 & 86 & 45 & 197 & 28.4 \\
Intelligent-Internet/ii-agent & 2.4 & 6 & 5 & 67 & 6.7 \\
griptape-ai/griptape & 2.3 & 37 & 256 & 1242 & 44.7 \\
dot-agent/nextpy & 2.3 & 12 & 88 & 828 & 87.1 \\
InternLM/lagent & 2.1 & 33 & 5 & 8 & 7.6 \\
trypromptly/LLMStack & 2.0 & 8 & 15 & 57 & 5.5 \\
melih-unsal/DemoGPT & 1.8 & 4 & 2 & 5 & 4.3 \\
openlit/openlit & 1.6 & 36 & 20 & 43 & 7.1 \\
agentuniverse-ai/agentUniverse & 1.5 & 27 & 30 & 70 & 28.9 \\
fetchai/uAgents & 1.4 & 60 & 15 & 136 & 11.6 \\
patched-codes/patchwork & 1.4 & 15 & 18 & 65 & 11.5 \\
Div99/agent-protocol & 1.2 & 15 & 11 & 22 & 6.1 \\
\midrule
\textbf{Total} & -- & -- & \textbf{1,981} & \textbf{15,964} & -- \\
\bottomrule
\end{tabularx}
\end{table}

\subsection{Collecting Open-Source Agentic Applications' Test Functions} \label{method:user:function-collection}

After collecting test functions from open-source AI agent frameworks, as described in Section~\ref{method:framework:function-collection}, the next step is to mine test functions from open-source agentic applications, i.e., standalone projects built using one or more of the AI agent frameworks. To identify such agentic applications, we perform dependency analysis and repository mining using the curated framework list shown in \tabref{tab:framework:repo-stats} as a seed. Specifically, we look for open-source repositories that import these frameworks through explicit mentions in their source code.

In the remainder of this section, we describe our process for mining test functions from agentic applications.

\subsubsection{GitHub Search} \label{method:user:mining:sub:01}
To discover open-source agentic applications, we perform GitHub Code Search using framework-specific import patterns. From the 39 agent frameworks listed in \tabref{tab:framework:repo-stats}, we observe that six do not expose software development kits (SDKs) or importable libraries. These frameworks primarily support GUI-based agent construction (e.g., drag-and-drop interfaces) and do not facilitate direct code-level integration. As a result, applications built on top of these frameworks could not be reliably identified via code analysis and are therefore omitted from this phase.

For each of the remaining 33 frameworks, we manually derive the Python import statements by inspecting their documentation, source code, and published examples. These statements are then used as search queries via the GitHub Code Search API,\footnote{\url{https://docs.github.com/en/rest/search/search}} which returns matches at the file level along with associated repository metadata. \tabref{table:user:framework_wise_states} summarizes the frameworks used, the corresponding import statements employed in our GitHub code search, and the initially identified repository count and the eventual number of agentic application repositories downloaded after applying different filters (see Section~\ref{rq2:mining:sub:02}).
% \hao{For this table: add `,' to numbers, sort the table by \# Filtered repos or \# Search results}
\begin{table}[t]
\footnotesize
\centering
\caption{Framework-wise import patterns, GitHub Code Search results, number of unique repositories in the search result, and downloaded repository counts after filtering toy repositories sorted by \#Search result.}
\label{table:user:framework_wise_states}
\begin{tabular}{>{\raggedright\arraybackslash}p{3.5cm} >{\raggedright\arraybackslash}p{2.9cm} r r r}
\toprule
\textbf{Framework name} & \textbf{\# Import statement(s)} & 
\textbf{\begin{tabular}[c]{@{}r@{}}\# Search\\results\end{tabular}} & 
\textbf{\# Repos} & 
\textbf{\begin{tabular}[c]{@{}r@{}}\# Filtered\\repos\end{tabular}} \\
\midrule
langchain-ai/langgraph       & \texttt{from langgraph.}                       & 6{,}720  & 904 & 67 \\
openai/openai-agents-python  & \texttt{from agents import}                    & 5{,}208  & 771 & 41 \\
camel-ai/camel               & \texttt{from camel.}                           & 3{,}488  & 154 & 17 \\
crewAIInc/crewAI             & \texttt{from crewai}                           & 2{,}964  & 832 & 45 \\
huggingface/smolagents       & \texttt{from smolagents}                       & 2{,}672  & 536 & 39 \\
microsoft/autogen            & \shortstack[l]{\texttt{from autogen\_}\\\texttt{agentchat.}\\\texttt{from autogen\_ext.}} & 2{,}572  & 426 & 40 \\
agno-agi/agno                & \texttt{from agno.}                            & 2{,}228  & 343 & 15 \\
pydantic/pydantic-ai         & \texttt{from pydantic\_ai import}              & 1{,}428  & 652 & 51 \\
cheshire-cat-ai/core         & \texttt{from cat.}                             & 1{,}656  & 342 & 19 \\
FoundationAgents/MetaGPT     & \texttt{from metagpt.}                         & 1{,}024  & 122 & 14 \\
fetchai/uAgents              & \texttt{from uagents import}                  & 912   & 340 & 9 \\
QwenLM/Qwen-Agent            & \texttt{from qwen\_agent.}                     & 808   & 108 & 6 \\
griptape-ai/griptape         & \texttt{from griptape.}                        & 820   & 78  & 13 \\
kyegomez/swarms              & \texttt{from swarms import}                   & 616   & 127 & 9 \\
MervinPraison/PraisonAI      & \texttt{from praisonaiagents import}          & 636   & 42  & 1 \\
livekit/agents               & \texttt{from livekit.agents}                  & 588   & 492 & 19 \\
AgentOps-AI/agentops         & \texttt{from agentops}                         & 528   & 126 & 14 \\
SylphAI-Inc/AdalFlow         & \texttt{from adalflow.}                        & 520   & 25  & 3 \\
agentuniverse-ai/agentUniverse & \texttt{from agentuniverse.}               & 502   & 4   & 0 \\
openlit/openlit              & \texttt{import openlit}                        & 468   & 68  & 9 \\
InternLM/lagent              & \texttt{from lagent.}                          & 454   & 106 & 12 \\
letta-ai/letta               & \texttt{from letta\_client}                   & 240   & 29  & 3 \\
lavague-ai/LaVague           & \texttt{from lavague.}                         & 225   & 16  & 1 \\
superduper-io/superduper     & \texttt{from superduper import}               & 170   & 2   & 0 \\
Div99/agent-protocol         & \texttt{from agent\_protocol import}          & 107   & 43  & 5 \\
awslabs/agent-squad          & \texttt{from agent\_squad.}                   & 104   & 6   & 0 \\
Upsonic/Upsonic              & \texttt{from upsonic import}                  & 82    & 20  & 4 \\
raga-ai-hub/RagaAI-Catalyst  & \texttt{from ragaai\_catalyst}                & 76    & 1   & 0 \\
melih-unsal/DemoGPT          & \texttt{from demogpt\_agenthub}               & 27    & 0   & 0 \\
\bottomrule
\end{tabular}
\end{table}

\subsubsection{Filtering Toy Repositories}\label{rq2:mining:sub:02}

We collect a list of 7,397 repositories importing at least one agent framework of \tabref{table:user:framework_wise_states} from an initial code search in GitHub.  Similar to Section~\ref{sec:method:framework:metadata-filtering}, we observe toy projects and personal one-time proof-of-concept (PoC) repositories in this list. Since the number of repositories is significantly higher than that of framework repositories, we apply additional metadata-based criteria, e.g., fork status, repository lifetime \citep{hassan2022code}, and commit frequency \citep{munaiah2017curating}, to filter out toy repositories in this set. We use the following threshold for each of the criteria:

\begin{itemize}
    \item To avoid duplicate test functions, we exclude all repository forks or clones of another repository, following previous empirical studies about testing~\citep{hassan2022code}.
    \item Similar to agent frameworks, we apply the contributor count filter and retain repositories with at least two contributors.
    \item To ensure the repositories are not one-time activities, we include repositories with a lifetime of at least one month. We measure the lifetime of a repository by calculating the difference between the last commit date and the creation date for the repository, following the previous studies in the testing domain~\citep{hassan2022code}
    \item To ensure regular development activity, we include repositories with a commit frequency of at least two per month throughout their lifetime~\citep{munaiah2017curating}. 
    \item Finally, we retain only repositories containing at least two test files, identified via filename prefixes (e.g., \texttt{test\_*.py}).
\end{itemize}

For the filtering process, we first collect metadata using the GitHub API sequentially and store the metadata in a PostgreSQL database for each repository. We then apply the filtering conditions individually, ensuring each step refines the dataset for the following filtering condition. 

\tabref{table:user:filter-wise-repository} presents the repository count at each filtering step. From our initial 7,397 repositories, these filters remove 6,958 repositories, and we prepare the remaining list of 439 repositories for the next step. To assess whether these repositories are representative of real-world agentic systems in terms of size and complexity, we measure their KLoC (thousands of lines of code), following prior empirical studies~\citep{barb2014statistical}. The median repository size is 30 KLoC, suggesting that the retained projects are non-trivial and comparable to production-grade software systems.

\begin{figure}[ht]
    \centering
    \includegraphics[width=0.75\linewidth]{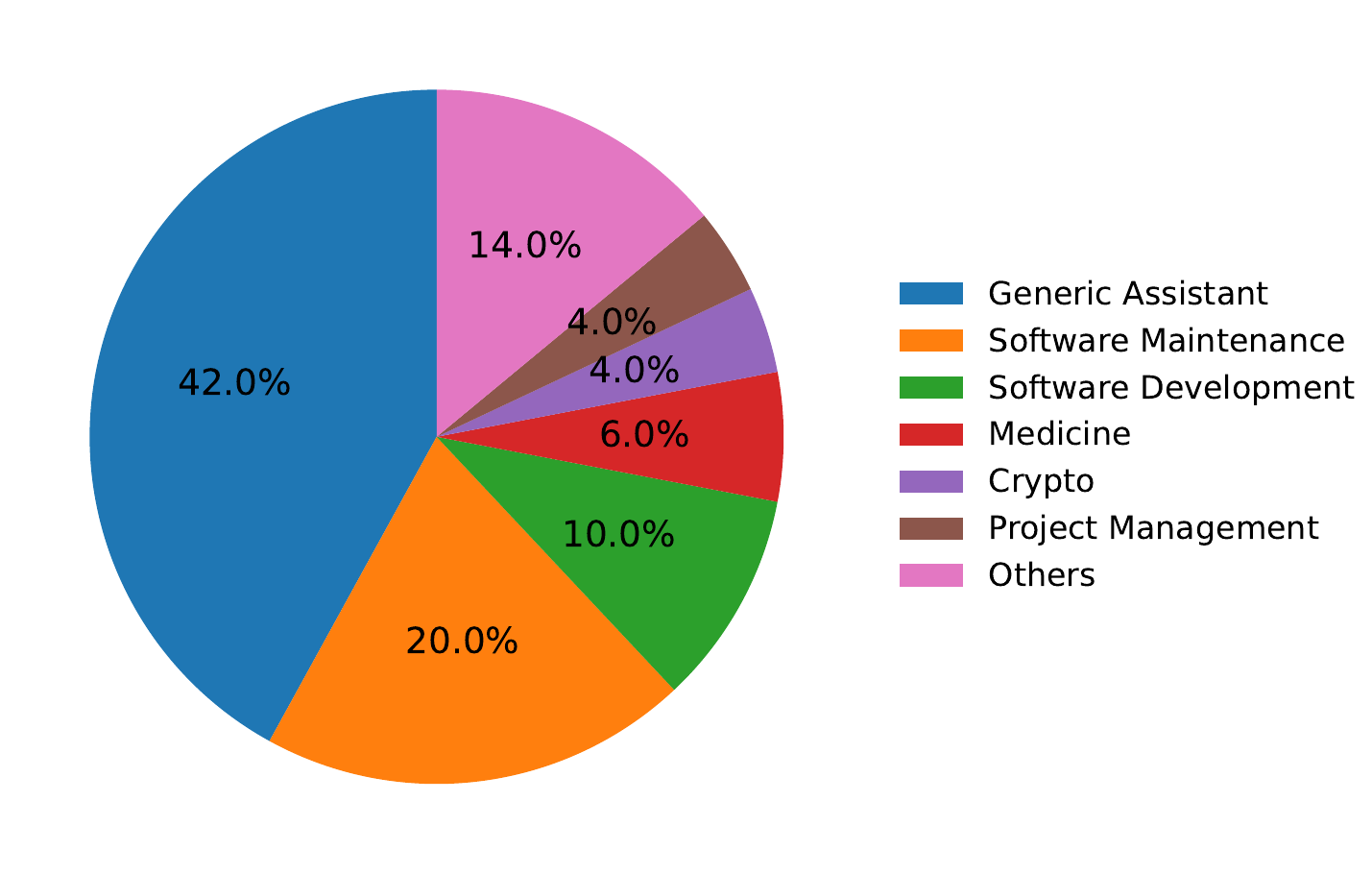}
    \caption{Distribution of agent application categories in our dataset. Each slice shows the proportion of agents assigned to a high-level category. Categories with more than one instance are reported individually: \emph{Generic Assistant} (application, workflow manager or builder), \emph{Software Maintenance} (monitoring, cloudops, review, testing, inferencing), \emph{Software Development} (coding, documentation), \emph{Medicine}, \emph{Crypto}, and \emph{Project Management}. All single-instance categories are aggregated into \emph{Others} comprising assistants capable of \emph{data analysis}, \emph{memory}, \emph{blogger}, \emph{ecommerce}, \emph{hrmis}, \emph{physics}, and \emph{research}.}
    \label{fig:method:agent-category-distribution}
\end{figure}

To further examine the diversity of application domains covered by our dataset, we randomly sample 50 agentic applications from the final set and manually inspect their GitHub tags and descriptions. This analysis reveals a broad range of application categories, including generic assistants, software development and maintenance tools, medical assistants, crypto-related agents, and project management systems. Figure~\ref{fig:method:agent-category-distribution} summarizes this distribution, indicating that the dataset spans diverse domains rather than being concentrated in a single application area.

% mean KLoC: Mean (Average):           432801.64

\begin{table}[t]
\caption{AI Agentic Applications: Filtering Toy Repositories}
\centering
\begin{tabular}{lr}
\toprule
\textbf{Filter} & \textbf{Repository Count} \\ 
\midrule
Non-fork      & 7,397                      \\
Distinct      & 7,074    \\
Contributor count $\geq$ 2            & 1,957  \\ 
Lifetime $\geq$ 1 month               & 1,222 \\ 
Commit frequency $>$ 2 per month      & 1,144 \\ 
Number of test files $>$ 1            & 439 \\ 
\bottomrule
\end{tabular}

\label{table:user:filter-wise-repository}
\end{table}

\subsubsection{Mining and Deduplicating Test Functions}\label{rq2:mining:sub:03}

To extract test functions from the 439 filtered agentic application repositories, we apply the same AST-based mining approach described in Section~\ref{methodology:subsection:framework:mining-test-function}. During this process, we exclude files located under \textit{site-packages} directories to avoid importing test functions originating from the agent frameworks.

However, upon inspection of the extracted data, we still observe duplicated test functions across different repositories. Many agentic applications have directly copied internal directories, including test files, from the frameworks they built upon, resulting in the same test function from the agent frameworks appearing in multiple agentic applications. To ensure analytical validity, we employed a SHA-256-based hashing technique~\citep{gueron2011sha} to eliminate such duplicates.

A test function is considered a duplicate if its hash exactly matches a function hash from one of the agent frameworks it imported. The de-duplication process involves the following steps:

\begin{itemize}
    \item For each test function extracted from an AI agent framework, we compute a SHA-256 hash (\texttt{HF}) based on its normalized function signature (including name and body). These are stored in Elasticsearch alongside the function metadata.
    \item For each test function extracted from an agentic application, we similarly compute a SHA-256 hash (\texttt{HA}).
    \item Given that an agentic application may depend on multiple frameworks, we compare each \texttt{HA} against all \texttt{HF} values from the relevant frameworks.
    \item If a match is found, the corresponding test function is marked as a duplicate and we exclude it.
    \item Otherwise, the function is retained and indexed as a unique test case.
\end{itemize}

This de-duplication procedure removes the redundant test functions, resulting in the final dataset of 110,891 unique test functions sourced from agentic applications, all stored in our Elasticsearch index for further analysis. In~\tabref{tab:file-function-metrics} we provide a count-level statistics for each of these steps.

\begin{table}[H]
\centering
\caption{File- and function-level statistics of the AI agentic applications}
\label{tab:file-function-metrics}
\begin{tabular}{lr}
\toprule
\textbf{Metric} & \textbf{Count} \\
\midrule
\# of total agentic application repositories & 439 \\
\# of total files across all repositories & 767,490 \\
\# of total test files across all repositories & 40,831 \\
\# of total functions in all test files & 789,343 \\
\# of extracted functions after de-duplication & 110,891 \\
\bottomrule
\end{tabular}
\end{table}

\subsubsection{Stratified Random Sampling}
Given the size of the agentic application dataset, e.g., 110,891 unique test functions, manually analyzing every function is impractical. Moreover, we observe that the distribution of test functions is highly skewed: the top 10 repositories alone account for approximately 70\% of the total test functions. As with the framework repositories (\secref{methodology:subsection:framework:stratified-random-sampling}), applying simple random sampling in this context risks disproportionately sampling from larger repositories, thereby obscuring practices in smaller ones.

To mitigate this imbalance, we again apply \textit{stratified random sampling}, treating each agentic application repository as a separate stratum and allocating sample size proportionally based on the number of test functions in each repository. Following the same Cochran formula~\citep{chaokromthong2021sample, ajibode2025towards} with finite population correction, and the same parameter settings ($Z = 1.96$, $p = 0.5$, and $\varepsilon = 0.05$), we compute the initial required sample size for $N = 110{,}891$ as approximately 383. We then randomly select test functions without replacement within each repository stratum. During manual inspection, we found that one sampled test function had an empty implementation and therefore excluded it from further analysis, resulting in a final valid sample of 382 test functions. This approach ensures representative coverage of testing practices across agentic applications, while maintaining consistency with the sampling strategy used for framework repositories.

\subsection{Card Sorting} \label{subsec:coding}

After filtering and sampling, we obtain two representative datasets of test functions: one from AI agent frameworks and another from agentic applications. To investigate testing patterns within these datasets, we need to systematically analyze how test functions are structured, what they aim to validate, and which components of the agent systems they target.

However, as test functions are written in code and vary widely in style, naming, and structure, they represent a form of textual data that is not easily categorized through automated means. To address this, we employ qualitative card sorting techniques, which are widely used in empirical software engineering to extract themes and patterns from code artifacts~\citep{spencer2009card}.

We first apply \textit{hybrid card-sorting} to identify emergent testing patterns, i.e., recurring structures, strategies, or practices embedded in the test code. This inductive approach allows patterns to emerge organically from the data while also allowing a set of predefined categories.~\citep{zampetti2022empirical}. Next, to identify which parts of the agent architecture are being tested, we extract the \textit{subject under test} (SUT) and apply a \textit{closed card sorting}~\citep{zhao2024empirical} to map those SUTs to the core agentic components defined in Section~\ref{sec:background}.

The following subsections describe these three-step procedures: hybrid card-sorting for testing patterns, extraction of SUTs, and closed card sorting for mapping SUTs to components.

\subsubsection{Hybrid Card Sorting for Testing Patterns}\label{subsubsec:open-coding}
To analyze testing patterns in test functions extracted from AI agent frameworks and agentic applications, we apply hybrid card-sorting, following methodologies established in prior empirical studies~\citep{zampetti2022empirical, nayebi2018anatomy, conrad2019making}. In a hybrid card-sorting, elements of both open and closed card sorting are combined, e.g., participants are given predefined categories into which to sort content (as in closed sorting), but are also free to create new categories when existing ones are insufficient (as in open sorting). Before detailing our card-sorting procedure, we first define what constitutes a testing pattern in this context and describe the underlying structure of unit tests.

In unit testing, a test function typically follows the AAA (Arrange-Act-Assert) structure~\citep{wei2025developers}. First, a test fixture is initialized to establish the desired state (\texttt{Arrange}); next, a command is executed on the subject under test (\texttt{Act}); and finally, the output is verified using assertions (\texttt{Assert})~
\citep{van2008exploring, tao2009introduction}. Among these stages, the \texttt{Act} phase tends to be straightforward and consistent, e.g., involving a direct invocation of a SUT function. In contrast, the \texttt{Arrange} and \texttt{Assert} phases often vary significantly depending on developer strategy, testing framework, and system complexity.

To illustrate how practitioners set up a test function and verify the output of the SUT, we present a sample unit test function in Figure~\ref{fig:framework:code:sample_test} that validates the behavior of the \texttt{oas3\_openai\_text\_to\_embedding} function. This function internally interacts with the OpenAI API, and the test isolates it by mocking external dependencies. Specifically, the test uses the \texttt{patching} technique to intercept and override the behavior of the \texttt{aiohttp.ClientSession.post} method (lines 3-4). The mocked response is configured to return a predefined JSON payload, simulating a successful API call (lines 5-8). This setup phase demonstrates two common structural patterns, i.e., \texttt{Mocking} and \texttt{Patching} to handle the dependencies. Additionally, the test includes verification logic that checks both the existence and structure of the returned embedding (lines 14-17). These lines reflect a combination of basic presence assertions (e.g., \texttt{assert result}) and threshold assertions (e.g., \texttt{assert len(result.data[0].\linebreak embedding) > 0}), which are often used to confirm the correctness and completeness of test outputs.

\begin{figure}[t]
\centering
\begin{tikzpicture}[remember picture]

\node[anchor=north west, inner sep=0] (code) at (0,0){
    \begin{minipage}{.9\linewidth}
\begin{lstlisting}[style=emsepython]
async def test_embedding(mocker):
    config = Config.default()
    mock_post = mocker.patch('aiohttp.ClientSession.post')
    mock_response = mocker.AsyncMock()
    mock_response.status = 200
    data = await aread(Path(__file__).parent / '../../data/openai/embedding.json')
    mock_response.json.return_value = json.loads(data)
    mock_post.return_value.__aenter__.return_value = mock_response
    type(config.get_openai_llm()).proxy = mocker.PropertyMock(
        return_value='http://mock.proxy'
    )
    llm_config = config.get_openai_llm()
    assert llm_config
    assert llm_config.proxy
    result = await oas3_openai_text_to_embedding(
        'Panda emoji',
        openai_api_key=llm_config.api_key,
        proxy=llm_config.proxy
    )
    assert result
    assert result.model
    assert len(result.data) > 0
    assert len(result.data[0].embedding) > 0
\end{lstlisting}
    \end{minipage}
};

\draw[decorate,decoration={brace,amplitude=8pt},thick] 
    (11,-0.4) -- (11,-4.5)
    node[midway,xshift=0.7cm,rotate=270]{\small \textbf{Arrange}};

\draw[decorate,decoration={brace,amplitude=8pt},thick] 
    (11,-5.2) -- (11,-6.7)
    node[midway,xshift=0.7cm,rotate=270]{\small \textbf{Act}};

\draw[decorate,decoration={brace,amplitude=8pt},thick] 
    (11,-6.8) -- (11,-8.2)
    node[midway,xshift=0.7cm,rotate=270]{\small \textbf{Assert}};

\end{tikzpicture}
\caption{A sample test function marked with Arrange--Act--Assert blocks.}
\label{fig:framework:code:sample_test}
\end{figure}

Testing patterns are these recurring strategies for structuring setup and verification logic in test functions. They improve the maintainability, diagnosability, and reliability of tests~\citep{meszaros2003test, gonzalez2017large}. In practice, these strategies manifest as different techniques practitioners use; for example, they may use mocking, patching, parameterization, or custom assertions to isolate behavior, simulate dependencies, or evaluate correctness. To identify these patterns, we focus on analyzing the source code of the test functions and ground our findings through qualitative analysis, e.g., hybrid card sorting.

For card sorting, the first and fourth authors of this study acted as independent raters. The first author has 10 years of industry experience and 3 years of research experience, and the fourth author has 12 years of industry and 9 years of research experience. Prior to initiating the card-sorting, all raters reviewed five foundational studies on testing in software engineering and machine learning~\citep{gonzalez2017large, openja2024empirical, zhang2015assertions, zhu2025understanding, zamprogno2022dynamic}. These studies provided representative examples of known patterns but were not treated as a closed card-sorting framework, meaning raters were encouraged to identify novel patterns that may emerge from AI agent test functions.

At the beginning, we randomly chose 15\% of the sampled test functions from both agent frameworks and agent applications (58 test functions from each), which were independently sorted by the first and fourth authors. Each rater spent approximately 20 hours uncovering the testing patterns from the sampled functions. Each rater recorded their findings in a structured spreadsheet containing:

\begin{itemize}
    \item Repository name
    \item Test function name
    \item Full test function source code
    \item Observed testing patterns (arrange and assertion level)
\end{itemize}

A single test function can exhibit multiple patterns simultaneously. Hence, we employed a multi-label annotation strategy to account for the presence of such concurrent patterns within individual test functions. To guarantee a comprehensive understanding of the context, raters examined the test code, the System Under Test (SUT), and relevant README documentation. Additionally, they assessed the novelty of these patterns by cross-referencing them with the taxonomies established in the five selected studies.

After completing this sorting, the raters held a dedicated meeting to align their taxonomies. Importantly, this discussion focused exclusively on the naming and granularity of pattern categories and not on individual card-sorting results to avoid bias. During this calibration, the raters identified and resolved three types of discrepancies:
\begin{enumerate}[label=(\roman*)]
    \item Inconsistent naming of equivalent patterns, leading to redundant entries.
    \item Overly broad pattern categories that conflated multiple strategies.
    \item Missing distinctions where raters overlooked subtle but meaningful pattern variations.
\end{enumerate}

To address these issues, the raters developed a consensus taxonomy by merging redundancies, refining naming conventions in alignment with existing literature, and agreeing on consistent granularity levels. Once the taxonomy was finalized, the raters re-sorted the 15\% dataset. Traditional agreement coefficients such as Fleiss’ Kappa and  Krippendorff’s $\alpha$ assume that each item receives a single categorical label, whereas our annotation allows multiple labels per test function. Hence, applying these coefficients is not suitable for such multi-label cases~\citep{marchal2022establishing}. Consequently, following prior multi-label annotation studies~\citep{parker2024large, bitew2023distractor}, we use Jaccard similarity, which directly measures overlap between the two raters’ label sets. Specifically, each test function may be assigned multiple pattern labels by each rater; therefore, we treat annotations as sets of labels and compute the Jaccard coefficient as the ratio of the intersection of the two raters’ label sets to their union. To further explain, let $A_i$ and $B_i$ denote the sets of labels assigned by reviewer 1 and reviewer 2 to item $i$. 
The Jaccard similarity for item $i$ is defined as:

\[
J(A_i,B_i) = \frac{|A_i \cap B_i|}{|A_i \cup B_i|}
\]

where $|A_i \cap B_i|$ represents the number of labels assigned by both reviewers 
and $|A_i \cup B_i|$ represents the total number of unique labels assigned by either reviewer.
The overall inter-rater agreement is computed as the average Jaccard similarity across all annotated items:

\[
J_{avg} = \frac{1}{N}\sum_{i=1}^{N} J(A_i,B_i)
\]

where $N$ is the number of items with at least one assigned label.

We observed agreement scores of 0.88 for agent framework patterns and 0.85 for agentic applications, both exceeding the 0.80 threshold for strong agreement. Though the original version of Krippendorff's $\alpha$ does not handle multiple labels, a variant of Krippendorff's $\alpha$ using the MASI distance~\citep{passonneau2006measuring} can be used to handle them. A prior study of disagreement in natural language inference used this MASI distance to calculate the inter-rater agreement~\citep{jiang2022investigating}. Hence, to complement the Jaccard similarity, we also measured Krippendorff's $\alpha$, with the MASI distance, which is 0.80 for agent frameworks and 0.75 for agentic applications. While Jaccard similarity captures raw set overlap between raters, Krippendorff’s $\alpha$ with MASI distance adjusts for chance agreement and partial label matches; thus, both metrics together provide a more comprehensive assessment of inter-rater agreement.

 % \hao{the tense is not consistent in this section}
\subsubsection{Extracting SUTs}\label{subsubsec:extracting-suts} 
To understand which parts of the agent architecture are being tested, we first identified the \textit{Subject Under Test} (SUT) in each sampled test function. The SUT refers to the primary object, function, or module that the test function aims to execute and validate~\citep{van2008exploring}. While every test function must target at least one SUT, extracting this information is non-trivial, as test code often lacks explicit references to the name of the SUT.

Similar to the process used for identifying testing patterns, we sampled 15\% of the test functions from both the agent framework and agentic application datasets for component-level labeling. Two authors (first and fourth) independently served as raters and categorized each test function by recording the repository name, function name, and its corresponding System Under Test (SUT) in a spreadsheet. For instance, in the test function shown in Figure~\ref{fig:framework:code:sample_test}, the primary method under test is \texttt{oas3\_openai\_text\_to\_embedding}, which generates embeddings from input text. An embedding is a representation of an input text in a numerical format, typically as vectors or matrices in a high-dimensional space~\citep{patil2023survey}. Based on this functional purpose and domain context, we labeled the SUT as \texttt{Embedding}.

After completing the initial labeling round, the raters compared their annotations and assessed inter-rater reliability using Cohen’s Kappa~\citep{landis1977measurement}, as each test function corresponds to a single SUT. The resulting agreement score was 0.52, indicating moderate agreement and highlighting the presence of labeling conflicts. To resolve these conflicts, a third rater, the second author of the paper, was introduced as an adjudicator. Following conflict resolution, the agreement score improved to 0.87, demonstrating strong consistency in the finalized taxonomy. Based on this high agreement, the first author completed the remaining labeling using the consensus definitions.

% \hao{fix this}
\subsubsection{Closed Card Sorting for Components}\label{subsubsec:closed-coding-components}
With the SUTs identified in the previous step (\secref{subsubsec:extracting-suts}), we performed a closed card-sorting to uncover the higher-level architectural components of agent frameworks and applications being tested. We considered the components described in~\secref{sec:background:canonical-agent-architecture} and Appendix-\ref{sec:appendix-canonical-architecture} as the source of components and grounded each SUT to the relevant component through the closed card sorting process. 

% \hao{fix this}
Following the methodology outlined in \secref{subsubsec:open-coding}, the first and second authors jointly map the SUTs to a component label. For example, in Figure~\ref{fig:framework:code:sample_test}, the extracted SUT is an embedding created and used for various purposes, e.g., similarity search, context retrieval, knowledge storage, agent frameworks, and agentic applications. As defined in \secref{sec:background:canonical-agent-architecture}, embeddings fall under \component{Resource Artifacts}.

% !TeX root = 0-main.tex

\section{RQ1: What testing practices are commonly adopted by practitioners to evaluate open-source AI agent frameworks and agentic applications?}
\label{sec:rq1-results}
% \subsection{Motivation}
As established in the Introduction, the prevailing benchmark-driven evaluation culture often fails to capture the practical reliability of agentic systems, overlooking critical failure modes like prompt decay~\citep{bhargava2024prompt} and tool-use errors that are already documented in the field~\citep{ma2024my, kokane2025toolscan}. While software testing offers a path to bridge this reliability gap, the inherent non-determinism and non-reproducibility of foundation models (FMs) fundamentally challenge the application of traditional, deterministic testing techniques~\citep{hassan2024rethinking}. Hence, this RQ seeks to uncover what testing strategies actually emerge in real-world agent development, focusing on how practitioners structure tests and assert correctness despite non-deterministic outputs. The goal is to produce a foundational catalog of these real-world strategies, providing the first empirical baseline on how the open-source community is negotiating the unique quality assurance hurdles of the agentic paradigm.

\begin{figure}[t]
    \centering
    \includegraphics[width=\linewidth]{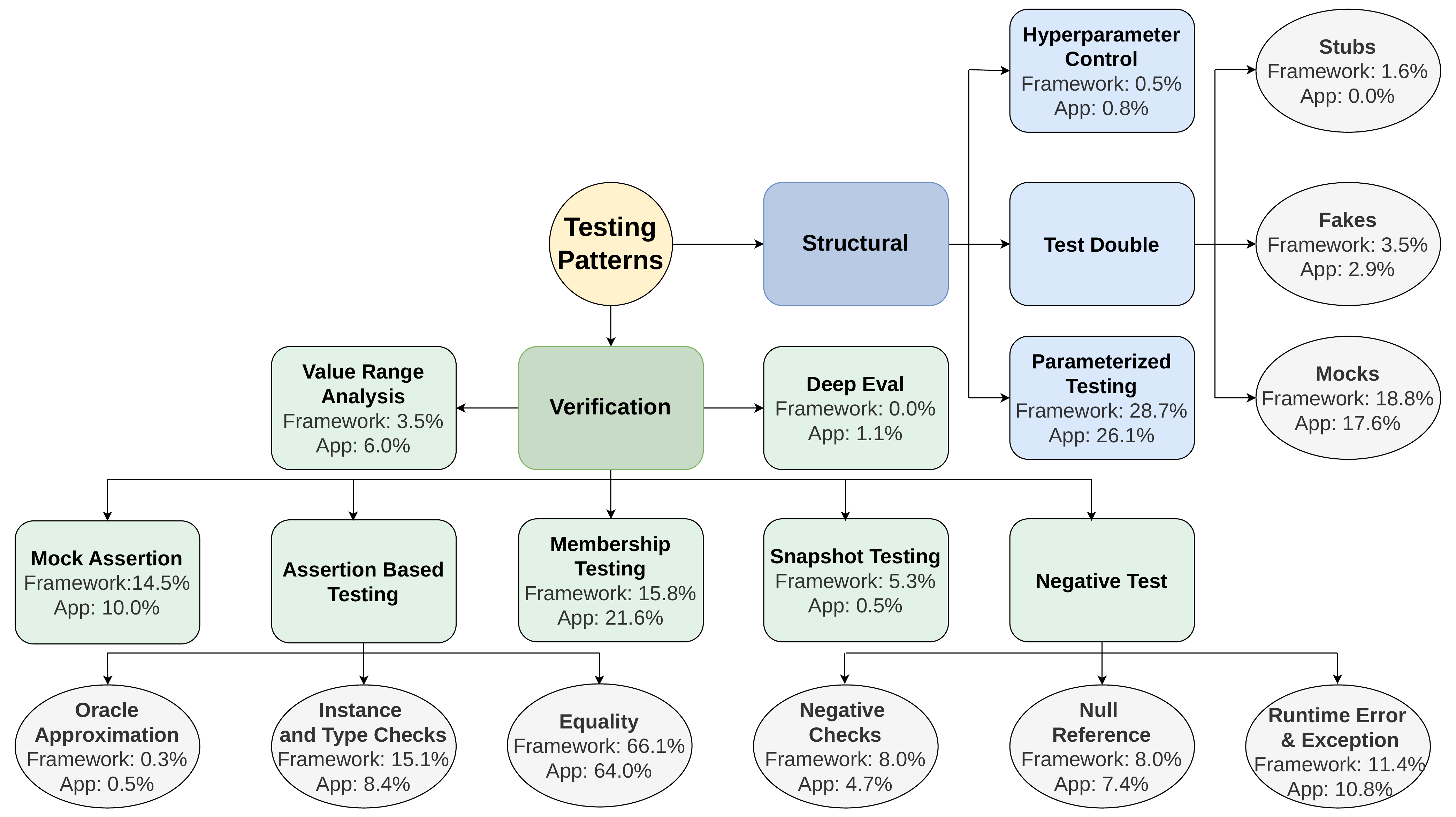}
    \caption{Overview of testing patterns observed in agent frameworks and applications. The three structural patterns (highlighted in blue) and seven verification patterns (highlighted in green) are organized under the top-level testing patterns. Elliptical nodes represent sub-patterns grouped under their corresponding high-level categories.
    }
    % \hao{0.0\% in deep eval? fix it in the figure}
    \label{fig:result:rq1:all-patterns}
\end{figure}

\subsection{Approach}
To identify testing patterns in AI agent frameworks and agentic applications, we perform \texttt{hybrid card-sorting} on test functions mined from open-source repositories, as described in~\secref{subsubsec:open-coding}. Hybrid card-sorting allows us to systematically extract recurring testing patterns related to test structure, e.g., how to \texttt{Arrange} the test function and verification strategies and how to \texttt{Assert}. This process not only focuses on the existing patterns but also enables uncovering emergent patterns that have not been seen in the literature before. We provide an example of each pattern in Appendix~\ref{sec:appendix-1-testing-patterns}. We also compare the prevalence of each testing pattern between the top-10 popular repositories and the remaining studied repositories using Fisher’s exact test~\citep{kim2017statistical} on binary per-test-function indicators, and controlled for multiple hypothesis testing using the Benjamini–Hochberg false discovery rate (FDR)~\citep{thissen2002quick} adjustment. The FDR correction limits the expected proportion of false positives among statistically significant findings, ensuring robust inference under multiple comparisons.

% Next, we apply \texttt{closed coding} to classify each test function into test categories. The three most prevalent and commonly used software testing
% categories for detecting errors are white-box testing, black-box testing, and gray-box testing \citep{khan2012comparative}. We perform closedcoding on each test function to identify which category it belongs to as outlined in Section~\ref{subsubsec:closed-coding}. 

\subsection{Findings}
\textbf{We identify three structural patterns for ``arranging'' test functions and seven verification patterns for ``asserting'' test outcomes}. These patterns, uncovered through the hybrid card-sorting process,  along with their frequencies in both agent frameworks and agentic applications are illustrated in Figure~\ref{fig:result:rq1:all-patterns}. Additionally, we discovered three sub-patterns under the \pattern{Test Double} structural pattern, and three sub-patterns each under the verification patterns  \pattern{Assertion Based Testing} and \pattern{Negative Test}.

\textbf{Despite the novelty of AI agent frameworks and agentic applications, 80\% of their testing patterns are directly inherited from classical software engineering or ML applications, reaffirming the robustness of classical testing strategies.} To clarify this landscape, we distinguish between traditional adaptations, which extend classical paradigms to structure and verify the agents, and non-determinism-specific innovations, which emerge uniquely to address the non-deterministic characteristics of LLM-driven systems in~\tabref{tab:rq1:result:testing-patterns-by-type}. We find that 8 out of 10 patterns have been previously reported in either ML application testing~\citep{openja2024empirical}, or in general software engineering literature~\citep{zhu2025understanding, fujita2023empirical, zamprogno2022dynamic, lam2018characteristic}, or in both. While ML testing research has focused more on verification patterns than on structural patterns, prior studies have still documented the use of structural patterns, e.g., test doubles~\citep{wan2019does} in ML applications. As the majority of observed patterns fall into the former category, this indicates strong continuity with established testing traditions.  Overall, our findings suggest that practitioners are not reinventing the wheel but are strategically adapting battle-tested patterns to a new domain. The key shift is not in the patterns themselves, but in their frequency and combination.

\begin{table}[t]
\footnotesize
\centering
% \captionsetup{justification=justified, singlelinecheck=false}
\caption{Distribution of testing patterns across agent frameworks, agentic applications, machine learning (ML) applications, and traditional (trad.) software, expressed as percentages and sorted alphabetically by pattern name. Green-highlighted cells denote patterns newly identified in agent-based systems that were not previously reported in either ML or traditional software domains. “NS” indicates the pattern was not studied in the corresponding domain. ML application data (marked with blue superscript\supblue{1}) is from~\citep{openja2024empirical}; traditional software data is sourced from~\citep{lam2018characteristic}\supblue{2}, \citep{zhu2025understanding}\supblue{3}, \citep{zamprogno2022dynamic}\supblue{4}, and \citep{fujita2023empirical}\supblue{5}.}
\label{tab:rq1:result:testing-patterns-by-type}
\begin{tabular}{>{\raggedright\arraybackslash}p{1.4cm} >{\raggedright\arraybackslash}p{2.6cm} 
                >{\raggedleft\arraybackslash}p{1.4cm} 
                >{\raggedleft\arraybackslash}p{1.2cm} 
                >{\raggedleft\arraybackslash}p{1.4cm} 
                >{\raggedleft\arraybackslash}p{1.7cm}}
\toprule
\textbf{Testing type} & \textbf{Testing patterns} & 
\begin{tabular}[c]{@{}r@{}}\textbf{\% Agent}\\\textbf{framework}\end{tabular} & 
\begin{tabular}[c]{@{}r@{}}\textbf{\% Agentic}\\\textbf{apps}\end{tabular} & 
\begin{tabular}[c]{@{}r@{}}\textbf{\% ML}\\\textbf{apps\supblue{1}}\end{tabular} & 
\begin{tabular}[c]{@{}r@{}}\textbf{\% Trad.}\\\textbf{software}\end{tabular} \\
\midrule
\multirow{3}{*}{Structural}
    & \cellcolor{green!50}Hyperparameter Control      & 0.5  & 0.8    & 0   & 0 \\
    & Parameterized Testing       & 28.7 & 26.1 & NS  & 9.0\supblue{2} \\
    & Test Double                 & 23.9 & 20.5 & NS  & 14.7\supblue{3}
 \\
\midrule
\multirow{7}{*}{Verification}
    & Assertion Based Testing     & 81.5 & 72.9 & 24.9 & 56.9\supblue{4} \\
    & \cellcolor{green!50}DeepEval                    & 0.0    & 1.1  & 0.0    & 0.0 \\
    & Membership Testing          & 15.9 & 21.6 & 8.8  & 13.8\supblue{4} \\
    & Mock Assertion              & 14.5 & 10.0   & 1.0    & 4.0\supblue{4} \\
    & Negative Test               & 27.4 & 22.9 & 24.5 & 8.1\supblue{4} \\
    & Snapshot Testing            & 5.3  & 0.5  & 0.0    & 7.0\supblue{5} \\
    & Value Range Analysis        & 3.5  & 6.0  & 9.2  & 2.2\supblue{4} \\
\bottomrule
\end{tabular}
\end{table}

\textbf{To evaluate the non-deterministic response of underlying FM in agentic applications, practitioners leverage a new state-of-the-art verification pattern, i.e., \pattern{DeepEval}.} This pattern represents a necessary non-determinism-specific innovation, as classical assertion mechanisms (e.g., exact matching) cannot accommodate the semantic variability inherent in LLM outputs. Unlike conventional test oracles, DeepEval is designed to orchestrate the evaluation of interactive and non-deterministic model behavior (e.g., creativity) under task-specific criteria, making it particularly suitable for agentic systems~\citep{kim2025evaluating, mohammadi2025evaluation}. In general, DeepEval integrates multiple evaluation techniques, including \textit{G-Eval}~\citep{liu2023g} for assessing answer relevancy, task success, and hallucination detection, and \textit{RAGAS}~\citep{es2024ragas} for evaluating faithfulness and precision. Both approaches leverage the \textit{LLM-as-a-judge} paradigm~\citep{zheng2023judging}, in which a language model interprets test outputs using a set of criteria generated through \textit{Chain-of-Thought prompting}~\citep{wei2022chain}. Additionally, \pattern{DeepEval} supports user-defined logic through a rule composition mechanism based on directed acyclic graphs known as \textit{DAGMetric}~\citep{confident2024deepeval}.

We illustrate a test function that uses \pattern{DeepEval} in Figure~\ref{fig:rq1:result:deepeval} to explain how \pattern{DeepEval} helps agentic applications to tackle the uncertainty from FM. In this example, \texttt{test\_relevant\_content\_retrieved} leverages \pattern{DeepEval} to evaluate a resume retriever agent’s output against a job description and target company. By assessing whether the agent output satisfies a semantic objective rather than matching a fixed string, this pattern directly targets the kind of non-deterministic behavior that arises in agent execution. The test function invokes the G-Eval metric, which operationalizes the LLM-as-a-judge to handle non-deterministic responses. The user-provided evaluation\_steps (line 18) serve as explicit evaluation criteria, instructing the judge to assess the answer's relevancy by verifying that the correct company name is included. This approach validates the semantic correctness of the agent's output, with the test passing if the LLM-as-a-judge's confidence score meets the specified 0.7 threshold.

\begin{figure*}[!t]
\centering
\begin{lstlisting}[
    style=emsepython,
    emph={
        assert_test,
        GEval,
        evaluation_params,
        threshold,
        evaluation_steps
    }
]
def test_relevant_content_retrieved(
    user_proxy: UserProxyAgent,
    resume_retriever_agent: ResumeRetriever,
    job_description: str,
    company: str
) -> None:
    message = "Here is the job description: " + job_description
    chat_outcome = get_chat_outcome(user_proxy, resume_retriever_agent, message)
    assert_test(
        LLMTestCase(
            input=message,
            actual_output=chat_outcome,
            context=[company]
        ),
        [
            GEval(
                name="Inclusion",
                evaluation_params=[
                    LLMTestCaseParams.INPUT,
                    LLMTestCaseParams.ACTUAL_OUTPUT,
                ],
                threshold=0.7,
                evaluation_steps=[
                    f"Check that the output contains the company name '{company}'",
                    f"Check that the output does not contain any company name other than '{company}'",
                ],
            )
        ],
    )
\end{lstlisting}
\caption{Example DeepEval test case that verifies whether the retrieved output includes only the correct company name. Pattern starts by triggering \textit{assert\_test}, and \textit{GEval} is invoked subsequently. Evaluation parameters, threshold, and validation steps are configured in the highlighted lines.}
\label{fig:rq1:result:deepeval}
\end{figure*}

\textbf{To achieve test reproducibility, practitioners employ a hyperparameter control mechanism as a structural pattern while testing agent frameworks.} We illustrate a real-world test function in Figure~\ref{fig:rq1:result:hyperparameter-controle} where ``temperature'', a hyperparameter through which randomness of FM's output can be controlled, is set as 0 (line 6). In general, a higher temperature during generation allows FM to explore more and produce more diverse outputs \citep{xu2022systematic}. However, for the test function where practitioners need reproducible output every time, they set the temperature to 0 or a very low value. Unlike traditional configuration management, this pattern specifically targets the stochastic nature of the underlying model, marking it as an innovation designed to enforce determinism in an inherently probabilistic environment.

\begin{figure*}[!t]
\centering
\begin{lstlisting}[
    style=emsepython,
    emph={temperature}
]
def test_10_concurrent_API_calls(self):
    tools = []
    with open('./data/schemas/get-headers-params.json', 'r') as f:
        tools = ToolFactory.from_openapi_schema(f.read(), {})
    ceo = Agent(
        name='CEO',
        tools=tools,
        instructions="You are an agent that tests concurrent API calls. You must say 'success' if the output contains headers, and 'error' if it does not and **nothing else**."
    )
    agency = Agency([ceo], temperature=0)
    result = agency.get_completion(
        "Please call PrintHeaders tool TWICE at the same time in a single message. "
        "If any of the function outputs does not contain headers, please say 'error'."
    )
    self.assertTrue(
        result.lower().count('error') == 0,
        agency.main_thread.thread_url
    )
\end{lstlisting}
\caption{Test function where temperature is set as 0 for forcing FMs to always select the most probable next token during text generation, eliminating any randomness in the output.}
\label{fig:rq1:result:hyperparameter-controle}
\end{figure*}

\textbf{Adoption of emerging patterns like \pattern{DeepEval} and \pattern{Hyperparameter Control} is negligible (around 1\%), revealing a significant gap between state-of-the-art techniques and practitioner awareness.} Figure~\ref{fig:result:rq1:all-patterns} shows that a mere 1.1\% of test functions of agentic applications leverage DeepEval to validate foundation model outputs, while just 0.5\% and 0.8\% of test functions from agent frameworks and agentic applications utilize hyperparameter control mechanisms, respectively, during test setup. This low adoption rate likely reflects their recent introduction, lack of awareness, or the associated steep learning curve to use these patterns.

\textbf{Parameterized testing is the most heavily adopted structural pattern to handle input variability in agent-based systems, being nearly three times more frequent than in traditional software.} As shown in~\tabref{tab:rq1:result:testing-patterns-by-type}, this pattern appears in 28.7\% of framework tests and 26.1\% of application tests. This widespread use of executing a single test with multiple input values~\citep{tillmann2005parameterized} starkly contrasts with the 9\% adoption rate previously reported in traditional software engineering~\citep{lam2018characteristic}. This pronounced reliance on parameterization suggests that developers of agentic systems require scalable testing practices to validate logic against the wide range of dynamic data and probabilistic outputs inherent to systems driven by foundation models.

\begin{figure*}[t]
\centering

\subfloat[Agent Frameworks]{
  \includegraphics[width=0.48\textwidth]{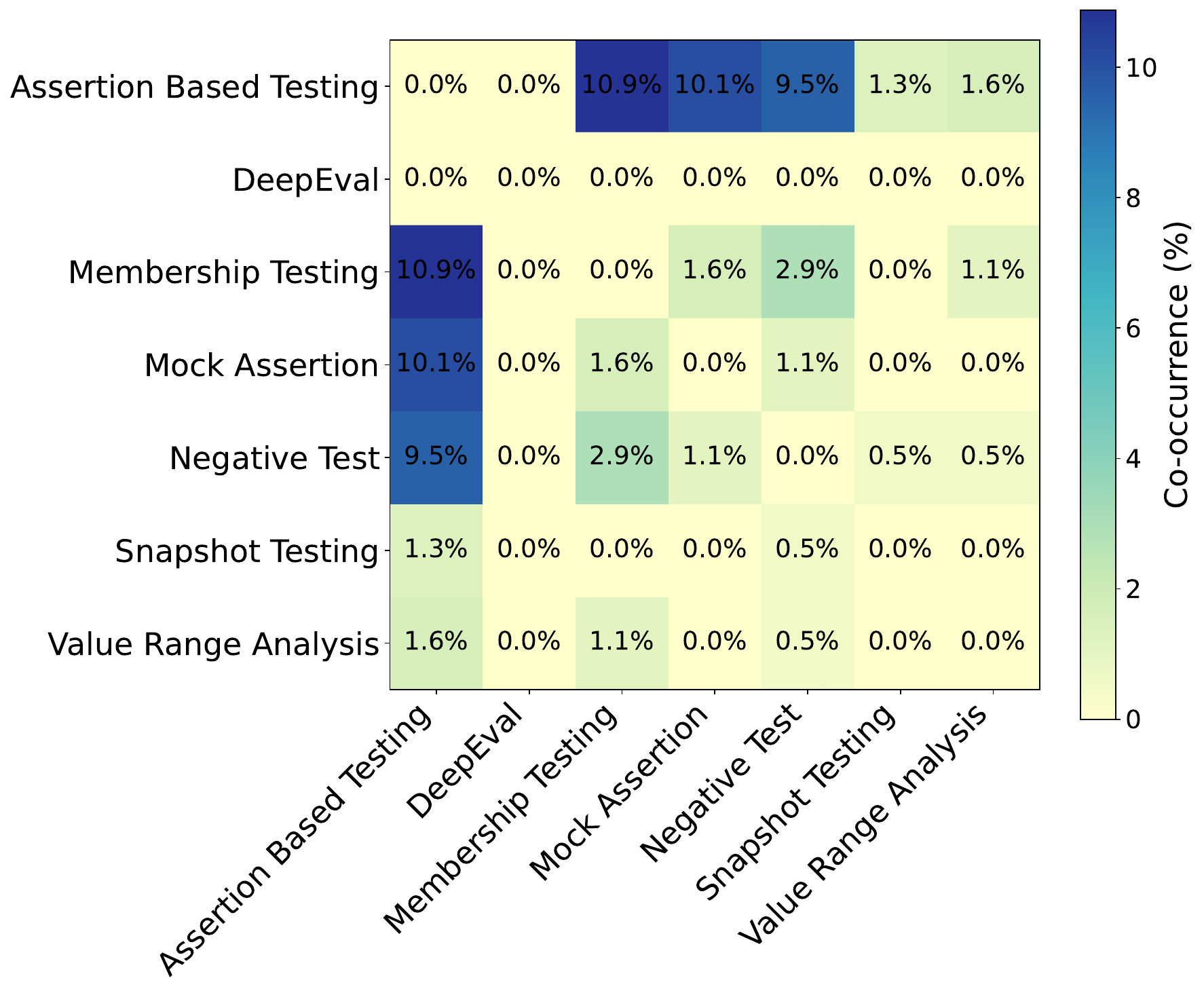}
  \label{fig:rq1:heatmap_of_verification_pattern_cooccurence:framework}
}
\hfill
\subfloat[Agentic Applications]{
  \includegraphics[width=0.48\textwidth]{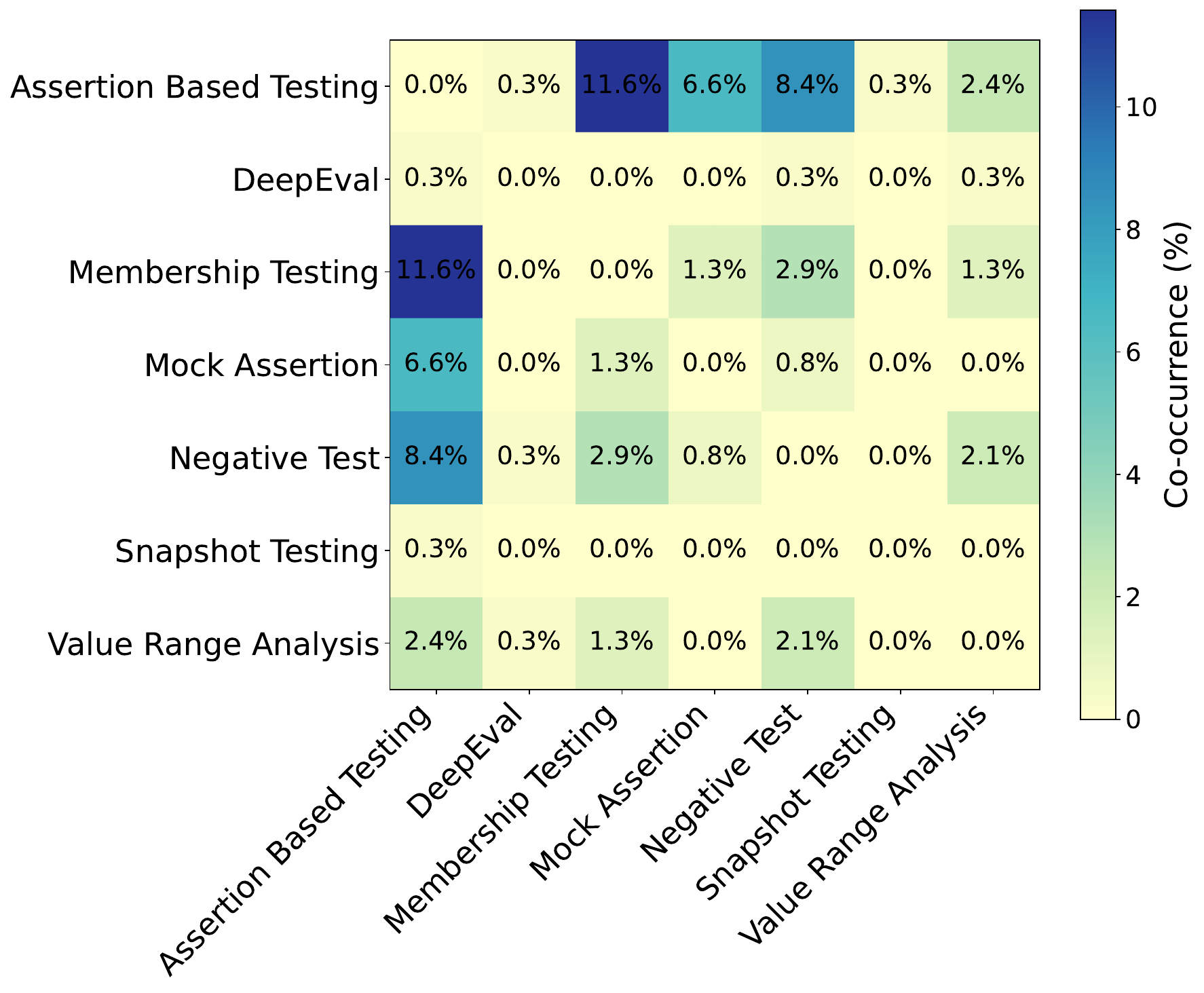}
  \label{fig:rq1:heatmap_of_verification_pattern_cooccurence:application}
}

\caption{Co-occurrence frequency of verification patterns in the same test function in agent frameworks and agentic applications.}
\label{fig:rq1:heatmap_of_verification_pattern_cooccurence}
\end{figure*}

\textbf{While strict assertion-based testing remains the primary verification method (72--82\%), its frequent co-occurrence with more flexible patterns such as membership testing, mock assertion, and negative testing signals a strategic adaptation to the inherent uncertainty in agent frameworks and agentic applications}. As shown in~\tabref{tab:rq1:result:testing-patterns-by-type}, assertion-based testing is employed 3--4 times more frequently than the next most common pattern. Yet, as the co-occurrence matrix in Figure~\ref{fig:rq1:heatmap_of_verification_pattern_cooccurence} reveals, assertion-based tests are often augmented with other verification patterns. In agent frameworks, 40.2\% of test functions using assertion-based verification also include at least one flexible verification strategy, compared to 36.7\% in agentic applications. These complementary patterns offer more adaptable validation: membership testing checks for the presence of a pattern in the output~\citep{openja2024empirical}, mock assertions verify interactions with test doubles, e.g., mock objects~\citep{zhu2025understanding}, and negative testing ensures robust error handling~\citep{openja2024empirical}. This trend highlights a necessary evolution in verification, as the probabilistic and semantically variable responses from FM-based agents demand approaches that balance traditional rigor with the flexibility needed to validate correctness in the face of uncertainty.

\textbf{Highly popular agent frameworks do not adopt testing patterns differently from the rest of the agent ecosystem.} To examine this, we compare the ten most popular agent frameworks (ranked by GitHub star count) against all remaining repositories using Fisher’s Exact Test for each testing pattern, and present our results in Table~\ref{tab:testing-pattern-significance}. The analysis shows that, after correcting for multiple comparisons using FDR, only \pattern{Mock Assertion}-based testing is significantly more prevalent in top-10 frameworks (22.9\% vs.\ 9.1\%, $p=0.0004$, $p_{\text{adj}}=0.0049$). In contrast, all other testing patterns exhibit no statistically significant differences. Similarly, for agentic applications, we do not find any pattern significantly dominating in the top-10 agentic applications.

\begin{table}[t]
\footnotesize
\centering
\caption{Comparison of testing pattern prevalence between top-10 and other repositories. Statistical significance is evaluated using p-value and adjusted p-value (FDR correction).}
\label{tab:testing-pattern-significance}
\begin{tabular}{lrrrrr}
\toprule
\textbf{Testing Pattern} & \textbf{Overall (\%)} & \textbf{Top10 (\%)} & \textbf{Others (\%)} & \textbf{p-value} & \textbf{p\_adj} \\
\midrule
Mock Assertion           & 14.9 & 22.9 & 9.1  & 0.0004 & 0.0049 \\
Test Double              & 24.4 & 29.9 & 20.5 & 0.0389 & 0.1428 \\
Parameterized Testing    & 28.7 & 33.8 & 25.0 & 0.0658 & 0.1449 \\
Assertion Based Testing  & 70.0 & 72.6 & 68.2 & 0.3644 & 0.6293 \\
Functional               & 2.9  & 5.1  & 1.4  & 0.0577 & 0.1449 \\
Value Range Analysis     & 3.4  & 4.5  & 2.7  & 0.4004 & 0.6293 \\
DeepEval                 & 0.0  & 0.0  & 0.0  & 1.0000 & 1.0000 \\
Hyperparameter Control   & 0.5  & 0.0  & 0.9  & 0.5126 & 0.7049 \\
Membership Testing       & 15.9 & 14.7 & 16.8 & 0.6686 & 0.7355 \\
Negative Test            & 24.4 & 22.9 & 25.5 & 0.6272 & 0.7355 \\
Snapshot Testing         & 5.3  & 1.9  & 7.7  & 0.0176 & 0.0972 \\
\bottomrule
\end{tabular}
\end{table}

\begin{footnotesize}
\begin{mybox}{Summary of RQ1}
\begin{itemize}
    \item While practitioners predominantly rely on traditional testing patterns to accommodate the challenges of non-determinism and non-reproducibility in agentic systems, specialized patterns (i.e., \pattern{DeepEval} and \pattern{Hyperparameter Control}) remain rarely adopted despite being specifically designed to address these issues. 
    \item This limited adoption might highlight either a lack of awareness within the developer community, the steep learning curve associated with these novel patterns, or insufficient integration with existing testing frameworks and tools. 
\end{itemize}
\end{mybox}
\end{footnotesize}

\section{RQ2: How do these testing practices map to the architectural components of the agent frameworks and agentic applications?}
\label{sec:rq2-results}

Identifying testing patterns alone is insufficient without understanding where they are applied within agent architectures. FM-based agents comprise heterogeneous components with fundamentally different failure modes, ranging from deterministic infrastructure to non-deterministic decision logic. Consequently, uniform testing effort across components would be both impractical and suboptimal.

This RQ examines how practitioners distribute testing effort across architectural components and whether testing strategies align with component-specific risks. By mapping observed tests to a stable canonical architecture, we aim to reveal systematic testing blind spots and prioritization biases that would otherwise remain obscured by framework-specific implementations.
\subsection{Approach}
We begin by manually identifying the subjects under test (SUTs) from each test function, as outlined in~\secref{subsubsec:extracting-suts}. These SUTs are then mapped to canonical agent architectural components using a closed card-sorting process based on the taxonomy described in~\secref{subsubsec:closed-coding-components}. Finally, we analyze the co-occurrence of testing patterns (\tabref{tab:rq1:result:testing-patterns-by-type}) within these canonical components to determine which testing strategies are most commonly applied to which parts of the system.

\subsection{Findings}
In this subsection, we first define all the SUTs observed in the sampled test functions during our manual review process and map them to the canonical components of the JaCaMo framework. Next, we present the prevalence of these components in the test functions. Finally, we report the component-wise testing patterns, including those defined in~\tabref{tab:rq1:result:testing-patterns-by-type}.

\begin{figure}[ht]
    \centering
    \includegraphics[width=0.98\linewidth]{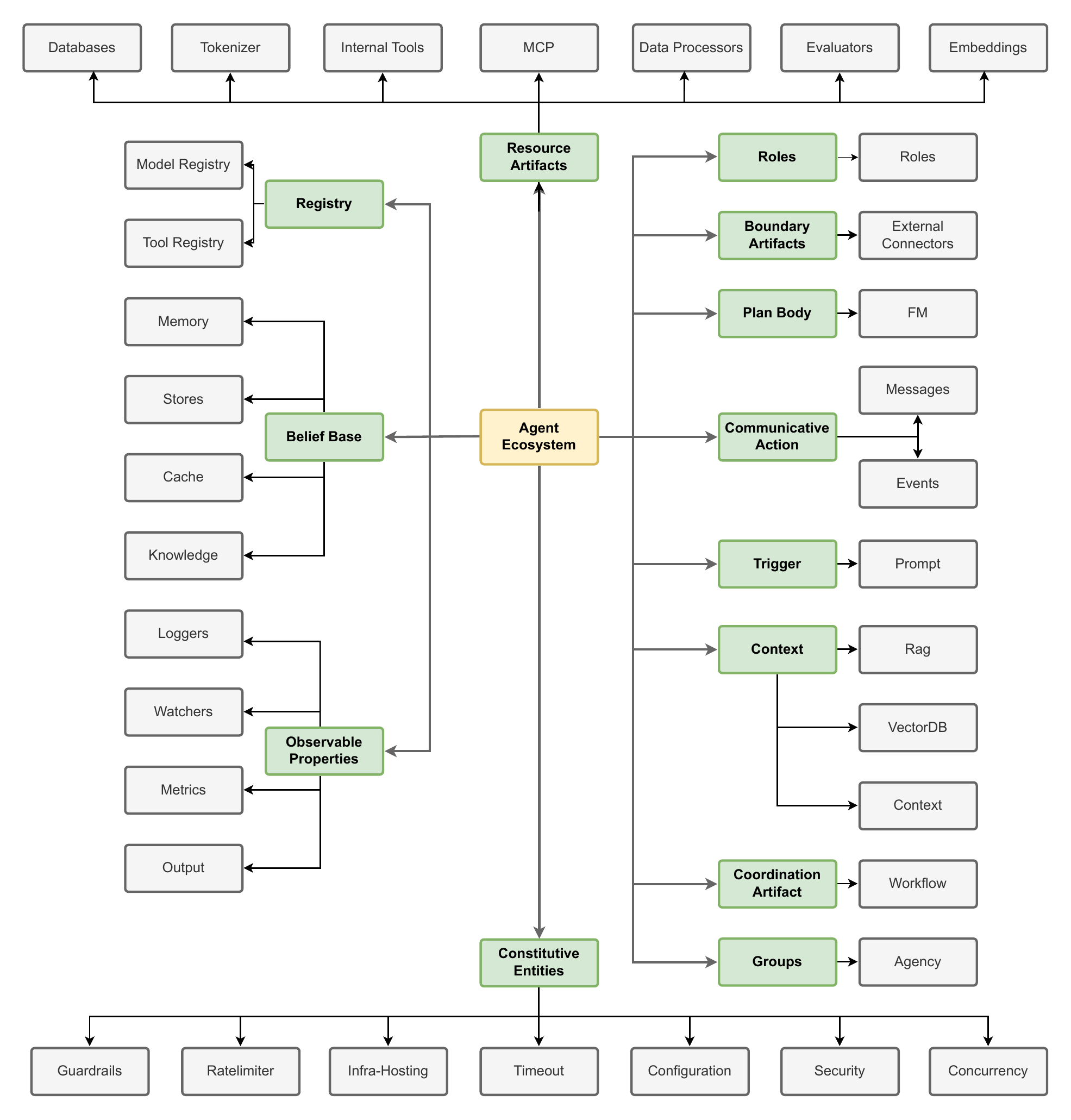}
    \caption{An overview illustrating the mapping between SUTs and canonical agent architectural components extracted from the sample dataset. The green nodes represent architectural components derived from existing literature, while the gray leaf nodes correspond to specific SUTs identified from the sampled test functions.}
    \label{fig:result:rq2:sut-component-mapping}
\end{figure}
\subsubsection{Component Mapping With SUT}\label{sec:result:rq2:component_list}
\textbf{We found that a total of 35 different SUTs are being tested, and mapped these SUTs to 13 canonical agent components}. An overview of the mapping is illustrated in Figure~\ref{fig:result:rq2:sut-component-mapping}. We identify at least one and a maximum of seven SUTs that fall under different architectural components.~\tabref{tab:sut_components} provides a high-level description of the SUTs under each component.

\begin{longtable}{>{\raggedright\arraybackslash}p{2.5cm} >{\raggedright\arraybackslash}p{2cm} >{\raggedright\arraybackslash}p{6.4cm}}
    \caption{Summary of System Under Test (SUT) Components and Characteristics ordered in the same order the components are described in \secref{sec:background:canonical-agent-architecture}}\label{tab:sut_components} \\
    \toprule
    \textbf{Component} & \textbf{SUT} & \textbf{Characteristics} \\
    \midrule
    \endfirsthead

    \toprule
    \textbf{Component} & \textbf{SUT} & \textbf{Characteristics} \\
    \midrule
    \endhead

    \bottomrule
    \endlastfoot
     % Belief Base (Merged Rows)
    \multirow[t]{4}{2.5cm}{Belief Base} 
    & Memory & Persistent storage of trial history, observations, and learned knowledge~\citep{zhang2025survey}. \\
    & Cache & Short-term semantic cache that holds recently accessed results for reuse~\citep{bang2023gptcache, zhang2025cost}. \\
    & Filestore \& Document Store & Structured or unstructured document repositories used for long-term reference. \\
    & Knowledge Base & Runtime-accessible structured knowledge repositories integrated into the agent's memory framework. \\ \midrule

    % Trigger
    Trigger & Prompt & Text-based inputs that provide instructions to foundation models (FMs) for generating outputs~\citep{marvin2023prompt}. \\ \midrule

    % Context
    \multirow[t]{3}{2.5cm}{Context} & RAG & Agents supply domain-specific context to FMs via retrieval-augmented generation (RAG), improving relevance and factuality~\citep{liu2025agent}. \\ 
    & VectorDB & Vector databases store contextual and domain knowledge for retrieval during task execution~\citep{barron2024domain}. \\
    & Context & Some test cases provide context through plain text, such as conversation history, apart from formal RAG and vector-based storage. \\ \midrule

    % Plan Body
    Plan Body & Foundation Models & Agents use FMs (e.g., LLMs) to interpret goals, reason, and generate actionable plans. Chain-of-thought reasoning enables structured multi-step task execution~\citep{liu2025agent, shen2024hugginggpt, wei2022chain}. \\ \midrule
    
    % Communicative Action
    Communicative Action & Messages \& Events & Agents receive and dispatch messages or events to collaborate with other agents and systems~\citep{nascimento2023self}. \\ \midrule

    % Resource Artifacts (Merged Rows)
    \multirow[t]{8}{2.5cm}{Resource Artifacts} 
    & Internal Tools & Shared resources that different agents can (re)use to solve common problems, such as document managers, calculators, and command-line tools. \\
    & MCP & Model Context Protocol (MCP) enables standardized discovery and invocation of tools via external MCP servers~\citep{hasan2025modelcontextprotocolmcp}. \\
    & Ranker & Tools that rank or score retrieved results to improve the accuracy of the responses  and decisions~\citep{liu2025agent}. \\
    & Evaluators \& Validators & Modules that evaluate and validate the outputs, prompts, or intermediate artifacts generated by agents or tools. \\
    & Tokenizer & Converts strings into token sequences (IDs) used internally by FMs~\citep{schmidt2024tokenization} and also used for knowledge and retrieving context. \\
    & Embeddings & Semantic vector representations of various inputs (e.g., text, images) used for retrieval and reasoning~\citep{liu2025agent}. \\
    & Databases & Data repositories used for storing structured information during agent execution. \\
    & Data Processor & Serializers, deserializers, parse, or transformers used to prepare or normalize input/output data formats. \\ \midrule

    % Coordination Artifacts
    Coordination Artifacts & Workflow & Defined sequences of actions including perception, planning, action, and adaptation, guiding agent behavior~\citep{li2024survey}. \\ \midrule

    % Boundary Artifacts
    Boundary Artifacts & External Connectors & Interfaces for invoking external services or tools as part of the agent workflow~\citep{liu2025agent}. \\ \midrule
  
    % Registry
    Registry & Tool Registry & Central repository for registering, discovering, and invoking tools and agents; one test function also involved a model registry~\citep{liu2025agent}. \\ \midrule

    % Observable Property (Merged Rows)
    \multirow[t]{4}{2.5cm}{Observable Property} 
    & Loggers & Components that collect telemetry and debugging data from agent executions~\citep{hassan2024rethinking}. \\
    & Watchers & Monitors that detect changes in files, databases, or environments and notify agents accordingly. \\
    & Metrics & Quantitative indicators used to evaluate agent behavior, accuracy, or performance~\citep{liu2025agent}. \\
    & Output & Results produced by agents, such as code, text, reports, or files. \\ \midrule

    % Role
   Role & Roles & Assigned responsibilities or personas that guide agent behavior, tool selection, and orchestration~\citep{liu2025agent}. \\ \midrule

    % Groups
    Groups & Agency & Multiple agents with different roles can collaborate to form an agency. \\ \midrule
    % Constitutive Entity (Merged Rows)
    \multirow[t]{7}{2.5cm}{Constitutive Entity}
    & Rate Limiter & Mechanism that constrains frequency of resource or API access (e.g., FM calls). \\
    & Guardrails & Constraints on inputs/outputs to ensure safety, alignment, and compliance~\citep{liu2025agent}. \\
    & Timeout & Fail-safes that interrupt long-running or unresponsive tasks. \\
    & Security & Authentication and authorization controls for safe agent-environment interactions. \\
    & Infrastructure Hosting & Deployment environments (e.g., Docker, VMs) that host agent services. \\
    & Configuration & Schema-driven configuration files used to tune agent behavior. \\
    & Concurrency & Controls limiting the number of simultaneous actions or requests made by an agent. \\
\end{longtable}

\subsubsection{Component-wise Test Frequency} \label{sec:result:rq2:component_wise_test_freq}

\begin{figure}[ht]
\centering

\subfloat[Distribution of test function count across architectural components.]{
  \includegraphics[width=0.48\linewidth]{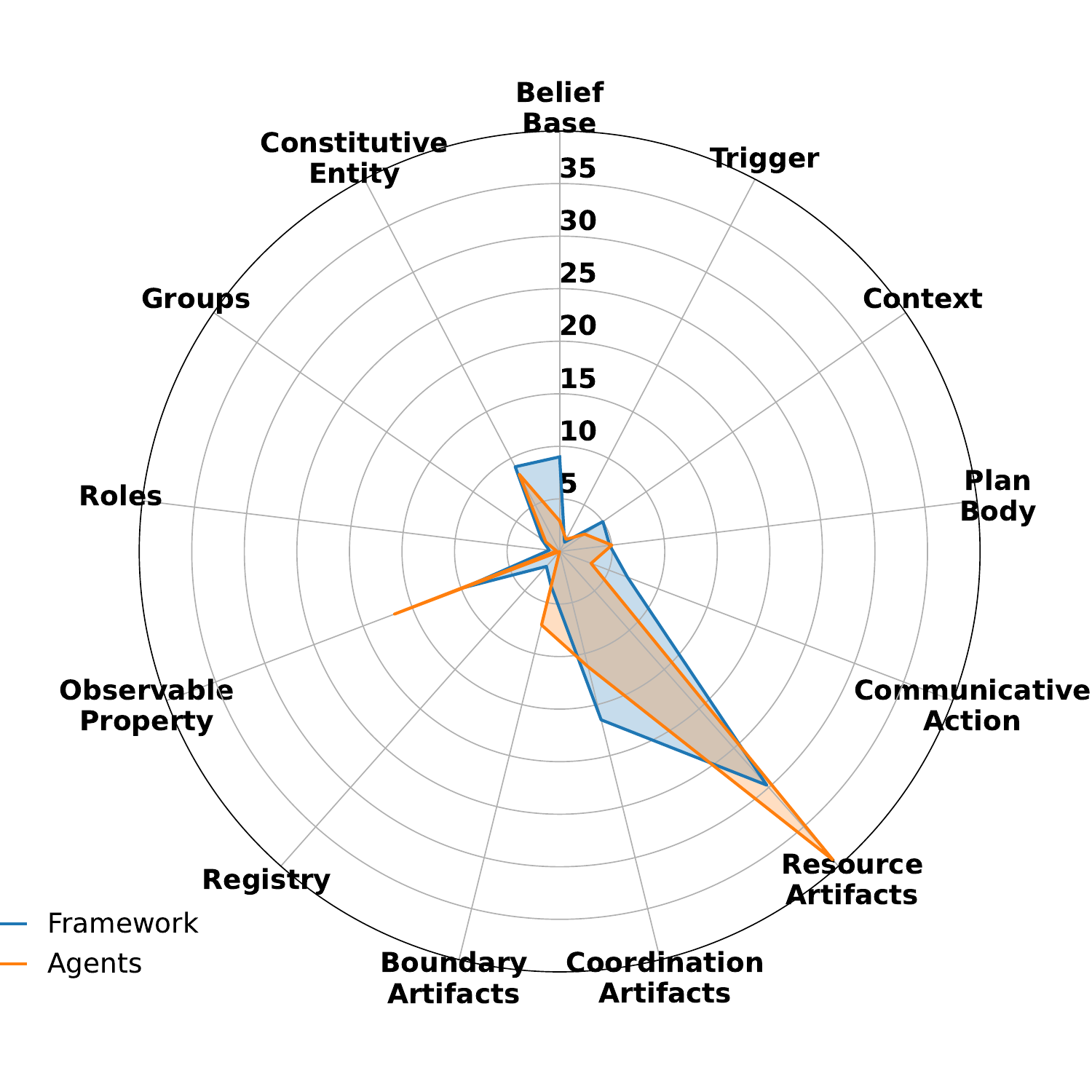}
  \label{fig:result:components:count}
}
\hfill
\subfloat[Distribution of test Lines of Code (LoC) across architectural components.]{
  \includegraphics[width=0.48\linewidth]{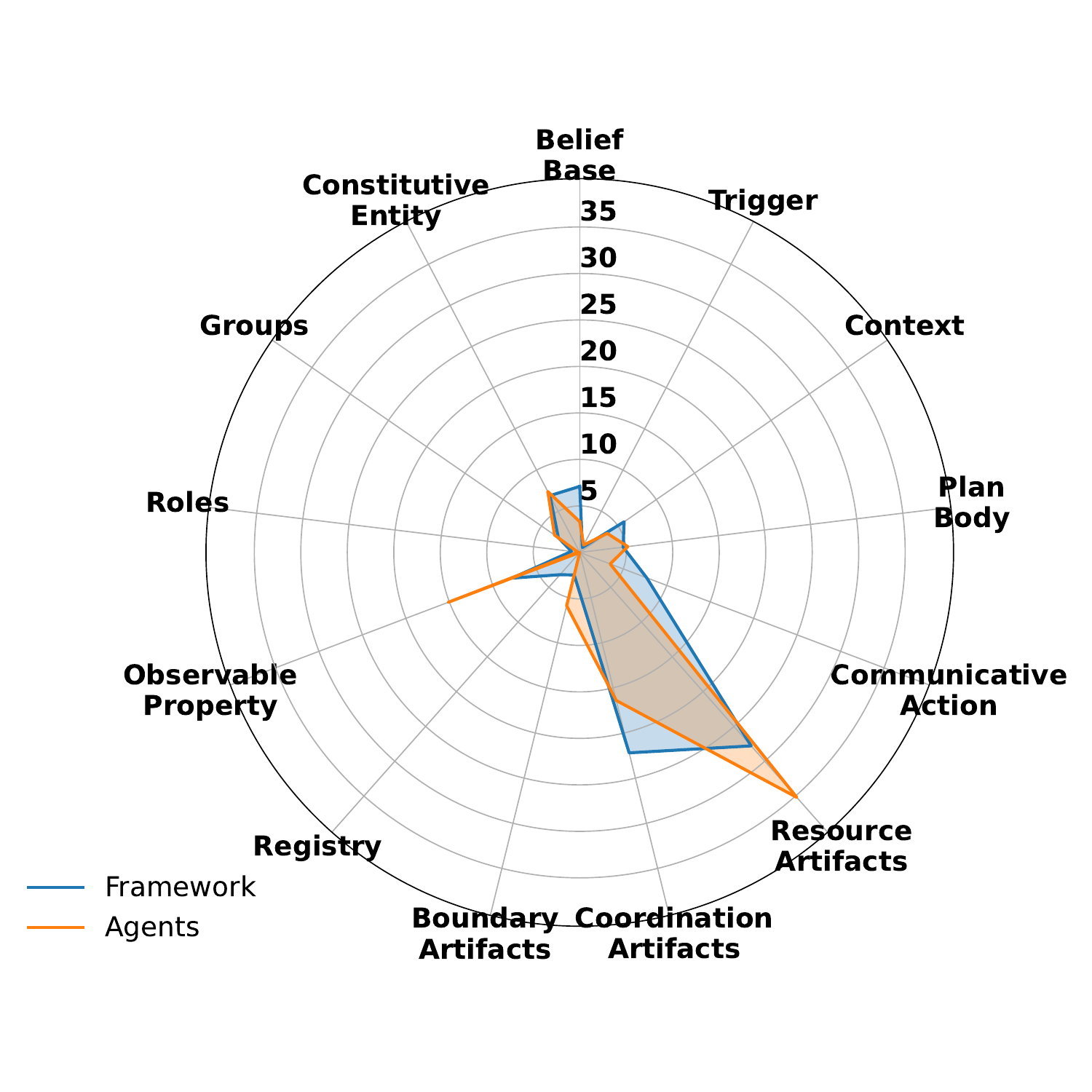}
  \label{fig:result:components:loc}
}

\caption{Components tested in the agent ecosystem.}
\label{fig:result:combined:components}
\end{figure}

\textbf{AI agent testing inverts the traditional ML paradigm, shifting testing effort from the non-deterministic model to the deterministic artifacts around it}. Figure~\ref{fig:result:combined:components} presents the distribution of test functions across architectural components within agent frameworks and agentic applications. To quantify the testing effort, we triangulate our findings using two complementary metrics: (i) the raw count of test functions, a standard proxy in ML testing research~\citep{openja2024empirical}, and (ii) the Lines of Code (LoC) within those tests, a widely accepted measure of implementation effort in software engineering~\citep{labuschagne2017measuring}. Together, these metrics capture both how often components are tested and how much coding effort is invested in testing them. Our analysis shows where the testing effort is invested: \component{Resource Artifacts} receive the lion's share of testing effort, accounting for 29.7\% (tested in 112 functions out of 377 sampled functions) of test functions in frameworks (27.8\% LoC, consuming 1,209 lines out of 4,354 lines of test code) and 39.2\% (tested in 149 functions out of 382 sampled functions) in applications (35.2\% LoC, 1480 out of 4,197 LoC). This is in stark contrast to the \component{Plan Body} (containing the FM), which receives much lower direct testing effort accounting for 4.8\% (tested in 18 functions out of 377 sampled functions) of test functions in frameworks (4.7\% LoC, 203 out of 4,354 LoC) and 5.0\% (tested in 19 functions out of 382 sampled functions) in applications (5.2\% LoC, 220 out of 4,197 LoC). The disparity highlights a dramatic reallocation of engineering effort compared to traditional ML, where model testing typically consumes 24--30\% of the effort~\citep{openja2024empirical}. This suggests that developers are strategically allocating their limited testing resources to components they can reliably control and verify.

\textbf{Despite the Foundation Model’s strong dependency on \component{Triggers} (prompts), testing this component remains critically under-addressed, highlighting a substantial blind spot in ensuring the robustness of agents frameworks and agentic applications}. As shown in Figure~\ref{fig:result:combined:components}, around 1\% of test functions across both frameworks and applications are dedicated to the \component{Trigger} component. This low investment is deeply concerning, as prior work has demonstrated that prompt quality, e.g., example selection, ordering, and contextual richness, significantly affects FM performance~\citep{wu2024prompt}. Moreover, prompt brittleness poses a growing risk as remote foundation models continue to evolve, often breaking existing prompts and degrading agent behavior unless regression testing is in place~\citep{ma2024my}. The near-total absence of prompt regression testing leaves agentic systems highly vulnerable to unexpected decay in functionality, highlighting an urgent need for systematic prompt validation.

\textbf{The top four architectural components tested in both agent frameworks and agentic applications are identical, suggesting potential overlapping of testing effort across development layers}. As shown in Figure~\ref{fig:result:combined:components}, the most frequently tested components in agent frameworks are \component{Resource Artifacts}, \component{Coordination Artifacts}, \component{Observable Property}, and \component{Constitutive Entity}, which together account for 65\% of test functions. Interestingly, these same four components dominate in agentic applications as well, comprising 75.6\% of test functions. This overlap indicates that both framework and application developers are heavily investing in testing similar subsystems, potentially leading to redundant validation across layers. In contrast, other critical components receive limited attention, e.g., \component{Context}, which grounds agent behavior through retrieved knowledge, and \component{Group}, which enables multi-agent collaboration, are seldom tested. This uneven distribution of testing effort suggests a need for further investigation into how redundancy can be minimized and whether resources should be reallocated to under-tested yet crucial components in the agent ecosystem.

\subsubsection{Component Wise Testing Patterns}\label{result:rq2:component_wise_testing_patterns}

\begin{figure*}
    \centering
    \includegraphics[width=0.98\textwidth]{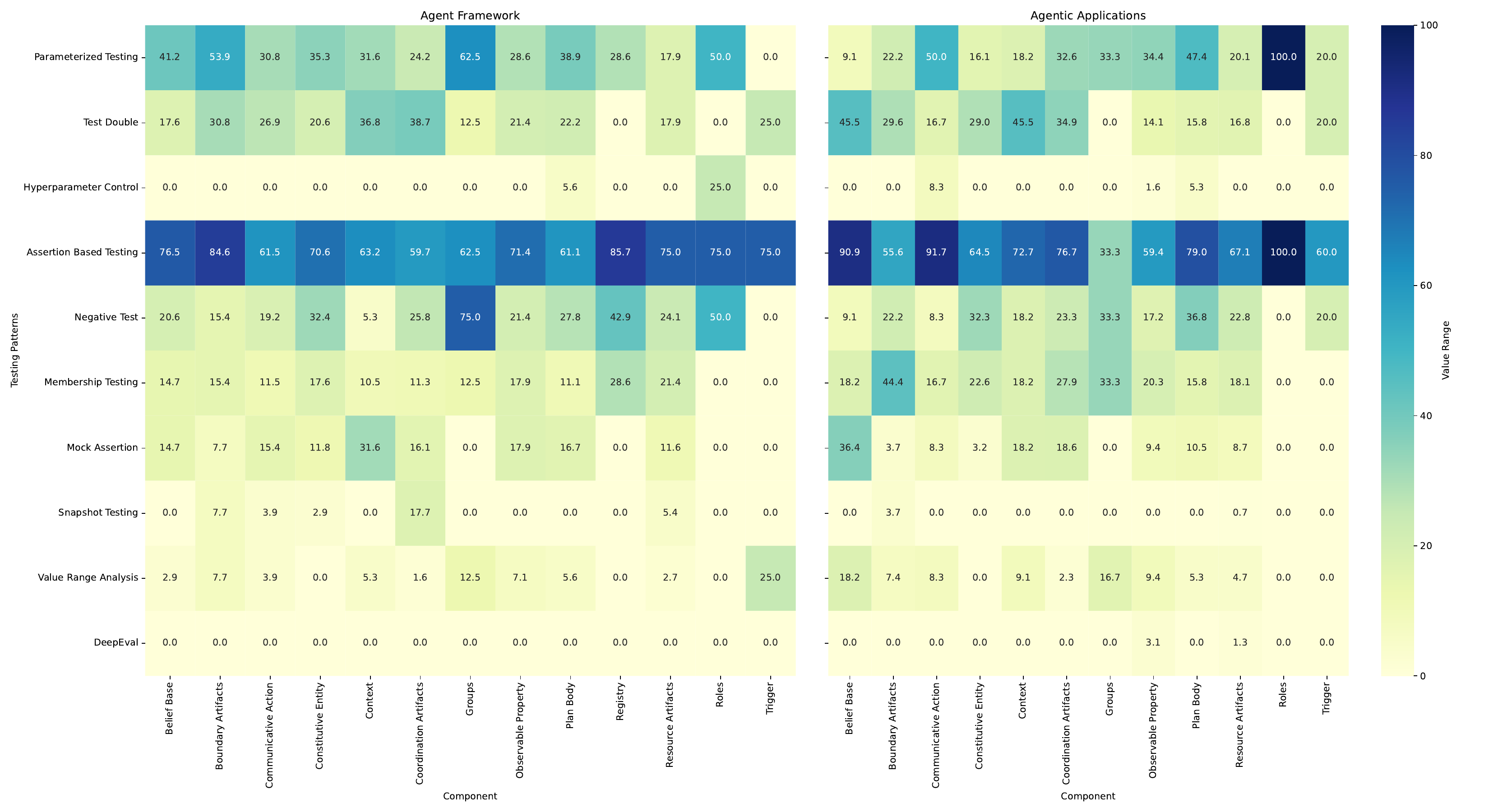}
    \caption{Component-wise Testing Patterns}
    \label{fig:combined:component_wise_testing_patterns}
\end{figure*}

\textbf{\pattern{Assertion-Based Testing} and \pattern{Parameterized Testing} are consistently used across all components in both agent frameworks and agentic applications, highlighting their flexibility and strong practitioner preference}. As shown in Figure~\ref{fig:combined:component_wise_testing_patterns},  these two patterns show consistently high prevalence across the entire component landscape, unlike other testing patterns that show more component-specific adoption. This universal applicability stems from their fundamental utility, e.g., parameterized tests efficiently cover diverse inputs with minimal code, while assertion-based tests provide straightforward, deterministic checks. Both are well-established techniques in traditional software engineering, as discussed in~\secref{sec:rq1-results}, and their widespread usage in the agent ecosystem suggests that practitioners favor these familiar, battle-tested strategies when adapting to the complexities of FM-based agent systems.

\textbf{Although agent frameworks and agentic applications test overlapping architectural components, they adopt distinctly different testing patterns for 69\% of the canonical components, revealing divergent testing priorities and practices}. Figure~\ref{fig:combined:component_wise_testing_patterns} shows that this divergence is prominent in components \component{Belief Base}, \component{Boundary Artifacts}, \component{Communicative Action}, \component{Constitutive Entity}, \component{Coordination Artifacts}, \component{Groups}, \component{Observable Property}, \component{Registry}, and \component{Roles}. The fundamental split in testing philosophy separates agent frameworks and applications: frameworks are tested for universal robustness, while applications are tested for specific, contextual correctness. 

\textbf{For \component{Coordination Artifacts}, this means that frameworks prioritize validating the entire workflow state while applications focus on verifying final outcomes}. Both agent frameworks and agentic applications rely heavily on \pattern{Test Double} for testing \component{Coordination Artifact}. Agent frameworks complement this pattern by using \pattern{Negative Testing}, \pattern{Snapshot Assertion}, and \pattern{Mock Assertion} to a markedly greater extent, reflecting a focus on validating intermediate workflow states. By contrast, agentic applications entirely forgo \pattern{Snapshot Assertion} and instead rely on \pattern{Membership Testing}, adopting it approximately 2.5 times more frequently to accommodate flexible, outcome-oriented validation for this component. This contrast suggests that frameworks emphasize preserving workflow integrity by capturing intermediate states, whereas applications prioritize checking artifact presence with more flexible, less rigid criteria.

\textbf{This diverging behavior continues with \component{Constitutive Entities}, where framework testing aims for comprehensive edge-case coverage, while application testing validates specific, isolated interactions}. Agent frameworks more frequently use \pattern{Parameterized Testing}, \pattern{Negative Testing}, \pattern{Membership Testing}, and \pattern{Mock Assertion}, while agentic applications employ \pattern{Parameterized Testing} and \pattern{Mock Assertion} at 2.2x and 3.5x lower rates, respectively, but use \pattern{Test Double} 1.4x more often. This reflects a difference in emphasis: frameworks favor systematic exploration of edge-case behaviors in environment-defining components, whereas applications focus on simulating and verifying isolated behaviors through mocking and test doubles.

\textbf{This contrast is further evident in testing \component{Boundary Artifacts}, where the use of \pattern{Assertion-Based Testing} drops from 84.6\% in frameworks to 55.6\% in applications, while \pattern{Membership Testing} increases nearly 2.9x in applications.} Both agent frameworks and agentic applications rely on \pattern{Test Double} to isolate third-party integrations. Agent frameworks complement this pattern with greater use of \pattern{Mock Assertion} and \pattern{Snapshot Testing}, reflecting an emphasis on stricter validation of integration behavior. In contrast, agentic applications de-emphasize these techniques and instead favor lighter, context-specific checks aligned with their operational needs.

\textbf{The novel patterns \pattern{DeepEval} and \pattern{Hyperparameter Control} are each confined to just one or two components, explaining how their current use is highly specialized and not yet generalized across the agent ecosystem}. The heatmap in Figure~\ref{fig:combined:component_wise_testing_patterns} shows that these patterns have found their initial niche use-cases. \pattern{DeepEval}, for instance, is used only in agentic applications for testing \component{Observable Property} and \component{Resource Artifacts}. This is its ideal niche: validating ambiguous, FM-generated artifacts, e.g., output like code, summaries, where traditional assertions fail. Likewise, \pattern{Hyperparameter Control} is primarily used to test \component{Roles} within frameworks, where practitioners use hyperparameters (e.g., temperature) to ensure reproducible role-wise agent behavior during testing. It is also applied in the \component{Plan Body} to obtain more consistent FM outputs. While not yet in broad use ( less than 1\% in both agent frameworks and agentic application), their presence in these key areas shows that novel patterns are successfully solving problems that were previously intractable, signaling their potential to expand as the agent ecosystem matures.

\begin{footnotesize}
\begin{mybox}{Summary of RQ2}
\begin{itemize}
    \item Practitioners strategically allocate testing efforts, prioritizing the test of deterministic infrastructure components such as Resource and Coordination Artifacts rather than the testing of non-deterministic models like Plan Bodies, indicating an inversion from traditional ML practices.
    \item While agent frameworks and agentic applications test overlapping canonical components, their testing philosophies diverge across 69\% of components: frameworks emphasize general robustness through rigorous checks, whereas applications focus on context-specific correctness using more adaptive and relaxed patterns.
    \item There are critical testing blind spots in the form of the near-total neglect of the \component{Trigger} (prompt) component, exposing agents to significant risks of performance decay and silent failures as underlying foundation models evolve.
\end{itemize}
\end{mybox}
\end{footnotesize}

\section{Implications}
\label{sec:implications}
Our empirical study reveals a rapidly evolving testing landscape for FM-based AI agent ecosystems. While practitioners are adapting established software engineering principles, significant gaps and strategic misalignments are evident. The findings from our investigation into testing patterns (RQ1) and component-level focus (RQ2) carry profound implications for the key stakeholders in this ecosystem. In this section, we discuss these implications for practitioners and researchers, outlining actionable pathways to foster a more mature, reliable, and secure agentic ecosystem.

\subsection{Implications for Practitioners}
Our findings offer a clear roadmap for both the developers of agent frameworks and the developers building applications upon them. The core message is a call for a more conscious and stratified approach to testing, acknowledging the unique challenges posed by FMs.

\subsubsection{For Agent Framework Developers}
\textbf{Framework developers should integrate advanced semantic verification capabilities, e.g., \pattern{DeepEval}, into their testing infrastructure to address the inherent non-determinism of foundation models}. Our results in Section~\ref{sec:rq1-results} show that while assertion-based testing is dominant, it is fundamentally ill-equipped to handle the semantically variable yet valid outputs from agentic applications in terms of relevancy, accuracy, truthfulness and precision. This forces practitioners into a brittle testing paradigm. The emergence of patterns like \pattern{DeepEval}, despite its low adoption ($\approx1\%$), offers a clear path forward. Frameworks should provide built-in support for such LLM-as-a-judge techniques~\citep{zheng2023judging}. By incorporating evaluators like G-Eval for relevancy and RAGAS for faithfulness~\citep{liu2023g, es2024ragas}, developers can offer users a robust mechanism to validate the intent and semantic correctness of an agent's output, rather than relying on fragile, exact-match string comparisons. This would shift the quality assurance burden from the application developer to the framework, significantly enhancing the reliability of the entire ecosystem.

However, the reluctance to adopt these patterns can be understood in light of practical constraints. The underlying use of LLM-as-a-Judge in these patterns introduces substantial overhead. Unlike traditional unit tests, which execute in milliseconds, semantic evaluation requires invoking a secondary LLM to judge the output produced by the first LLM. This creates a ``token tax,'' i.e., the judge requires the full context plus the generated output, effectively doubling the number of tokens consumed and making such setups impractical for large-scale or production-grade testing~\citep{rajbahadur2024cool}. When reasoning models are used as LLM-as-a-Judge, costs can escalate drastically; recent benchmarks indicate that using top-tier models (e.g., GPT-5-reasoning) for evaluation can cost nearly USD-$79.00$ per 1,000 evaluations~\citep{huang2025llm}. Consequently, latency can become a bottleneck as multiple inference cycles might be required to complete the testing~\citep{huyen2024ai}. 

Beyond resource constraints, reliability remains an issue. AI judges are also probabilistic and prone to criteria ambiguity along with specific biases, e.g., verbosity bias, ordering or self-preference, and often require a model stronger than the application itself to provide a valid critique~\citep{eval2025chip}. Consequently, these judges still require periodic human oversight to ensure the evaluator does not drift~\citep{pitfall2025chip}.

To address the ``token tax’’ and high inference latency, frameworks can facilitate the use of smaller, specialized judge models using curriculum engineering~\citep{rajbahadur2024cool}. Several teams in practice have used specialized, smaller, and less expensive models to effectively validate the outputs of stronger, more expensive models without compromising validity~\citep{huyen2024ai}. Additionally, frameworks can support ``spot-checking'' capabilities by leveraging established test prioritization techniques and hierarchical test trees\linebreak~\citep{yoo2012regression, cruciani2019scalable}. This may allow developers to evaluate only a statistically significant subset of responses rather than the entire test suite, achieving a pragmatic balance between cost and confidence. Furthermore, to mitigate latency bottlenecks, testing tools should support asynchronous architectures, e.g., offloading expensive semantic evaluations to nightly runs or background processes rather than blocking immediate deployment pipelines. Finally, to enhance reliability, frameworks can explicitly manage evaluator bias by adopting strategies such as automatic input permutation, prompt management, and comparative evaluation. This will shift the complexity of robust, cost-effective testing from application developers to the framework itself.

\textbf{Framework developers should establish and promote a testing contract to delineate testing responsibilities between framework and application layers}. The substantial overlap in testing effort, as detailed in Section~\ref{sec:result:rq2:component_wise_test_freq}, suggests redundant work being carried out across the ecosystem. Framework developers can mitigate this by defining a clear testing contract. This contract would specify the robustness guarantees that a compliant framework must provide (e.g., ``all \component{Coordination Artifact}s have been validated against concurrency issues''). In return, it would guide application developers to focus their efforts on their specific logic, trusting the framework's certified guarantees. By publishing these guidelines, creating best-practice documentation, and potentially even offering a ``framework certification'', framework developers can foster specialization, reduce redundant work, and optimize the allocation of testing effort across the entire ecosystem.

\textbf{Framework developers should mitigate the gap between testing of deterministic components and non-deterministic agent artifacts}. In our study, we observed that non-deterministic FMs receive comparatively less attention in testing effort. Beyond tooling limitations, this pattern may be a consequence of deeper cultural and cognitive factors. Prior studies show that practitioners prioritize the controllability and observability of the subject under test when deciding which modules to test~\citep{garousi2019survey}. Controllability ensures consistent outputs for controlled inputs, while observability enables straightforward inspection and interpretation of test results. Deterministic components naturally satisfy these criteria, whereas non-deterministic FMs violate long-standing testing norms rooted in reproducibility and oracle-based validation. In this sense, the shift of testing effort can be understood as a strategic response to current capability limitations rather than a contradiction of them, as prior work also shows that testing strategies often emerge from practical constraints~\citep{whyte2011mitigating}. Because non-deterministic components are harder to control and observe, developers seem to adapt by prioritizing artifacts whose behavior can be verified more feasibly under existing testing constraints.

This mismatch is further reinforced by developers’ implicit acceptance of LLM non-determinism as an inherent system property rather than a testable fault. Recent surveys indicate that practitioners struggle to integrate agents into CI/CD pipelines precisely because these pipelines presuppose deterministic behavior~\citep{pan2025measuring}. As a result, developers often adopt compensatory practices, such as limiting the number of reasoning steps or introducing human-in-the-loop validation, rather than attempting to address or test the non-determinism~\citep{pan2025measuring}. Further, qualitative evidence from open-source development environments reinforces this interpretation and illustrates how practitioners attempt to address these challenges in practice. For example, framework developers explicitly discuss “flaky” behavior arising from non-determinism and attempt to mitigate it~\footnote{\url{https://github.com/langchain-ai/langgraph/pull/319}}, often by mocking LLM clients to enforce consistent reproducibility during testing~\footnote{\url{https://github.com/microsoft/autogen/pull/4096}}. Similarly, practitioners discuss the difficulty of verifying whether agents actually executed tools or merely hallucinated results~\footnote{\url{https://github.com/crewAIInc/crewAI/pull/3388}}. Together, these observations suggest that the lower testing effort devoted to non-deterministic components is not solely a consequence of technical limitations. It also reflects a long-standing culture among developers of expecting deterministic behavior in tests, as well as the implicit acceptance of LLM non-determinism as a given system property. This combination highlights the need for framework developers to actively mitigate the gap between testing of deterministic components and non-deterministic agent artifacts.

\subsubsection{For Agentic Application Developers}
\textbf{Agentic application developers must establish systematic prompt regression suites to mitigate the risks of model evolution and prompt brittleness}. Our analysis in Section~\ref{sec:result:rq2:component_wise_test_freq} exposed a critical blind spot: the \component{Trigger} component (i.e., the prompt) is tested in around 1\% of cases. This is a ticking time bomb for system stability. Because foundation models are frequently updated by their providers, often without detailed information about version differences~\citep{ajibode2025towards}, prompts that worked perfectly yesterday may fail silently tomorrow~\citep{ma2024my}. Agentic applications are uniquely positioned to address this by creating regression testing harnesses for prompts. These harnesses could maintain a curated set of ``golden'' prompts and corresponding semantic outputs, automatically validating them against new or updated FM versions. By treating prompts as first-class, testable artifacts, agentic application developers can provide a crucial layer of stability, safeguarding applications from unforeseen behavioral degradation and ensuring long-term dependability.

\textbf{Developers should adopt hyperparameter control as a fundamental debugging technique to isolate the subject under test from FM-induced non-reproducibility}. The negligible adoption of Hyperparameter Control (0.5\% in frameworks, 0\% in apps) shown in Section~\ref{sec:rq1-results} indicates a missed opportunity. When a test fails, developers face a critical ambiguity: is the defect in their own code, or is it a result of the FM's random output? By setting hyperparameters like temperature to 0, developers can motivate deterministic outputs from the FM, effectively reducing the randomness to create a reproducible testing environment~\citep{xu2022systematic}. This allows them to systematically debug their application's \component{Coordination Artifacts} (workflows) and \component{Resource Artifacts} (tools) without the confounding variable of FM unpredictability. Once the application logic is confirmed to be sound, the temperature can be increased to test the system's behavior under more realistic, non-deterministic conditions.

\textbf{Application developers should shift their testing focus towards business logic and integration points by relying on the robustness guarantees of the underlying framework}. Our analysis in Section~\ref{sec:result:rq2:component_wise_test_freq} revealed a duplication of effort: both framework and application developers concentrate their testing on the same four components (\component{Resource Artifacts}, \component{Coordination Artifacts}, \component{Observable Property}, and \component{Constitutive Entity}). While seemingly diligent, this can be inefficient at times. As shown in Section~\ref{result:rq2:component_wise_testing_patterns}, frameworks test these components for general-purpose robustness (e.g., using Snapshot Testing on workflows), whereas applications test them for specific use cases (e.g., using Membership Testing on outputs). Application developers should offload the responsibility of foundational robustness to the framework. This allows them to redirect their limited testing resources to what truly matters at the application layer: the correctness of their custom tools, the reliability of their Boundary Artifacts (external connectors), and the integrity of their unique business workflows.

\subsection{Implications for Researchers}

\textbf{Researchers should conduct empirical studies to diagnose the barriers preventing the adoption of novel, non-determinism-specific testing patterns}. The contrast between the potential of patterns like \pattern{DeepEval} and \pattern{Hyperparameter Control} and their near-zero adoption (Section~\ref{sec:rq1-results}) points to a significant gap between the state-of-the-art and the state-of-the-practice. Is this gap caused by a lack of awareness, tooling immaturity, perceived complexity, or a cultural adherence to traditional testing paradigms? Future research, employing qualitative methods like surveys and interviews with practitioners, could uncover these root causes. The resulting insights would be invaluable for designing more intuitive testing tools, developing targeted educational materials, and creating theories of technology adoption tailored to the rapidly evolving FM-based AI agent engineering landscape.

\textbf{Researchers should formalize a comprehensive testing methodology for agentic systems that accounts for their unique architectural and behavioral characteristics}. Our study revealed two foundational shifts in testing practice: the strategic adaptation of ``less strict'' patterns (RQ1) and the inversion of testing effort away from the non-deterministic model and towards the deterministic artifacts and infrastructure (RQ2). To enhance the implications of our findings, we synthesized best practices from traditional Software Engineering (SE), classical Machine Learning (ML), and Large Language Model (LLM) application domains. We queried Google Scholar using terms such as ``unit testing best practices'', ``machine learning system testing best practices'', ``testing llm applications'', and ``testing large language model based software''.~\tabref{tab:recommendations_synthesis} maps these recommendations to the adoption behaviors observed in agent frameworks and agentic applications in our study. Our empirical results show that agent frameworks and agentic applications fully adopt, partially adopt, adopt in a simplified form, or do not adopt these inherited practices.

From traditional SE, we observe partial adoption of deterministic testing, code coverage, and systematic test design in both RQ1 and RQ2. These practices are applied primarily to deterministic system shell components, but are not used across full agent workflows. The strict recommendation of deterministic oracles from SE~\citep{daka2014survey} is often infeasible for non-deterministic model components in FM-based agents.

From classical ML recommendations, regression testing practices are partially adopted in agent systems, whereas metamorphic and differential testing are not, as property-based validation remains difficult to define for open-ended agent outputs~\citep{chandrasekaran2023test}. Negative testing is partially adopted, particularly for constitutive entities, whereas adversarial fuzzing remains rare in practice.

From the LLM application domain, several recommendations are either fully or partially adopted. For instance, probabilistic assertions on top of deterministic tests are being partially adopted in agent frameworks and agentic applications. For example, in~\secref{appendix-deepeval}, we illustrate a real-world test function that adopts probabilistic testing, which represents a paradigm shift over deterministic testing. Also, layered testing is fully adopted across most systems. Curriculum-based AI judges and controlled execution for repeatability are adopted in simplified form, for example, through patterns such as \pattern{DeepEval}, 
\pattern{Parameterized Testing}, and \pattern{Mock Assertion}. In contrast, aggregated oracles and prompt perturbation for robustness have not yet been adopted. Component-aware testing strategies are adopted in simplified form, often implicitly, without formal guidance.

Overall, this comparison suggests that agent testing is best understood as a hybrid regime. It adopts mature deterministic testing practices for infrastructure, adapts evaluation and oracle strategies from LLM testing to accommodate expected non-determinism, and finds some classical ML testing strategies not directly transferable without component-specific reformulation.
\begin{table}[H]
\small
\centering
\caption{Comparison of academic testing recommendations and their adoption in agentic systems. The bracketed numbers in the \textbf{Domain} column denote the following sources: 
\textbf{Traditional SE}: [1]~\citep{daka2014survey}; 
\textbf{Classical ML}: [2]~\citep{chandrasekaran2023test}, [3]~\citep{openja2024empirical}; 
\textbf{LLM Applications}: [4]~\citep{dobslaw2025challenges}, [5]~\citep{ma2025rethinking}, [6]~\citep{rajbahadur2024cool}.}
\label{tab:recommendations_synthesis}
\begin{tabular}{p{4cm} l p{2cm} p{5cm}}
\toprule
\textbf{Recommendation} & \textbf{Domain} & \textbf{Adoption Status} & \textbf{Remarks} \\
\midrule
% --- SE Section ---
Use deterministic tests & SE [1] & Adopted Partially & Strictly applied to deterministic infrastructure (System Shell), but infeasible for the non-deterministic model core. \\ \addlinespace

Code coverage and systematic test design & SE [1] & Adopted Partially & High coverage is pursued for \component{Resource Artifacts}, but widely ignored for \component{Triggers} \\ \addlinespace

% --- ML Section ---
Metamorphic and differential testing & ML [2] & Not Adopted & Formal property-based relations are difficult to define for open-ended agent outputs. \\ \addlinespace

Regression testing practices & ML [3] & Adopted Partially & Observed in deterministic artifacts, though often missing critical regression selection strategies for prompts. \\ \addlinespace

Negative and adversarial testing & ML [2] & Adopted Partially & Negative testing is common for constitutive entity, but adversarial fuzzing is rare. \\ \addlinespace

% --- LLM Section ---
Aggregated oracles (voting, variance) & LLM [4] & Not Adopted & Systematic aggregation (e.g., majority voting) is never formalized. \\ \addlinespace

Prompt perturbation for robustness & LLM [4] & Not Adopted & Systematic perturbation to test prompt fragility is theoretically recommended but absent in practice. \\ \addlinespace

Probabilistic assertions & LLM [5] & Adopted Fully & Widely adopted as ``less strict'' assertions (semantic similarity, embedding checks). \\ \addlinespace

Layered testing (System Shell vs.\ Model) & LLM [5] & Adopted Fully & Traditional testing patterns for deterministic components (system shell) and less strict patterns for non-deterministic components (model) \\ \addlinespace

Curriculum-based AI judges & LLM [6] & Adopted in Simplified Form & Advanced judge training is not adopted; however, simpler judge implementations (e.g., DeepEval) are emerging. \\ \addlinespace

Controlled execution for repeatability & LLM [6] & Adopted in Simplified Form & Repeatability is approximated through parameterized testing and controlled configuration through test doubles, mock assertion and hyper-parameter control.
 \\ \addlinespace

Component-aware testing strategies & LLM [5] & Adopted in Simplified Form & Developers implicitly tailor testing to agent components (e.g., boundary artifacts). \\
\bottomrule
\end{tabular}
\end{table}
We therefore call for a unified ``Theory of Agent Testing'' that makes these transfer boundaries explicit and provides prescriptive guidance linking canonical agent components to suitable testing strategies, thereby moving practice from ad hoc solutions toward disciplined, evidence-based testing.

% !TeX root = 0-main.tex

\section{Threats to Validity}
\label{sec:threats}
\subsection{External Threats}
Our initial framework identification process was dependent on the results of GitHub’s keyword-based search API. This method is susceptible to search biases and may not capture all relevant repositories, posing a threat of an incomplete sample. To address this, we supplemented automated discovery with manual, domain-expert verification. For instance, the popular \texttt{langchain} framework was not returned by our initial queries. However, our investigation revealed that the \texttt{langchain-ai} organization now advocates for \texttt{langgraph}\footnote{\url{https://github.com/langchain-ai/langchain?tab=readme-ov-file\#langchains-ecosystem}} for building agentic systems, which was included in our final dataset. Similarly, \texttt{babyagi}, another popular framework in the research domain, was not included in our search results, and our manual analysis suggests that the framework’s repository explicitly warns against production use.\footnote{\url{https://github.com/yoheinakajima/babyagi}} These evidence suggest that while we cannot guarantee our list is exhaustive, this two-stage process of automated search followed by expert curation significantly reduces the risk of omitting key, production-relevant frameworks and strengthens the representativeness of our sample.

Our method for identifying agentic applications by searching for specific import statements presents two potential validity threats: (i) Any project that utilizes an agent framework through unconventional means, such as dynamic imports or custom wrappers that obscure the standard import statements, would not be included in our initial dataset. (ii) Noisy projects, e.g., projects that imported one or more agent frameworks for a minor, non-agentic utility, can be part of the dataset. 

Our study focuses on identifying how different components of agentic systems are tested in practice, rather than empirically evaluating how variations or failures in those components affect system behavior. For example, while we report limited prompt testing in current practice, analyzing how prompt variations influence agent behavior is outside the scope of this study. Instead, we refer to prior component-specific research that investigates such effects.

\subsection{Construct Threats}
To ground the architectural components of agent frameworks and agentic applications, we mapped each System Under Test (SUT) extracted from test functions to one of the canonical components defined in prior literature\linebreak(\secref{sec:background:canonical-agent-architecture}). This mapping task introduces the possibility of rater bias. To mitigate this threat, we employed a closed card-sorting methodology with two independent raters assigning SUTs to predefined components. Disagreements were resolved through discussion, and inter-rater reliability was measured using Cohen’s Kappa, yielding a substantial agreement score of $\kappa = 0.83$.

In comparing testing patterns across domains (e.g., software engineering, and ML applications), we faced inconsistencies in pattern granularity across source studies. Some testing patterns reported in the literature were more fine-grained than others (e.g., \texttt{equality} vs. \texttt{Assertion Based Testing}). To ensure a valid comparison, we normalized frequencies by converting raw counts to relative proportions within each domain, and we aggregated semantically similar sub-patterns into broader parent categories based on prior taxonomies\linebreak~\citep{openja2024empirical}. This allowed for consistent, meaningful comparison while preserving the underlying structure of each domain’s testing practices.

Finally, our reliance on the \texttt{test\_*} filename convention to identify test files constitutes a construct validity threat. While this aligns with standard practices in Python ecosystems (e.g., pytest discovery rules) and prior empirical studies~\citep{jebnoun2020scent, openja2024empirical}, it may exclude repositories that employ other naming schemes or embed tests within production code. As a result, our findings are scoped to repositories that follow explicit prefix-based test file naming conventions. Investigating alternative testing structures and assessing whether repositories that follow other conventions exhibit different testing practices remain important directions for future work.

\section{Conclusion}
\label{sec:conclusion}

This study provides the first large-scale empirical evidence of how practitioners are navigating the complex and uncertain terrain of testing FM-based AI agents. By analyzing a large-scale dataset of 39 agent frameworks and 439 agentic applications, we reveal that testing patterns in this domain reflect a pragmatic, yet incomplete, adaptation to the unique challenges posed by non-determinism and evolving architectural complexity.

Our findings show that practitioners predominantly rely on adapted traditional testing strategies, e.g., assertion-based testing is complemented by negative testing and membership assertions in 40.2\% of agent framework test functions and 36.7\% of agentic application test functions. However, the limited adoption of domain-specific techniques (e.g., \texttt{DeepEval}, used in only 1.1\% of agentic application test functions) indicates a notable gap in awareness and usability. Furthermore, we provide a list of 13 canonical components of agent architecture that can accommodate both existing and future components of agent frameworks and agentic applications. Mapping the testing patterns to these components, we reveal a critical blind spot, namely the near-total neglect of the \component{Trigger} component, which is covered in only around 1\% of test functions. This component, serving as the prompt interface that directly governs the FM, exposes the entire ecosystem to risks of semantic drift, performance decay, and silent failures. 

The significance of these findings is that the path to reliable AI agents does not lie in discarding established engineering principles, but in augmenting them with targeted, new methodologies. Our work lays the groundwork for more resilient agent development by offering a concrete empirical baseline and highlighting where testing effort must be rebalanced.

\section*{Declaration}
\label{sec:conflict}

\subsection*{Funding}
Not applicable.

\subsection*{Ethical Approval}
This study does not involve human participants or animals.

\subsection*{Informed Consent}
Not applicable. No human subjects were involved in this study.

\subsection*{Data Availability Statement}
The dataset, experiment code, and experiment results of this study are available in our replication package.\footnote{
\url{https://github.com/SAILResearch/replication-25-agent-testing-empirical-study}}

\subsection*{Conflict of Interest}
The authors declared that they have no known competing interests or personal relationships that could have (appeared to) influenced the work reported in this article.

\subsection*{Clinical Trial Number in the Manuscript}
Not applicable.

\subsection*{Author Contributions}
\begin{itemize}
    \item Mohammed Mehedi Hasan: Conceptualization, Data Collection, Methodology, Data Analysis, Writing – Original
Draft.
    \item Hao Li: Methodology, Implication, Data Validation, Writing – Review \& Editing.
    \item  Emad Fallahzadeh: Writing – Review \& Editing, Conceptual Guidance, Research Direction.
    \item  Gopi Krishnan Rajbahadu: Data Validation, Writing – Review \& Editing,  Conceptual Guidance, Research Direction.
    \item Bram Adams: Writing – Review, Supervision, Research Direction.
    \item Ahmed E. Hassan: Supervision, Research Direction.
\end{itemize}

\section{Data Availability Statement}
\label{sec:availability}

The datasets generated and analyzed during this study are available in the replication package.

\bibliographystyle{IEEEtranN} %sorted by order of appearance
\bibliography{references} 

% Generated by IEEEtranN.bst, version: 1.14 (2015/08/26)
\begin{thebibliography}{119}
\providecommand{\natexlab}[1]{#1}
\providecommand{\url}[1]{#1}
\csname url@samestyle\endcsname
\providecommand{\newblock}{\relax}
\providecommand{\bibinfo}[2]{#2}
\providecommand{\BIBentrySTDinterwordspacing}{\spaceskip=0pt\relax}
\providecommand{\BIBentryALTinterwordstretchfactor}{4}
\providecommand{\BIBentryALTinterwordspacing}{\spaceskip=\fontdimen2\font plus
\BIBentryALTinterwordstretchfactor\fontdimen3\font minus \fontdimen4\font\relax}
\providecommand{\BIBforeignlanguage}[2]{{%
\expandafter\ifx\csname l@#1\endcsname\relax
\typeout{** WARNING: IEEEtranN.bst: No hyphenation pattern has been}%
\typeout{** loaded for the language `#1'. Using the pattern for}%
\typeout{** the default language instead.}%
\else
\language=\csname l@#1\endcsname
\fi
#2}}
\providecommand{\BIBdecl}{\relax}
\BIBdecl

\bibitem[Liu et~al.(2023{\natexlab{a}})Liu, Yu, Zhang, Xu, Lei, Lai, Gu, Ding, Men, Yang, et~al.]{liu2023agentbench}
X.~Liu, H.~Yu, H.~Zhang, Y.~Xu, X.~Lei, H.~Lai, Y.~Gu, H.~Ding, K.~Men, K.~Yang \emph{et~al.}, ``Agentbench: Evaluating llms as agents,'' \emph{arXiv preprint arXiv:2308.03688}, 2023.

\bibitem[Hettiarachchi(2025)]{hettiarachchi2025exploring}
I.~Hettiarachchi, ``Exploring generative ai agents: Architecture, applications, and challenges,'' \emph{Journal of Artificial Intelligence General science (JAIGS) ISSN: 3006-4023}, vol.~8, no.~1, pp. 105--127, 2025.

\bibitem[Park et~al.(2023)Park, O'Brien, Cai, Morris, Liang, and Bernstein]{park2023generative}
J.~S. Park, J.~O'Brien, C.~J. Cai, M.~R. Morris, P.~Liang, and M.~S. Bernstein, ``Generative agents: Interactive simulacra of human behavior,'' in \emph{Proceedings of the 36th annual acm symposium on user interface software and technology}, 2023, pp. 1--22.

\bibitem[Wu et~al.(2024{\natexlab{a}})Wu, Bansal, Zhang, Wu, Li, Zhu, Jiang, Zhang, Zhang, Liu, et~al.]{wu2024autogen}
Q.~Wu, G.~Bansal, J.~Zhang, Y.~Wu, B.~Li, E.~Zhu, L.~Jiang, X.~Zhang, S.~Zhang, J.~Liu \emph{et~al.}, ``Autogen: Enabling next-gen llm applications via multi-agent conversations,'' in \emph{First Conference on Language Modeling}, 2024.

\bibitem[Mialon et~al.(2023)Mialon, Fourrier, Wolf, LeCun, and Scialom]{mialon2023gaia}
G.~Mialon, C.~Fourrier, T.~Wolf, Y.~LeCun, and T.~Scialom, ``Gaia: a benchmark for general ai assistants,'' in \emph{The Twelfth International Conference on Learning Representations}, 2023.

\bibitem[Zhou et~al.(2023)Zhou, Xu, Zhu, Zhou, Lo, Sridhar, Cheng, Ou, Bisk, Fried, et~al.]{zhou2023webarena}
S.~Zhou, F.~F. Xu, H.~Zhu, X.~Zhou, R.~Lo, A.~Sridhar, X.~Cheng, T.~Ou, Y.~Bisk, D.~Fried \emph{et~al.}, ``Webarena: A realistic web environment for building autonomous agents,'' \emph{arXiv preprint arXiv:2307.13854}, 2023.

\bibitem[Wang et~al.(2023)Wang, Wang, Liu, Chen, Yuan, Peng, and Ji]{wang2023mint}
X.~Wang, Z.~Wang, J.~Liu, Y.~Chen, L.~Yuan, H.~Peng, and H.~Ji, ``Mint: Evaluating llms in multi-turn interaction with tools and language feedback,'' \emph{arXiv preprint arXiv:2309.10691}, 2023.

\bibitem[Gevers et~al.(2025)Gevers, De~Marez, Van~Nooten, Lemmens, Kosar, Lotfi, Banar, Fivez, De~Bruyne, and Daelemans]{gevers2025benchmarks}
I.~Gevers, V.~De~Marez, J.~Van~Nooten, J.~Lemmens, A.~Kosar, E.~Lotfi, N.~Banar, P.~Fivez, L.~De~Bruyne, and W.~Daelemans, ``In benchmarks we trust... or not?'' in \emph{Proceedings of the 2025 Conference on Empirical Methods in Natural Language Processing}, 2025, pp. 23\,673--23\,687.

\bibitem[Ma et~al.(2025)Ma, Yang, Hu, Ying, Jin, Du, Xing, Li, Shi, Liu, et~al.]{ma2025rethinking}
W.~Ma, Y.~Yang, Q.~Hu, S.~Ying, Z.~Jin, B.~Du, Z.~Xing, T.~Li, J.~Shi, Y.~Liu \emph{et~al.}, ``Rethinking testing for llm applications: Characteristics, challenges, and a lightweight interaction protocol,'' \emph{arXiv preprint arXiv:2508.20737}, 2025.

\bibitem[Niedermayr et~al.(2016)Niedermayr, Juergens, and Wagner]{niedermayr2016will}
R.~Niedermayr, E.~Juergens, and S.~Wagner, ``Will my tests tell me if i break this code?'' in \emph{Proceedings of the International Workshop on Continuous Software Evolution and Delivery}, 2016, pp. 23--29.

\bibitem[Mohammadi et~al.(2025)Mohammadi, Li, Lo, and Yip]{mohammadi2025evaluation}
M.~Mohammadi, Y.~Li, J.~Lo, and W.~Yip, ``Evaluation and benchmarking of llm agents: A survey,'' in \emph{Proceedings of the 31st ACM SIGKDD Conference on Knowledge Discovery and Data Mining V. 2}, 2025, pp. 6129--6139.

\bibitem[Ma et~al.(2024)Ma, Yang, and K{\"a}stner]{ma2024my}
W.~Ma, C.~Yang, and C.~K{\"a}stner, ``(why) is my prompt getting worse? rethinking regression testing for evolving llm apis,'' in \emph{Proceedings of the IEEE/ACM 3rd International Conference on AI Engineering-Software Engineering for AI}, 2024, pp. 166--171.

\bibitem[Razavi et~al.(2025)Razavi, Soltangheis, Arabzadeh, Salamat, Zihayat, and Bagheri]{razavi2025benchmarking}
A.~Razavi, M.~Soltangheis, N.~Arabzadeh, S.~Salamat, M.~Zihayat, and E.~Bagheri, ``Benchmarking prompt sensitivity in large language models,'' in \emph{European Conference on Information Retrieval}.\hskip 1em plus 0.5em minus 0.4em\relax Springer, 2025, pp. 303--313.

\bibitem[Zhuo et~al.(2024)Zhuo, Zhang, Fang, Duan, Lin, and Chen]{zhuo2024prosa}
J.~Zhuo, S.~Zhang, X.~Fang, H.~Duan, D.~Lin, and K.~Chen, ``Prosa: Assessing and understanding the prompt sensitivity of llms,'' in \emph{Findings of the Association for Computational Linguistics: EMNLP 2024}, 2024, pp. 1950--1976.

\bibitem[Rafi et~al.(2026)Rafi, Kim, Chen, and Wang]{rafi2024impact}
M.~N. Rafi, D.~J. Kim, T.-H. Chen, and S.~Wang, ``Order matters! an empirical study on large language models' input order bias in software fault localization,'' in \emph{Proceedings of the 48th IEEE/ACM International Conference on Software Engineering (ICSE)}, 2026, to appear.

\bibitem[Kokane et~al.(2025)Kokane, Zhu, Awalgaonkar, Zhang, Prabhakar, Hoang, Liu, RN, Yang, Yao, et~al.]{kokane2025toolscan}
S.~Kokane, M.~Zhu, T.~M. Awalgaonkar, J.~Zhang, A.~Prabhakar, T.~Q. Hoang, Z.~Liu, R.~RN, L.~Yang, W.~Yao \emph{et~al.}, ``Toolscan: A benchmark for characterizing errors in tool-use llms,'' in \emph{ICLR 2025 Workshop on Building Trust in Language Models and Applications}, 2025.

\bibitem[Hasan et~al.(2025)Hasan, Li, Fallahzadeh, Rajbahadur, Adams, and Hassan]{hasan2025modelcontextprotocolmcp}
\BIBentryALTinterwordspacing
M.~M. Hasan, H.~Li, E.~Fallahzadeh, G.~K. Rajbahadur, B.~Adams, and A.~E. Hassan, ``Model context protocol (mcp) at first glance: Studying the security and maintainability of mcp servers,'' 2025. [Online]. Available: \url{https://arxiv.org/abs/2506.13538}
\BIBentrySTDinterwordspacing

\bibitem[Huang and Hughes(2025)]{huang2025agentic}
K.~Huang and C.~Hughes, ``Agentic ai communication protocols and security,'' in \emph{Securing AI Agents: Foundations, Frameworks, and Real-World Deployment}.\hskip 1em plus 0.5em minus 0.4em\relax Springer, 2025, pp. 81--110.

\bibitem[Lukyanenko et~al.(2024)Lukyanenko, Samuel, Parsons, Storey, Pastor, and Jabbari]{lukyanenko2024universal}
R.~Lukyanenko, B.~M. Samuel, J.~Parsons, V.~C. Storey, O.~Pastor, and A.~Jabbari, ``Universal conceptual modeling: principles, benefits, and an agenda for conceptual modeling research,'' \emph{Software and Systems Modeling}, vol.~23, no.~5, pp. 1077--1100, 2024.

\bibitem[H{\"a}ndler(2023)]{handler2023taxonomy}
T.~H{\"a}ndler, ``A taxonomy for autonomous llm-powered multi-agent architectures.'' in \emph{KMIS}, 2023, pp. 85--98.

\bibitem[Boissier et~al.(2020)Boissier, Bordini, Hubner, and Ricci]{boissier2020multi}
O.~Boissier, R.~H. Bordini, J.~Hubner, and A.~Ricci, \emph{Multi-agent oriented programming: programming multi-agent systems using JaCaMo}.\hskip 1em plus 0.5em minus 0.4em\relax Mit Press, 2020.

\bibitem[Hassan et~al.(2024)Hassan, Lin, Rajbahadur, Gallaba, Cogo, Chen, Zhang, Thangarajah, Oliva, Lin, et~al.]{hassan2024rethinking}
A.~E. Hassan, D.~Lin, G.~K. Rajbahadur, K.~Gallaba, F.~R. Cogo, B.~Chen, H.~Zhang, K.~Thangarajah, G.~Oliva, J.~Lin \emph{et~al.}, ``Rethinking software engineering in the era of foundation models: A curated catalogue of challenges in the development of trustworthy fmware,'' in \emph{Companion Proceedings of the 32nd ACM International Conference on the Foundations of Software Engineering}, 2024, pp. 294--305.

\bibitem[At{\i}l et~al.(2025)At{\i}l, Aykent, Chittams, Fu, Passonneau, Radcliffe, Rajagopal, Sloan, Tudrej, T{\"u}re, et~al.]{atil2024non}
B.~At{\i}l, S.~Aykent, A.~Chittams, L.~Fu, R.~J. Passonneau, E.~Radcliffe, G.~R. Rajagopal, A.~Sloan, T.~Tudrej, F.~T{\"u}re \emph{et~al.}, ``Non-determinism of “deterministic” llm system settings in hosted environments,'' in \emph{Proceedings of the 5th Workshop on Evaluation and Comparison of NLP Systems}, 2025, pp. 135--148.

\bibitem[Krakowski(2025)]{krakowski2025human}
S.~Krakowski, ``Human-ai agency in the age of generative ai,'' \emph{Information and Organization}, vol.~35, no.~1, p. 100560, 2025.

\bibitem[Hexmoor et~al.(2026)Hexmoor, Lammens, Caicedo, and Shapiro]{hexmoor2026behaviour}
H.~Hexmoor, J.~Lammens, G.~Caicedo, and S.~C. Shapiro, \emph{Behaviour based AI, cognitive processes, and emergent behaviors in autonomous agents}.\hskip 1em plus 0.5em minus 0.4em\relax WIT Press, 2026, vol.~1.

\bibitem[Zhu et~al.(2025)Zhu, Terragni, Wei, Cheung, Wu, and Liu]{zhu2025understanding}
H.~Zhu, V.~Terragni, L.~Wei, S.-C. Cheung, J.~Wu, and Y.~Liu, ``Understanding and characterizing mock assertions in unit tests,'' \emph{Proceedings of the ACM on Software Engineering}, vol.~2, no. FSE, pp. 554--575, 2025.

\bibitem[Zamprogno et~al.(2022)Zamprogno, Hall, Holmes, and Atlee]{zamprogno2022dynamic}
L.~Zamprogno, B.~Hall, R.~Holmes, and J.~M. Atlee, ``Dynamic human-in-the-loop assertion generation,'' \emph{IEEE Transactions on Software Engineering}, vol.~49, no.~4, pp. 2337--2351, 2022.

\bibitem[Openja et~al.(2024)Openja, Khomh, Foundjem, Jiang, Abidi, and Hassan]{openja2024empirical}
M.~Openja, F.~Khomh, A.~Foundjem, Z.~M. Jiang, M.~Abidi, and A.~E. Hassan, ``An empirical study of testing machine learning in the wild,'' \emph{ACM Transactions on Software Engineering and Methodology}, 2024.

\bibitem[Lewis et~al.(2020)Lewis, Perez, Piktus, Petroni, Karpukhin, Goyal, K{\"u}ttler, Lewis, Yih, Rockt{\"a}schel, et~al.]{lewis2020retrieval}
P.~Lewis, E.~Perez, A.~Piktus, F.~Petroni, V.~Karpukhin, N.~Goyal, H.~K{\"u}ttler, M.~Lewis, W.-t. Yih, T.~Rockt{\"a}schel \emph{et~al.}, ``Retrieval-augmented generation for knowledge-intensive nlp tasks,'' \emph{Advances in neural information processing systems}, vol.~33, pp. 9459--9474, 2020.

\bibitem[Myers(2006)]{myers2006art}
G.~J. Myers, \emph{The art of software testing}.\hskip 1em plus 0.5em minus 0.4em\relax John Wiley \& Sons, 2006.

\bibitem[Meszaros(2007)]{meszaros2007xunit}
G.~Meszaros, \emph{xUnit test patterns: Refactoring test code}.\hskip 1em plus 0.5em minus 0.4em\relax Pearson Education, 2007.

\bibitem[Van~Rompaey and Demeyer(2008)]{van2008exploring}
B.~Van~Rompaey and S.~Demeyer, ``Exploring the composition of unit test suites,'' in \emph{2008 23rd IEEE/ACM International Conference on Automated Software Engineering-Workshops}.\hskip 1em plus 0.5em minus 0.4em\relax IEEE, 2008, pp. 11--20.

\bibitem[Tao(2009)]{tao2009introduction}
Y.~Tao, ``An introduction to assertion-based verification,'' in \emph{2009 IEEE 8th International Conference on ASIC}.\hskip 1em plus 0.5em minus 0.4em\relax IEEE, 2009, pp. 1318--1323.

\bibitem[Wei et~al.(2022{\natexlab{a}})Wei, Xiao, Yu, Chen, Wang, Wong, and Clune]{wei2022automatically}
C.~Wei, L.~Xiao, T.~Yu, X.~Chen, X.~Wang, S.~Wong, and A.~Clune, ``Automatically tagging the “aaa” pattern in unit test cases using machine learning models,'' in \emph{Proceedings of the 37th IEEE/ACM International Conference on Automated Software Engineering}, 2022, pp. 1--3.

\bibitem[Rao et~al.(1995)Rao, Georgeff, et~al.]{rao1995bdi}
A.~S. Rao, M.~P. Georgeff \emph{et~al.}, ``Bdi agents: from theory to practice.'' in \emph{Icmas}, vol.~95, 1995, pp. 312--319.

\bibitem[Li et~al.(2023)Li, Hammoud, Itani, Khizbullin, and Ghanem]{li2023camel}
G.~Li, H.~Hammoud, H.~Itani, D.~Khizbullin, and B.~Ghanem, ``Camel: Communicative agents for" mind" exploration of large language model society,'' \emph{Advances in Neural Information Processing Systems}, vol.~36, pp. 51\,991--52\,008, 2023.

\bibitem[Liu et~al.(2025)Liu, Lo, Lu, Zhu, Zhao, Xu, Harrer, and Whittle]{liu2025agent}
Y.~Liu, S.~K. Lo, Q.~Lu, L.~Zhu, D.~Zhao, X.~Xu, S.~Harrer, and J.~Whittle, ``Agent design pattern catalogue: A collection of architectural patterns for foundation model based agents,'' \emph{Journal of Systems and Software}, vol. 220, p. 112278, 2025.

\bibitem[Wei et~al.(2025)Wei, Xiao, Yu, Wong, and Clune]{wei2025developers}
C.~Wei, L.~Xiao, T.~Yu, S.~Wong, and A.~Clune, ``How do developers structure unit test cases? an empirical analysis of the aaa pattern in open source projects,'' \emph{IEEE Transactions on Software Engineering}, 2025.

\bibitem[Zhu et~al.(2023)Zhu, Wei, Terragni, Liu, Cheung, Wu, Sheng, Zhang, and Song]{zhu2023stubcoder}
H.~Zhu, L.~Wei, V.~Terragni, Y.~Liu, S.-C. Cheung, J.~Wu, Q.~Sheng, B.~Zhang, and L.~Song, ``Stubcoder: Automated generation and repair of stub code for mock objects,'' \emph{ACM Transactions on Software Engineering and Methodology}, vol.~33, no.~1, pp. 1--31, 2023.

\bibitem[Kampmann and Zeller(2019)]{kampmann2019carving}
A.~Kampmann and A.~Zeller, ``Carving parameterized unit tests,'' in \emph{2019 IEEE/ACM 41st International Conference on Software Engineering: Companion Proceedings (ICSE-Companion)}.\hskip 1em plus 0.5em minus 0.4em\relax IEEE, 2019, pp. 248--249.

\bibitem[Fontes and Gay(2023)]{fontes2023integration}
A.~Fontes and G.~Gay, ``The integration of machine learning into automated test generation: A systematic mapping study,'' \emph{Software Testing, Verification and Reliability}, vol.~33, no.~4, p. e1845, 2023.

\bibitem[Wang et~al.(2024)Wang, Huang, Chen, Liu, Wang, and Wang]{wang2024software}
J.~Wang, Y.~Huang, C.~Chen, Z.~Liu, S.~Wang, and Q.~Wang, ``Software testing with large language models: Survey, landscape, and vision,'' \emph{IEEE Transactions on Software Engineering}, vol.~50, no.~4, pp. 911--936, 2024.

\bibitem[Zhang et~al.(2020)Zhang, Harman, Ma, and Liu]{zhang2020machine}
J.~M. Zhang, M.~Harman, L.~Ma, and Y.~Liu, ``Machine learning testing: Survey, landscapes and horizons,'' \emph{IEEE Transactions on Software Engineering}, vol.~48, no.~1, pp. 1--36, 2020.

\bibitem[Pei et~al.(2017)Pei, Cao, Yang, and Jana]{pei2017deepxplore}
K.~Pei, Y.~Cao, J.~Yang, and S.~Jana, ``Deepxplore: Automated whitebox testing of deep learning systems,'' in \emph{proceedings of the 26th Symposium on Operating Systems Principles}, 2017, pp. 1--18.

\bibitem[Nishi et~al.(2018)Nishi, Masuda, Ogawa, and Uetsuki]{nishi2018test}
Y.~Nishi, S.~Masuda, H.~Ogawa, and K.~Uetsuki, ``A test architecture for machine learning product,'' in \emph{2018 IEEE International Conference on Software Testing, Verification and Validation Workshops (ICSTW)}.\hskip 1em plus 0.5em minus 0.4em\relax IEEE, 2018, pp. 273--278.

\bibitem[Tramer et~al.(2017)Tramer, Atlidakis, Geambasu, Hsu, Hubaux, Humbert, Juels, and Lin]{tramer2017fairtest}
F.~Tramer, V.~Atlidakis, R.~Geambasu, D.~Hsu, J.-P. Hubaux, M.~Humbert, A.~Juels, and H.~Lin, ``Fairtest: Discovering unwarranted associations in data-driven applications,'' in \emph{2017 IEEE European Symposium on Security and Privacy (EuroS\&P)}.\hskip 1em plus 0.5em minus 0.4em\relax IEEE, 2017, pp. 401--416.

\bibitem[Nejadgholi and Yang(2019)]{nejadgholi2019study}
M.~Nejadgholi and J.~Yang, ``A study of oracle approximations in testing deep learning libraries,'' in \emph{2019 34th IEEE/ACM International Conference on Automated Software Engineering (ASE)}.\hskip 1em plus 0.5em minus 0.4em\relax IEEE, 2019, pp. 785--796.

\bibitem[Sahoo et~al.(2021)Sahoo, Zhao, Chen, and Ermon]{sahoo2021reliable}
R.~Sahoo, S.~Zhao, A.~Chen, and S.~Ermon, ``Reliable decisions with threshold calibration,'' \emph{Advances in Neural Information Processing Systems}, vol.~34, pp. 1831--1844, 2021.

\bibitem[Gonzalez et~al.(2020)Gonzalez, Zimmermann, and Nagappan]{gonzalez2020state}
D.~Gonzalez, T.~Zimmermann, and N.~Nagappan, ``The state of the ml-universe: 10 years of artificial intelligence \& machine learning software development on github,'' in \emph{Proceedings of the 17th International conference on mining software repositories}, 2020, pp. 431--442.

\bibitem[Li and Bezemer(2025)]{li2025bridging}
H.~Li and C.-P. Bezemer, ``Bridging the language gap: an empirical study of bindings for open source machine learning libraries across software package ecosystems,'' \emph{Empirical Software Engineering}, vol.~30, no.~1, p.~6, 2025.

\bibitem[Kalliamvakou et~al.(2014)Kalliamvakou, Gousios, Blincoe, Singer, German, and Damian]{kalliamvakou2014promises}
E.~Kalliamvakou, G.~Gousios, K.~Blincoe, L.~Singer, D.~M. German, and D.~Damian, ``The promises and perils of mining github,'' in \emph{Proceedings of the 11th working conference on mining software repositories}, 2014, pp. 92--101.

\bibitem[Munaiah et~al.(2017)Munaiah, Kroh, Cabrey, and Nagappan]{munaiah2017curating}
N.~Munaiah, S.~Kroh, C.~Cabrey, and M.~Nagappan, ``Curating github for engineered software projects,'' \emph{Empirical Software Engineering}, vol.~22, pp. 3219--3253, 2017.

\bibitem[Han et~al.(2019)Han, Deng, Xia, Wang, and Yin]{han2019characterization}
J.~Han, S.~Deng, X.~Xia, D.~Wang, and J.~Yin, ``Characterization and prediction of popular projects on github,'' in \emph{2019 IEEE 43rd annual computer software and applications conference (COMPSAC)}, vol.~1.\hskip 1em plus 0.5em minus 0.4em\relax IEEE, 2019, pp. 21--26.

\bibitem[Zou et~al.(2019)Zou, Zhang, Xia, Holmes, and Chen]{zou2019branch}
W.~Zou, W.~Zhang, X.~Xia, R.~Holmes, and Z.~Chen, ``Branch use in practice: A large-scale empirical study of 2,923 projects on github,'' in \emph{2019 ieee 19th international conference on software quality, reliability and security (qrs)}.\hskip 1em plus 0.5em minus 0.4em\relax IEEE, 2019, pp. 306--317.

\bibitem[Jebnoun et~al.(2020)Jebnoun, Ben~Braiek, Rahman, and Khomh]{jebnoun2020scent}
H.~Jebnoun, H.~Ben~Braiek, M.~M. Rahman, and F.~Khomh, ``The scent of deep learning code: An empirical study,'' in \emph{Proceedings of the 17th International Conference on Mining Software Repositories}, 2020, pp. 420--430.

\bibitem[Okken(2022)]{okken2022python}
B.~Okken, \emph{Python Testing with pytest}.\hskip 1em plus 0.5em minus 0.4em\relax Pragmatic Bookshelf, 2022.

\bibitem[Bodea(2022)]{bodea2022pytest}
A.~Bodea, ``Pytest-smell: a smell detection tool for python unit tests,'' in \emph{Proceedings of the 31st ACM SIGSOFT International Symposium on Software Testing and Analysis}, 2022, pp. 793--796.

\bibitem[Cui et~al.(2010)Cui, Li, Guo, Wang, and Ma]{cui2010code}
B.~Cui, J.~Li, T.~Guo, J.~Wang, and D.~Ma, ``Code comparison system based on abstract syntax tree,'' in \emph{2010 3rd IEEE International Conference on Broadband Network and Multimedia Technology (IC-BNMT)}.\hskip 1em plus 0.5em minus 0.4em\relax IEEE, 2010, pp. 668--673.

\bibitem[Spirin et~al.(2021)Spirin, Bogomolov, Kovalenko, and Bryksin]{spirin2021psiminer}
E.~Spirin, E.~Bogomolov, V.~Kovalenko, and T.~Bryksin, ``Psiminer: A tool for mining rich abstract syntax trees from code,'' in \emph{2021 IEEE/ACM 18th International Conference on Mining Software Repositories (MSR)}.\hskip 1em plus 0.5em minus 0.4em\relax IEEE, 2021, pp. 13--17.

\bibitem[Zhao et~al.(2024)Zhao, Chen, Bangash, Adams, and Hassan]{zhao2024empirical}
Z.~Zhao, Y.~Chen, A.~A. Bangash, B.~Adams, and A.~E. Hassan, ``An empirical study of challenges in machine learning asset management,'' \emph{Empirical Software Engineering}, vol.~29, no.~4, p.~98, 2024.

\bibitem[Chaokromthong et~al.(2021)Chaokromthong, Sintao, et~al.]{chaokromthong2021sample}
K.~Chaokromthong, N.~Sintao \emph{et~al.}, ``Sample size estimation using yamane and cochran and krejcie and morgan and green formulas and cohen statistical power analysis by g* power and comparisions,'' \emph{Apheit international journal of interdisciplinary social sciences and technology}, vol.~10, no.~2, pp. 76--86, 2021.

\bibitem[Barb et~al.(2014)Barb, Neill, Sangwan, and Piovoso]{barb2014statistical}
A.~S. Barb, C.~J. Neill, R.~S. Sangwan, and M.~J. Piovoso, ``A statistical study of the relevance of lines of code measures in software projects,'' \emph{Innovations in Systems and Software Engineering}, vol.~10, no.~4, pp. 243--260, 2014.

\bibitem[Hassan and Rahman(2022)]{hassan2022code}
M.~M. Hassan and A.~Rahman, ``As code testing: Characterizing test quality in open source ansible development,'' in \emph{2022 IEEE Conference on Software Testing, Verification and Validation (ICST)}.\hskip 1em plus 0.5em minus 0.4em\relax IEEE, 2022, pp. 208--219.

\bibitem[Gueron et~al.(2011)Gueron, Johnson, and Walker]{gueron2011sha}
S.~Gueron, S.~Johnson, and J.~Walker, ``Sha-512/256,'' in \emph{2011 Eighth International Conference on Information Technology: New Generations}.\hskip 1em plus 0.5em minus 0.4em\relax IEEE, 2011, pp. 354--358.

\bibitem[Ajibode et~al.(2025)Ajibode, Bangash, Cogo, Adams, and Hassan]{ajibode2025towards}
A.~Ajibode, A.~A. Bangash, F.~R. Cogo, B.~Adams, and A.~E. Hassan, ``Towards semantic versioning of open pre-trained language model releases on hugging face,'' \emph{Empirical Software Engineering}, vol.~30, no.~3, pp. 1--63, 2025.

\bibitem[Spencer(2009)]{spencer2009card}
D.~Spencer, \emph{Card sorting: Designing usable categories}.\hskip 1em plus 0.5em minus 0.4em\relax Rosenfeld Media, 2009.

\bibitem[Zampetti et~al.(2022)Zampetti, Kapur, Di~Penta, and Panichella]{zampetti2022empirical}
F.~Zampetti, R.~Kapur, M.~Di~Penta, and S.~Panichella, ``An empirical characterization of software bugs in open-source cyber--physical systems,'' \emph{Journal of Systems and Software}, vol. 192, p. 111425, 2022.

\bibitem[Nayebi et~al.(2018)Nayebi, Kuznetsov, Chen, Zeller, and Ruhe]{nayebi2018anatomy}
M.~Nayebi, K.~Kuznetsov, P.~Chen, A.~Zeller, and G.~Ruhe, ``Anatomy of functionality deletion: an exploratory study on mobile apps,'' in \emph{Proceedings of the 15th International Conference on Mining Software Repositories}, 2018, pp. 243--253.

\bibitem[Conrad and Tucker(2019)]{conrad2019making}
L.~Y. Conrad and V.~M. Tucker, ``Making it tangible: hybrid card sorting within qualitative interviews,'' \emph{Journal of Documentation}, vol.~75, no.~2, pp. 397--416, 2019.

\bibitem[Meszaros et~al.(2003)Meszaros, Smith, and Andrea]{meszaros2003test}
G.~Meszaros, S.~M. Smith, and J.~Andrea, ``The test automation manifesto,'' in \emph{Conference on extreme programming and agile methods}.\hskip 1em plus 0.5em minus 0.4em\relax Springer, 2003, pp. 73--81.

\bibitem[Gonzalez et~al.(2017)Gonzalez, Santos, Popovich, Mirakhorli, and Nagappan]{gonzalez2017large}
D.~Gonzalez, J.~C. Santos, A.~Popovich, M.~Mirakhorli, and M.~Nagappan, ``A large-scale study on the usage of testing patterns that address maintainability attributes: patterns for ease of modification, diagnoses, and comprehension,'' in \emph{2017 IEEE/ACM 14th International Conference on Mining Software Repositories (MSR)}.\hskip 1em plus 0.5em minus 0.4em\relax IEEE, 2017, pp. 391--401.

\bibitem[Zhang and Mesbah(2015)]{zhang2015assertions}
Y.~Zhang and A.~Mesbah, ``Assertions are strongly correlated with test suite effectiveness,'' in \emph{Proceedings of the 2015 10th Joint Meeting on Foundations of Software Engineering}, 2015, pp. 214--224.

\bibitem[Marchal et~al.(2022)Marchal, Scholman, Yung, and Demberg]{marchal2022establishing}
M.~Marchal, M.~Scholman, F.~Yung, and V.~Demberg, ``Establishing annotation quality in multi-label annotations,'' in \emph{Proceedings of the 29th international conference on computational linguistics}, 2022, pp. 3659--3668.

\bibitem[Parker et~al.(2024)Parker, Anderson, Stone, and Oh]{parker2024large}
M.~J. Parker, C.~Anderson, C.~Stone, and Y.~Oh, ``A large language model approach to educational survey feedback analysis,'' \emph{International journal of artificial intelligence in education}, pp. 1--38, 2024.

\bibitem[Bitew et~al.(2023)Bitew, Deleu, Develder, and Demeester]{bitew2023distractor}
S.~K. Bitew, J.~Deleu, C.~Develder, and T.~Demeester, ``Distractor generation for multiple-choice questions with predictive prompting and large language models,'' in \emph{Joint European Conference on Machine Learning and Knowledge Discovery in Databases}.\hskip 1em plus 0.5em minus 0.4em\relax Springer, 2023, pp. 48--63.

\bibitem[Passonneau(2006)]{passonneau2006measuring}
\BIBentryALTinterwordspacing
R.~Passonneau, ``Measuring agreement on set-valued items ({MASI}) for semantic and pragmatic annotation,'' in \emph{Proceedings of the Fifth International Conference on Language Resources and Evaluation ({LREC}{'}06)}.\hskip 1em plus 0.5em minus 0.4em\relax Genoa, Italy: European Language Resources Association (ELRA), May 2006. [Online]. Available: \url{https://aclanthology.org/L06-1392/}
\BIBentrySTDinterwordspacing

\bibitem[Jiang and Marneffe(2022)]{jiang2022investigating}
N.-J. Jiang and M.-C.~d. Marneffe, ``Investigating reasons for disagreement in natural language inference,'' \emph{Transactions of the Association for Computational Linguistics}, vol.~10, pp. 1357--1374, 2022.

\bibitem[Patil et~al.(2023)Patil, Boit, Gudivada, and Nandigam]{patil2023survey}
R.~Patil, S.~Boit, V.~Gudivada, and J.~Nandigam, ``A survey of text representation and embedding techniques in nlp,'' \emph{IEEE Access}, vol.~11, pp. 36\,120--36\,146, 2023.

\bibitem[Landis and Koch(1977)]{landis1977measurement}
J.~R. Landis and G.~G. Koch, ``The measurement of observer agreement for categorical data,'' \emph{biometrics}, pp. 159--174, 1977.

\bibitem[Bhargava et~al.(2024)Bhargava, Witkowski, Detkov, and Thomson]{bhargava2024prompt}
A.~Bhargava, C.~Witkowski, A.~Detkov, and M.~Thomson, ``Prompt baking,'' \emph{arXiv preprint arXiv:2409.13697}, 2024.

\bibitem[Kim(2017)]{kim2017statistical}
H.-Y. Kim, ``Statistical notes for clinical researchers: Chi-squared test and fisher's exact test,'' \emph{Restorative dentistry \& endodontics}, vol.~42, no.~2, p. 152, 2017.

\bibitem[Thissen et~al.(2002)Thissen, Steinberg, and Kuang]{thissen2002quick}
D.~Thissen, L.~Steinberg, and D.~Kuang, ``Quick and easy implementation of the benjamini-hochberg procedure for controlling the false positive rate in multiple comparisons,'' \emph{Journal of educational and behavioral statistics}, vol.~27, no.~1, pp. 77--83, 2002.

\bibitem[Fujita et~al.(2023)Fujita, Kashiwa, Lin, and Iida]{fujita2023empirical}
S.~Fujita, Y.~Kashiwa, B.~Lin, and H.~Iida, ``An empirical study on the use of snapshot testing,'' in \emph{2023 IEEE International Conference on Software Maintenance and Evolution (ICSME)}.\hskip 1em plus 0.5em minus 0.4em\relax IEEE, 2023, pp. 335--340.

\bibitem[Lam et~al.(2018)Lam, Srisakaokul, Bassett, Mahdian, Xie, Lakshman, and De~Halleux]{lam2018characteristic}
W.~Lam, S.~Srisakaokul, B.~Bassett, P.~Mahdian, T.~Xie, P.~Lakshman, and J.~De~Halleux, ``A characteristic study of parameterized unit tests in. net open source projects,'' in \emph{32nd European Conference on Object-Oriented Programming (ECOOP 2018)}.\hskip 1em plus 0.5em minus 0.4em\relax Schloss Dagstuhl--Leibniz-Zentrum f{\"u}r Informatik, 2018, pp. 5--1.

\bibitem[Wan et~al.(2019)Wan, Xia, Lo, and Murphy]{wan2019does}
Z.~Wan, X.~Xia, D.~Lo, and G.~C. Murphy, ``How does machine learning change software development practices?'' \emph{IEEE Transactions on Software Engineering}, vol.~47, no.~9, pp. 1857--1871, 2019.

\bibitem[Kim and Oh(2025)]{kim2025evaluating}
S.~Kim and D.~Oh, ``Evaluating creativity: can llms be good evaluators in creative writing tasks?'' \emph{Applied Sciences}, vol.~15, no.~6, p. 2971, 2025.

\bibitem[Liu et~al.(2023{\natexlab{b}})Liu, Iter, Xu, Wang, Xu, and Zhu]{liu2023g}
Y.~Liu, D.~Iter, Y.~Xu, S.~Wang, R.~Xu, and C.~Zhu, ``G-eval: Nlg evaluation using gpt-4 with better human alignment,'' in \emph{Proceedings of the 2023 conference on empirical methods in natural language processing}, 2023, pp. 2511--2522.

\bibitem[Es et~al.(2024)Es, James, Anke, and Schockaert]{es2024ragas}
S.~Es, J.~James, L.~E. Anke, and S.~Schockaert, ``Ragas: Automated evaluation of retrieval augmented generation,'' in \emph{Proceedings of the 18th Conference of the European Chapter of the Association for Computational Linguistics: System Demonstrations}, 2024, pp. 150--158.

\bibitem[Zheng et~al.(2023)Zheng, Chiang, Sheng, Zhuang, Wu, Zhuang, Lin, Li, Li, Xing, et~al.]{zheng2023judging}
L.~Zheng, W.-L. Chiang, Y.~Sheng, S.~Zhuang, Z.~Wu, Y.~Zhuang, Z.~Lin, Z.~Li, D.~Li, E.~Xing \emph{et~al.}, ``Judging llm-as-a-judge with mt-bench and chatbot arena,'' \emph{Advances in Neural Information Processing Systems}, vol.~36, pp. 46\,595--46\,623, 2023.

\bibitem[Wei et~al.(2022{\natexlab{b}})Wei, Wang, Schuurmans, Bosma, Xia, Chi, Le, Zhou, et~al.]{wei2022chain}
J.~Wei, X.~Wang, D.~Schuurmans, M.~Bosma, F.~Xia, E.~Chi, Q.~V. Le, D.~Zhou \emph{et~al.}, ``Chain-of-thought prompting elicits reasoning in large language models,'' \emph{Advances in neural information processing systems}, vol.~35, pp. 24\,824--24\,837, 2022.

\bibitem[Confident(2024)]{confident2024deepeval}
A.~Confident, ``Deepeval,'' 2024.

\bibitem[Xu et~al.(2022)Xu, Alon, Neubig, and Hellendoorn]{xu2022systematic}
F.~F. Xu, U.~Alon, G.~Neubig, and V.~J. Hellendoorn, ``A systematic evaluation of large language models of code,'' in \emph{Proceedings of the 6th ACM SIGPLAN International Symposium on Machine Programming}, 2022, pp. 1--10.

\bibitem[Tillmann and Schulte(2005)]{tillmann2005parameterized}
N.~Tillmann and W.~Schulte, ``Parameterized unit tests,'' \emph{ACM SIGSOFT Software Engineering Notes}, vol.~30, no.~5, pp. 253--262, 2005.

\bibitem[Zhang et~al.(2025{\natexlab{a}})Zhang, Dai, Bo, Ma, Li, Chen, Zhu, Dong, and Wen]{zhang2025survey}
Z.~Zhang, Q.~Dai, X.~Bo, C.~Ma, R.~Li, X.~Chen, J.~Zhu, Z.~Dong, and J.-R. Wen, ``A survey on the memory mechanism of large language model-based agents,'' \emph{ACM Transactions on Information Systems}, vol.~43, no.~6, pp. 1--47, 2025.

\bibitem[Bang(2023)]{bang2023gptcache}
F.~Bang, ``Gptcache: An open-source semantic cache for llm applications enabling faster answers and cost savings,'' in \emph{Proceedings of the 3rd Workshop for Natural Language Processing Open Source Software (NLP-OSS 2023)}, 2023, pp. 212--218.

\bibitem[Zhang et~al.(2025{\natexlab{b}})Zhang, Wornow, and Olukotun]{zhang2025cost}
Q.~Zhang, M.~Wornow, and K.~Olukotun, ``Cost-efficient serving of llm agents via test-time plan caching,'' in \emph{ES-FoMo III: 3rd Workshop on Efficient Systems for Foundation Models}, 2025.

\bibitem[Marvin et~al.(2023)Marvin, Hellen, Jjingo, and Nakatumba-Nabende]{marvin2023prompt}
G.~Marvin, N.~Hellen, D.~Jjingo, and J.~Nakatumba-Nabende, ``Prompt engineering in large language models,'' in \emph{International conference on data intelligence and cognitive informatics}.\hskip 1em plus 0.5em minus 0.4em\relax Springer, 2023, pp. 387--402.

\bibitem[Barron et~al.(2024)Barron, Grantcharov, Wanna, Eren, Bhattarai, Solovyev, Tompkins, Nicholas, Rasmussen, Matuszek, et~al.]{barron2024domain}
R.~C. Barron, V.~Grantcharov, S.~Wanna, M.~E. Eren, M.~Bhattarai, N.~Solovyev, G.~Tompkins, C.~Nicholas, K.~{\O}. Rasmussen, C.~Matuszek \emph{et~al.}, ``Domain-specific retrieval-augmented generation using vector stores, knowledge graphs, and tensor factorization,'' in \emph{2024 International Conference on Machine Learning and Applications (ICMLA)}.\hskip 1em plus 0.5em minus 0.4em\relax IEEE, 2024, pp. 1669--1676.

\bibitem[Shen et~al.(2024)Shen, Song, Tan, Li, Lu, and Zhuang]{shen2024hugginggpt}
Y.~Shen, K.~Song, X.~Tan, D.~Li, W.~Lu, and Y.~Zhuang, ``Hugginggpt: Solving ai tasks with chatgpt and its friends in hugging face,'' \emph{Advances in Neural Information Processing Systems}, vol.~36, 2024.

\bibitem[Nascimento et~al.(2023)Nascimento, Alencar, and Cowan]{nascimento2023self}
N.~Nascimento, P.~Alencar, and D.~Cowan, ``Self-adaptive large language model (llm)-based multiagent systems,'' in \emph{2023 IEEE International Conference on Autonomic Computing and Self-Organizing Systems Companion (ACSOS-C)}.\hskip 1em plus 0.5em minus 0.4em\relax IEEE, 2023, pp. 104--109.

\bibitem[Schmidt et~al.(2024)Schmidt, Reddy, Zhang, Alameddine, Uzan, Pinter, and Tanner]{schmidt2024tokenization}
C.~W. Schmidt, V.~Reddy, H.~Zhang, A.~Alameddine, O.~Uzan, Y.~Pinter, and C.~Tanner, ``Tokenization is more than compression,'' in \emph{Proceedings of the 2024 Conference on Empirical Methods in Natural Language Processing}, 2024, pp. 678--702.

\bibitem[Li et~al.(2024)Li, Wang, Zeng, Wu, and Yang]{li2024survey}
X.~Li, S.~Wang, S.~Zeng, Y.~Wu, and Y.~Yang, ``A survey on llm-based multi-agent systems: workflow, infrastructure, and challenges,'' \emph{Vicinagearth}, vol.~1, no.~1, p.~9, 2024.

\bibitem[Labuschagne et~al.(2017)Labuschagne, Inozemtseva, and Holmes]{labuschagne2017measuring}
A.~Labuschagne, L.~Inozemtseva, and R.~Holmes, ``Measuring the cost of regression testing in practice: A study of java projects using continuous integration,'' in \emph{Proceedings of the 2017 11th joint meeting on foundations of software engineering}, 2017, pp. 821--830.

\bibitem[Wu et~al.(2024{\natexlab{b}})Wu, Lin, Dai, Hu, Shu, Ng, Jaillet, and Low]{wu2024prompt}
Z.~Wu, X.~Lin, Z.~Dai, W.~Hu, Y.~Shu, S.-K. Ng, P.~Jaillet, and B.~K.~H. Low, ``Prompt optimization with ease? efficient ordering-aware automated selection of exemplars,'' \emph{Advances in Neural Information Processing Systems}, vol.~37, pp. 122\,706--122\,740, 2024.

\bibitem[Rajbahadur et~al.(2024)Rajbahadur, Oliva, Lin, and Hassan]{rajbahadur2024cool}
G.~K. Rajbahadur, G.~A. Oliva, D.~Lin, and A.~E. Hassan, ``From cool demos to production-ready fmware: Core challenges and a technology roadmap,'' \emph{arXiv preprint arXiv:2410.20791}, 2024.

\bibitem[Huang et~al.(2025)Huang, Chew, Dutkiewicz, and Wang]{huang2025llm}
D.~Huang, S.~Chew, A.~Dutkiewicz, and Z.~Wang, ``Llm-as-a-judge for scalable test coverage evaluation: Accuracy, operational reliability, and cost,'' \emph{arXiv preprint arXiv:2512.01232}, 2025.

\bibitem[Huyen(2024)]{huyen2024ai}
C.~Huyen, \emph{AI Engineering: Building Applications with Foundation Models}.\hskip 1em plus 0.5em minus 0.4em\relax O'Reilly Media, Incorporated, 2024.

\bibitem[Huyen(2025{\natexlab{a}})]{eval2025chip}
\BIBentryALTinterwordspacing
------, ``How to evaluate ai that's smarter than us,'' 2025, accessed: 2026-01-06. [Online]. Available: \url{https://queue.acm.org/detail.cfm?id=3722043}
\BIBentrySTDinterwordspacing

\bibitem[Huyen(2025{\natexlab{b}})]{pitfall2025chip}
\BIBentryALTinterwordspacing
------, ``Common pitfalls when building generative ai applications,'' 2025, accessed: 2026-03-13. [Online]. Available: \url{https://huyenchip.com/2025/01/16/ai-engineering-pitfalls.html}
\BIBentrySTDinterwordspacing

\bibitem[Yoo and Harman(2012)]{yoo2012regression}
S.~Yoo and M.~Harman, ``Regression testing minimization, selection and prioritization: a survey,'' \emph{Software testing, verification and reliability}, vol.~22, no.~2, pp. 67--120, 2012.

\bibitem[Cruciani et~al.(2019)Cruciani, Miranda, Verdecchia, and Bertolino]{cruciani2019scalable}
E.~Cruciani, B.~Miranda, R.~Verdecchia, and A.~Bertolino, ``Scalable approaches for test suite reduction,'' in \emph{2019 IEEE/ACM 41st International Conference on Software Engineering (ICSE)}.\hskip 1em plus 0.5em minus 0.4em\relax IEEE, 2019, pp. 419--429.

\bibitem[Garousi et~al.(2019)Garousi, Felderer, and K{\i}l{\i}{\c{c}}aslan]{garousi2019survey}
V.~Garousi, M.~Felderer, and F.~N. K{\i}l{\i}{\c{c}}aslan, ``A survey on software testability,'' \emph{Information and Software Technology}, vol. 108, pp. 35--64, 2019.

\bibitem[Whyte and Mulder(2011)]{whyte2011mitigating}
G.~Whyte and D.~L. Mulder, ``Mitigating the impact of software test constraints on software testing effectiveness,'' \emph{Electronic Journal of Information Systems Evaluation}, vol.~14, no.~2, pp. pp254--270, 2011.

\bibitem[Pan et~al.(2025)Pan, Arabzadeh, Cogo, Zhu, Xiong, Agrawal, Mao, Shen, Pallerla, Patel, et~al.]{pan2025measuring}
M.~Z. Pan, N.~Arabzadeh, R.~Cogo, Y.~Zhu, A.~Xiong, L.~A. Agrawal, H.~Mao, E.~Shen, S.~Pallerla, L.~Patel \emph{et~al.}, ``Measuring agents in production,'' \emph{arXiv preprint arXiv:2512.04123}, 2025.

\bibitem[Daka and Fraser(2014)]{daka2014survey}
E.~Daka and G.~Fraser, ``A survey on unit testing practices and problems,'' in \emph{2014 IEEE 25th International Symposium on Software Reliability Engineering}.\hskip 1em plus 0.5em minus 0.4em\relax IEEE, 2014, pp. 201--211.

\bibitem[Chandrasekaran et~al.(2023)Chandrasekaran, Cody, McCarthy, Lanus, and Freeman]{chandrasekaran2023test}
J.~Chandrasekaran, T.~Cody, N.~McCarthy, E.~Lanus, and L.~Freeman, ``Test \& evaluation best practices for machine learning-enabled systems,'' \emph{arXiv preprint arXiv:2310.06800}, 2023.

\bibitem[Dobslaw et~al.(2025)Dobslaw, Feldt, Yoon, and Yoo]{dobslaw2025challenges}
F.~Dobslaw, R.~Feldt, J.~Yoon, and S.~Yoo, ``Challenges in testing large language model based software: A faceted taxonomy,'' \emph{arXiv preprint arXiv:2503.00481}, 2025.

\bibitem[Masterman et~al.(2024)Masterman, Besen, Sawtell, and Chao]{masterman2024landscape}
T.~Masterman, S.~Besen, M.~Sawtell, and A.~Chao, ``The landscape of emerging ai agent architectures for reasoning, planning, and tool calling: A survey,'' \emph{arXiv preprint arXiv:2404.11584}, 2024.

\bibitem[Cheng et~al.(2024)Cheng, Zhang, Zhang, Meng, Hong, Li, Wang, Wang, Yin, Zhao, et~al.]{cheng2024exploring}
Y.~Cheng, C.~Zhang, Z.~Zhang, X.~Meng, S.~Hong, W.~Li, Z.~Wang, Z.~Wang, F.~Yin, J.~Zhao \emph{et~al.}, ``Exploring large language model based intelligent agents: Definitions, methods, and prospects,'' \emph{arXiv preprint arXiv:2401.03428}, 2024.

\end{thebibliography}

\appendix
\section{Example of Testing Patterns}\label{sec:appendix-1}
\label{sec:appendix-1-testing-patterns}
\subsection{Structural Patterns}
\subsubsection{Hyperparameter Control}
\begin{lstlisting}[style=mypython,caption={Hyperparameter Control: On line 6 hyperparameter temperature is set at 0.}]
def test_10_concurrent_API_calls(self):
    tools = []
    with open('./data/schemas/get-headers-params.json', 'r') as f:
        tools = ToolFactory.from_openapi_schema(f.read(), {})
    ceo = Agent(name=\'CEO\', tools=tools, instructions="You are an agent that tests concurrent API calls. You must say 'success' if the output contains headers, and 'error' if it does not and **nothing else**.")
    (*@\colorbox{yellow}{agency = Agency([ceo], temperature=0)}@*)
    result = agency.get_completion("Please call PrintHeaders tool TWICE at the same time in a single message. If any of the function outputs do not contains headers, please say 'error'.")
    self.assertTrue(result.lower().count('error') == 0, agency.main_thread.thread_url)
\end{lstlisting}
\subsubsection{Parameterized Testing}
\begin{lstlisting}[style=mypython,
                   caption={Parameterized Testing: different llm models have been passed to the same test function through parameter.},
                   label={lst:custom-kwargs}]
(*@\colorbox{yellow}{@pytest.mark.parametrize('llm', ['transformers:gpt2', 'transformers:facebook/opt-350m'])}@*)
def test_custom_kwargs_transformers(llm):
    """Test if we can pass model specific kwargs."""
    llm = get_llm(llm)
    program = engine("Repeat the following 10 times: Repeat this. Repeat this. Repeat this. Repeat this.{{gen 'completion' max_tokens=4 repetition_penalty=10.0}}", llm=llm)
    executed_program = program()
    assert not executed_program['completion'].startswith(' Repeat this.')
\end{lstlisting}

\subsubsection{Test Double}

\begin{lstlisting}[style=mypython,
                   caption={A testdouble or Mock is setup to replicate the behavior of an MCP server}]
@pytest.mark.asyncio
async def test_mcp_error_handling(self):
    """Test MCP connection error handling."""
    (*@\colorbox{yellow}{with patch('mcp.client.stdio.stdio\_client') as mock\_stdio\_client:}@*)
        mock_stdio_client.side_effect = ConnectionError('Failed to connect to MCP server')
        try:
            async with mock_stdio_client(Mock()) as (read, write):
                pass
            assert False, 'Should have raised ConnectionError'
        except ConnectionError as e:
            assert 'Failed to connect to MCP server' in str(e)
\end{lstlisting}

\subsection{Verification Patterns}
\subsubsection{Assertion Based Testing}\label{appendix-assertion-based-testing}
\begin{lstlisting}[style=mypython,
                   caption={Assertion Based Testing where equality is being verified.}]
def test_init(self, memory):
    (*@\colorbox{yellow}{assert memory.name == 'MyMemory'}@*)
\end{lstlisting}

\subsubsection{DeepEval}\label{appendix-deepeval}
\begin{lstlisting}[style=mypython,
                   caption={DeepEval: Example of probabilistic assertion using DeepEval to verify that the retrieved output contains the expected company name in a resume retrieval agent.}]
def test_relevant_content_retrieved(user_proxy: UserProxyAgent, resume_retriever_agent: ResumeRetriever, job_description: str, company: str) -> None:
    message = "Here is the job description: " + job_description
    chat_outcome = get_chat_outcome(user_proxy, resume_retriever_agent, message)
    assert_test(
        LLMTestCase(
            input=message,
            actual_output=chat_outcome,
            context=[company]
        ),
        [
            GEval(
                name="Inclusion",
                evaluation_params=[
                    LLMTestCaseParams.INPUT,
                    LLMTestCaseParams.ACTUAL_OUTPUT,
                ],
                (*@\colorbox{yellow}{threshold=0.7}@*),
                evaluation_steps=[
                (*@\colorbox{yellow}{f"Check that the output contains the company name `{company}'"}@*)
                (*@\colorbox{yellow}{f"Check that the output does not contain any company name other than '{company}'"}@*)
                ],
            )
        ],
    )
\end{lstlisting}
\subsubsection{Membership Testing}
\begin{lstlisting}[style=mypython,
                   caption={Membership Testing: verifying whether a string or pattern is member of the result.}]
def test_task_guardrail_process_output(task_output):
    guardrail = LLMGuardrail(description='Ensure the result has less than 10 words', llm=LLM(model='gpt-4o'))
    result = guardrail(task_output)
    assert result[0] is False
    (*@\colorbox{yellow}{assert 'exceeding the guardrail limit of fewer than' in result[1].lower()}@*)
    guardrail = LLMGuardrail(description='Ensure the result has less than 500 words', llm=LLM(model='gpt-4o'))
    result = guardrail(task_output)
    assert result[0] is True
    assert result[1] == task_output.raw
\end{lstlisting}
\subsubsection{Mock Assertion}
\begin{lstlisting}[style=mypython,
                   caption={Mock Assertion: verifying whether the a method from the mocked object is invoked while testing the process\_request method.},
                   label={lst:pii-detection}]
async def test_pii_detection_blocking(self):
    """Test that content with PII is blocked"""
    self.mock_comprehend_client.detect_pii_entities.return_value = {'Entities': [{'Type': 'EMAIL', 'Score': 0.99}, {'Type': 'PHONE', 'Score': 0.95}]}
    response = await self.agent.process_request(input_text='Contact me at test@email.com', user_id='test_user', session_id='test_session', chat_history=[])
    self.assertIsNone(response)
    (*@\colorbox{yellow}{\texttt{self.mock\_comprehend\_client.detect\_pii\_entities.assert\_called\_once()}}@*)

\end{lstlisting}
\subsubsection{Negative Test}
\begin{lstlisting}[style=mypython,
                   caption={Negative Test: test for handling wrong start node in Mermaid code},
                   label={lst:mermaid-start-wrong}]
def test_mermaid_code_start_wrong():
     (*@\colorbox{yellow}{\texttt{with pytest.raises(LookupError):}}@*)
        graph1.mermaid_code(start_node=Spam)
\end{lstlisting}
\subsubsection{Snapshot Testing}
\begin{lstlisting}[style=mypython,
                   caption={Snapshot Assertion: markdown slash-command testing by comparing with snapshots.}]
def test_handle_slash_command_markdown():
    io = StringIO()
   assert handle\_slash\_command('/markdown', [], False, Console(file=io), 'default') == (None, False)
    (*@\colorbox{yellow}{\texttt{assert io.getvalue() == snapshot('No markdown output available.\textbackslash n')}}@*)
    messages: list[ModelMessage] = [ModelResponse(parts=[TextPart('[hello](\# hello)'), ToolCallPart('foo', '\{\}')])]
    io = StringIO()
    assert handle_slash_command('/markdown', messages, True, Console(file=io), 'default') == (None, True)
    (*@\colorbox{yellow}{\texttt{assert io.getvalue() == snapshot('Markdown output of last question:\textbackslash n\textbackslash n[hello](\# hello)\textbackslash n')}}@*)
\end{lstlisting}
\subsubsection{Value Range Analysis}
\begin{lstlisting}[style=mypython,
                   caption={Value Range Analysis: verifying whether the output is within a predefined range},
                   label={lst:path-traversal}]
def test_path_traversal_characters():
    filename = '../../etc/passwd'
    sanitized = sanitize_filename(filename)
    assert sanitized.startswith('passwd_')
    (*@\colorbox{yellow}{\texttt{assert len(sanitized) <= MAX\_FILENAME\_LENGTH}}@*)
\end{lstlisting}

% \hao{This is not how we create the appendix. Check the official LaTeX template}

\section{Canonical Agent Architecture}\label{sec:appendix-canonical-architecture}
To create a stable and durable analytical foundation that transcends implementation-specific details, we map the components observed in our dataset to the canonical, multi-agent systems architecture defined by the JaCaMo framework~\citep{boissier2020multi}. This model organizes components into three dimensions: the Agent, the Environment, and the Organization. We supplement this classical model with newer concepts like registries, which have become central to modern agent architectures~\citep{liu2025agent}.

\subsection{Belief Base}
The Belief Base contains the set of information, or beliefs, that an agent holds about itself and its environment at a given time. These beliefs, which may not always be complete or correct, form the agent's knowledge foundation for reasoning and decision-making~\citep{boissier2020multi}. In FM-based agents, this corresponds to memory systems, vector stores, and other persistent data sources.

\subsection{Goal}
A Goal represents a desired state of the world that an agent aims to achieve. Goals are the primary drivers of an agent's behavior, guiding its deliberation and planning processes to move from its current state to a desired one~\citep{masterman2024landscape}.

\subsection{Plan}
A Plan is a course of action that an agent can execute to achieve a goal. A plan is typically composed of a trigger, a context, and a body.

\subsection{Trigger}
The triggering event that activates a plan. In FM-based agents, the initial user request or prompt is the primary trigger for invoking a plan~\citep{boissier2020multi}.

\subsection{Context}
Supplemental, just-in-time information that helps an agent select the most appropriate plan from a set of available options. This is analogous to the documents and data supplied in Retrieval-Augmented Generation (RAG) to ground an FM's response~\citep{boissier2020multi}.

\subsection{Plan Body}
The sequence of actions or reasoning steps that constitute the plan. In modern agents, the plan body is often dynamically generated by an FM, which uses its reasoning capabilities to break down a high-level goal into concrete, executable steps~\citep{cheng2024exploring, shen2024hugginggpt}.

\subsection{Internal Action}
A computation performed within the agent's own boundary, such as updating a counter or transforming data, with no external effect~\citep{boissier2020multi}.

\subsection{External Action}
An operation that affects the agent's external environment, such as making an API call, executing code, or controlling a physical device~\citep{boissier2020multi}.

\subsection{Communicative Action}
A message exchanged between agents or between an agent and a user. These actions are fundamental to coordination and collaboration in multi-agent systems~\citep{boissier2020multi}.

\subsection{Resource Artifact}
Resource Artifacts are passive, reusable entities in the environment that agents can utilize to perform their tasks. Examples include shared resources like a database connection, a file, or a wrapper for an external API or tool~\citep{boissier2020multi}.

\subsection{Coordination Artifact}
Coordination Artifacts are entities specifically designed to manage and synchronize the interactions among multiple agents. Examples include shared data structures like message queues, blackboards, or event buses that facilitate orderly collaboration~\citep{boissier2020multi}.

\subsection{Boundary Artifact}
Boundary Artifacts are components that serve as a bridge between the agent system and the outside world. They enable agents to interact with external entities that are not part of the multi-agent system, such as a human user via a graphical user interface (GUI) or another software system~\citep{boissier2020multi}.

\subsection{Observable Property \& Signal}
These are mechanisms through which an agent perceives the state of an artifact. An Observable Property is a piece of data representing an artifact's state, which an agent can read. A Signal is an event emitted by an artifact to notify observing agents of a significant change or occurrence~\citep{boissier2020multi}. In modern systems, these correspond to logs, metrics, and return values.

\subsection{Registry}
While not part of the original JaCaMo model, the Registry has emerged as a critical component in modern agent ecosystems~\citep{liu2025agent}. A Registry is a centralized catalog where developers can publish, discover, and reuse agents or tools. For instance, MCP registries allow for the standardized discovery and invocation of tools across different frameworks, promoting interoperability and reuse~\citep{hasan2025modelcontextprotocolmcp}.

\end{document}